\documentstyle[12pt]{article}
\newcommand{\be}{\begin{equation}}
\newcommand{\ee}{\end{equation}}
\newcommand{\bea}{\begin{eqnarray}}
\newcommand{\eea}{\end{eqnarray}}
\newcommand{\dbar}{d \hspace{-0.1em} \bar{ } \hspace{.1em}}
\newcommand{\eq}[1]{eq.(\ref{#1})}
\newcommand{\rom}{{[\omega]}}
\newcommand{\vom}{\vec{\omega}}
\newcommand{\om}{\omega}
\newcommand{\bo}{\varphi_\rom}
\newcommand{\hbo}{\hat{\varphi}_\rom}
\newcommand{\jom}[1]{{[\omega_{#1}]}}

\def\psmode{"}                       

\def\pscircle(#1,#2,#3,#4){\put(0,0){\special{\psmode
1 1 scale #4 setlinewidth
newpath #1 #2 #3 0 360 arc stroke}}}
\def\crossoncircle(#1,#2,#3,#4,#5,#6){\put(0,0){\special{\psmode
1 1 scale #6 setlinewidth
#1 #4 cos #3 mul add #2 #4 sin #3 mul add translate #4 rotate
#5 2 div neg dup moveto #5 #5 rlineto
#5 2 div dup neg moveto #5 neg #5 rlineto stroke}}}
\input{epsf.sty}
\def\axowidth{0.5 }
\def\axoscale{1.0 }
\def\axoxoff{0 }
\def\axoyoff{0 }
\def\axoxo{0 }
\def\axoyo{0 }
\def\firstcall{1}
\def\Gluon(#1,#2)(#3,#4)#5#6{
\put(\axoxoff,\axoyoff){
}
}
\def\Photon(#1,#2)(#3,#4)#5#6{
\put(\axoxoff,\axoyoff){
}
}
\def\ZigZag(#1,#2)(#3,#4)#5#6{
\put(\axoxoff,\axoyoff){
}
}
\def\PhotonArc(#1,#2)(#3,#4,#5)#6#7{
\put(\axoxoff,\axoyoff){
}
}
\def\GlueArc(#1,#2)(#3,#4,#5)#6#7{
\put(\axoxoff,\axoyoff){
}
}
\def\ArrowArc(#1,#2)(#3,#4,#5){
\put(\axoxoff,\axoyoff){
}
}
\def\LongArrowArc(#1,#2)(#3,#4,#5){
\put(\axoxoff,\axoyoff){
}
}
\def\DashArrowArc(#1,#2)(#3,#4,#5)#6{
\put(\axoxoff,\axoyoff){
}
}
\def\ArrowArcn(#1,#2)(#3,#4,#5){
\put(\axoxoff,\axoyoff){
}
}
\def\LongArrowArcn(#1,#2)(#3,#4,#5){
\put(\axoxoff,\axoyoff){
}
}
\def\DashArrowArcn(#1,#2)(#3,#4,#5)#6{
\put(\axoxoff,\axoyoff){
}
}
\def\ArrowLine(#1,#2)(#3,#4){
\put(\axoxoff,\axoyoff){
}
}
\def\LongArrow(#1,#2)(#3,#4){
\put(\axoxoff,\axoyoff){
}
}
\def\DashArrowLine(#1,#2)(#3,#4)#5{
\put(\axoxoff,\axoyoff){
}
}
\def\Line(#1,#2)(#3,#4){
\put(\axoxoff,\axoyoff){
}
}
\def\DashLine(#1,#2)(#3,#4)#5{
\put(\axoxoff,\axoyoff){
}
}
\def\CArc(#1,#2)(#3,#4,#5){
\put(\axoxoff,\axoyoff){
}
}
\def\DashCArc(#1,#2)(#3,#4,#5)#6{
\put(\axoxoff,\axoyoff){
}
}
\def\Vertex(#1,#2)#3{
\put(\axoxoff,\axoyoff){
}
}
\def\BCirc(#1,#2)#3{
\put(\axoxoff,\axoyoff){
}
}
\def\GCirc(#1,#2)#3#4{
\put(\axoxoff,\axoyoff){
}
}
\def\EBox(#1,#2)(#3,#4){
\put(\axoxoff,\axoyoff){
}
}
\def\BBox(#1,#2)(#3,#4){
\put(\axoxoff,\axoyoff){
}
}
\def\GBox(#1,#2)(#3,#4)#5{
\put(\axoxoff,\axoyoff){
}
}
\def\Boxc(#1,#2)(#3,#4){
\put(\axoxoff,\axoyoff){
}
}
\def\BBoxc(#1,#2)(#3,#4){
\put(\axoxoff,\axoyoff){
}
}
\def\GBoxc(#1,#2)(#3,#4)#5{
\put(\axoxoff,\axoyoff){
}
}

\def\SetOffset(#1,#2){\def\axoxoff{#1 } \def\axoyoff{#2 }}
\def\SetScaledOffset(#1,#2){\def\axoxo{#1 } \def\axoyo{#2 }}
\def\pfont{Times-Roman }
\def\fsize{10 }
\makeatletter
\def\fmode{4 }
\def\@l@{l} \def\@r@{r} \def\@t@{t} \def\@b@{b}
\def\mymodetest#1{\ifx#1\end \let\next=\relax \else {
\if#1\@r@\global\def\fmodeh{-3 }\fi
\if#1\@l@\global\def\fmodeh{3 }\fi
\if#1\@b@\global\def\fmodev{-1 }\fi
\if#1\@t@\global\def\fmodev{1 }\fi
} \let\next=\mymodetest\fi \next}
\makeatother
\def\PText(#1,#2)(#3)[#4]#5{
\def\fmodev{0 }
\def\fmodeh{0 }
\mymodetest#4\end
\put(\axoxoff,\axoyoff){\makebox(0,0)[]{\special{"/\pfont findfont \fsize
 scalefont setfont #1 \axoxo add #2 \axoyo add #3
\fmode \fmodev add \fmodeh add \fsize (#5) \axoscale ptext }}}
}
\def\GOval(#1,#2)(#3,#4)(#5)#6{
\put(\axoxoff,\axoyoff){
}
}
\def\Oval(#1,#2)(#3,#4)(#5){
\put(\axoxoff,\axoyoff){
}
}
\let\eind=]

\def\kromme(#1,#2)#3{#1 \axoxo add #2 \axoyo add \ifx #3\eind\else
\expandafter\kromme\fi#3}
\def\LogAxis(#1,#2)(#3,#4)(#5,#6,#7,#8){
\put(\axoxoff,\axoyoff){
}
}
\def\LinAxis(#1,#2)(#3,#4)(#5,#6,#7,#8,#9){
\put(\axoxoff,\axoyoff){
}
}
\input rotate.tex
\makeatletter
\def\Text(#1,#2)[#3]#4{
\dimen0=\axoxoff \unitlength
\dimen1=\axoyoff \unitlength
\advance\dimen0 by #1 \unitlength
\advance\dimen1 by #2 \unitlength
\makeatletter
\@killglue\raise\dimen1\hbox to \z@{\kern\dimen0\makebox(0,0)[#3]{#4}\hss}
\ignorespaces
\makeatother
}
\def\rText(#1,#2)[#3][#4]#5{
\ifnum\firstcall=1\global\def\firstcall{0}\rText(-10000,#2)[#3][]{#5}\fi
\dimen2=\axoxoff \unitlength
\dimen3=\axoyoff \unitlength
\advance\dimen2 by #1 \unitlength
\advance\dimen3 by #2 \unitlength
\@killglue\raise\dimen3\hbox to \z@{\kern\dimen2
\makebox(0,0)[#3]{
\ifx#4l{\setbox3=\hbox{#5}\rotl{3}}\else{
\ifx#4r{\setbox3=\hbox{#5}\rotr{3}}\else{
\ifx#4u{\setbox3=\hbox{#5}\rotu{3}}\else{#5}\fi}\fi}\fi}\hss}
\ignorespaces
}
\makeatother
\def\BText(#1,#2)#3{
\put(\axoxoff,\axoyoff){
}
}
\def\GText(#1,#2)#3#4{
\put(\axoxoff,\axoyoff){
}
}
\def\B2Text(#1,#2)#3#4{
\put(\axoxoff,\axoyoff){
}
}
\def\G2Text(#1,#2)#3#4#5{
\put(\axoxoff,\axoyoff){
}
}
\newdimen\rotdimen
\def\vspec#1{\special{ps:#1}}
\def\rotstart#1{\vspec{gsave currentpoint currentpoint translate
   #1 neg exch neg exch translate}}
\def\rotfinish{\vspec{currentpoint grestore moveto}}
\def\rotr#1{\rotdimen=\ht#1\advance\rotdimen by\dp#1%
   \hbox to\rotdimen{\hskip\ht#1\vbox to\wd#1{\rotstart{90 rotate}%
   \box#1\vss}\hss}\rotfinish}
\def\rotl#1{\rotdimen=\ht#1\advance\rotdimen by\dp#1%
   \hbox to\rotdimen{\vbox to\wd#1{\vskip\wd#1\rotstart{270 rotate}%
   \box#1\vss}\hss}\rotfinish}%
\def\rotu#1{\rotdimen=\ht#1\advance\rotdimen by\dp#1%
   \hbox to\wd#1{\hskip\wd#1\vbox to\rotdimen{\vskip\rotdimen
   \rotstart{-1 dup scale}\box#1\vss}\hss}\rotfinish}%
\def\rotf#1{\hbox to\wd#1{\hskip\wd#1\rotstart{-1 1 scale}%
   \box#1\hss}\rotfinish}%

\begin{document}

\baselineskip 6 mm

\begin{centering}
{\bf{\Huge
{Bosonization of Fermi Liquids}}}
\vskip 1cm
J. Fr\"{o}hlich$^1$, R. G\"{o}tschmann$^{1,2}$\\
\vskip 2cm
\end{centering}
\begin{enumerate}
\item[$^1$] Theoretical Physics, ETH-H\"{o}nggerberg, CH-8093
Z\"{u}rich
\item[$^2$] Institut de Physique Th\'{e}orique, Universit\'{e} de Fribourg 
CH-1700 Fribourg
\end{enumerate}
\date{ }
\vskip 1cm
\begin{abstract}
We consider systems of non-relativistic, interacting electrons at finite 
density and zero temperature in  $d = 2, 3, ...$ dimensions. Our main concern
is to characterize those systems that, under the renormalization flow, are 
driven away from the Landau Fermi liquid (LFL) renormalization group fixed 
point. We are especially interested in understanding under what circumstances 
such a system is a Marginal Fermi liquid (MFL)  when the dimension of 
space is $ d \geq 2 .$ 
The interacting electron system is analyzed by {\sl combining} renormalization 
group (RG) methods with so called "Luther-Haldane" bosonization techniques. T
RG calculations are organized as a double expansion in the inverse scale 
parameter, $\lambda^{-1} ,$ which is proportional to the width of the effective
momentum space around the Fermi surface and in the running coupling constant, 
$ g_\lambda , $ which measures the strength of electron interactions at energy 
scales $ \sim \frac{v_F k_F}{\lambda} .$
For systems with a strictly convex Fermi surface, superconductivity is the only
symmetry breaking instability. Excluding such an instability,  the system can 
be analyzed by means of bosonization. The RG and the underlying perturbation 
expansion in powers of $ \lambda^{-1} $ serve to characterize the 
{\sl approximations} involved by bosonizing the system.
We argue that systems with short-range interactions flow to the LFL fixed 
point. Within the approximations involved by bosonization, the same holds for 
systems with long-range, longitudinal, density-density interactions.
For electron systems interacting via long-range, transverse, current-current 
interactions a deviation from LFL behaviour is possible : if the exponent 
$\alpha$ parametrizing the singularity of the interaction potential in momentum
space by $\;\hat{V}(|\vec{p}|) \sim \frac{1}{|\vec{p}|^\alpha}\;$ is greater 
than or equal to $ d-1 ,$ the results of the bosonization calculation  are 
consistent with a MFL. 
                      
\end{abstract}

\pagebreak

\section{Introduction}

In this paper, we consider systems of non-relativistic electrons at finite 
density and zero temperature in  $d = 2, 3, ...$ dimensions. The interactions 
between electrons are described by two-body potentials or by current-current 
interactions.
The Cooper channel which drives the BCS instability is turned off; (e.g. by 
assuming that the Fermi sphere of the non-interacting system has a suitable 
geometry \cite{1,19}).
\\
Our main concern in this paper is to characterize those systems that, under the
renormalization flow, are driven away from the Landau Fermi liquid (LFL)
renormalization group (RG) fixed point. 
More concretely, we are interested in understanding under 
what circumstances such a system is a Marginal Fermi liquid (MFL)  when the 
dimension of space is $ d \geq 2 .$ This problem comes up, for example, in the 
study of single-layer quantum Hall fluids at filling fractions $ \nu = \frac{1}
{2},\frac{1}{4}, ... $  and, perhaps, in the theoretical description of 
materials related to anisotropic $ \mbox{HT}_c $ superconductors 
(see \cite{2,3,4,5} and references therein) . \\
A one-dimensional MFL (or Luttinger liquid)  at zero temperature can be
 characterized by the 
property that the electron propagator falls off like $ |\vec{x}|^{-(1+\eta)} ,$
at large distances $ |\vec{x}| ,$ for an exponent $ \eta > 0 $ that depends on 
the electron-electron interaction and characterizes a new  RG fixed point. 
When
$ \eta $ vanishes the system is a LFL . In the limit of large
distance scales and low frequencies (scaling limit) the properties of a LFL
 are identical to those of a free system of non-interacting electrons (up
to a renormalization of the residue of the one-particle pole and of the Fermi 
velocity). In one
dimension, MFL correspond to a line of RG fixed points in the space of
effective Hamiltonians (or actions) containing the fixed point corresponding to
the free system.
\\
In $ d \geq 2 $ dimensions, we define a MFL to be a Fermi liquid
with an electron propagator falling off more rapidly than the free electron
propagator by at least a fractional inverse power of the distance between the 
arguments, but not exponentially in the distance.  Contrary to the 
superconducting instability, the instabilities leading to a MFL are not 
accompanied by symmetry breaking, and there is no energy gap in the excitation 
spectrum of
such systems. Yet,  Landau's picture of non-interacting quasi-particles does not
apply to the physics of Luttinger liquids.\\
Historically, MFL were discovered in the context of one-dimensional
systems of interacting electrons \cite{6}, where they appear naturally, for a 
large class of two-body  interactions.
The experimental observation that the normal phase of anisotropic 
$ \mbox{HT}_c $ superconductors exhibits many 
non-conventional features, incompatible with
LFL theory,  leads to the question whether electron-electron interactions can 
drive a system in {\sl two} or {\sl more} dimensions away from LFL's. The same 
question arises in the context of  single-layer QH fluids at filling fractions 
$\nu$ with even denominators.
\\
Recently, there have been many investigations of this problem. Besides numerical
studies, two analytic techniques proved useful in attempts to understand it:
the RG method, involving an expansion in an inverse scale parameter $ \lambda
^{-1} $ \cite{7}, which has led to a wealth of rigorous results 
\cite{1,19,7,8}, and the bosonization techniques \cite{9} proposed, in this 
context, by Haldane \cite{10}.\\
In this paper, we study the stability and instability of the LFL by using the 
same
two methods. We propose to combine them to clarify the picture of Fermi liquids.
 We implement the bosonization technique in a manner
elucidating its standing in terms of the fermionic perturbation expansion and 
rendering
the calculation of the electron propagator at large distance scales quite 
transparent.
Our techniques apply to a broad class of two-body interactions. (We presented a
preliminary account of our results at the 1994 Les Houches summer school
"Fluctuating Geometries in Statistical Mechanics and Field Theories", \cite{15})

\noindent Next, we summarize the contents of the various sections of this paper.

In Sect.2, we review the RG method for non-relativistic electron systems at 
finite density
and zero temperature in a $d+1$-dimensional, euclidean space-time, with 
$ d > 1 .$
The underlying perturbation theory is organized as a double expansion  in an 
inverse
 scale parameter $\lambda^{-1} $ and  in the (dimensionless) running coupling
constant, $ g_\lambda ,$ of the two-body interactions. \\
The inverse scale parameter   $\lambda^{-1}$ is proportional to the width of a 
shell
$ \Omega_\lambda $ around the Fermi surface. Having integrated out the electron
 modes
with momenta lying outside the shell $ \Omega_\lambda ,$ the effective action, 
$ S_\lambda ,$
for the remaining modes describes the physical properties of the electron 
system at distance
scales $ \sim \frac{\lambda}{k_F} , $ or energy scales $ \sim \frac{v_F k_F}
{\lambda} ,\,$
$k_F$ and $v_F$ being the Fermi momentum and velocity, resp. .
The running coupling constant $g_\lambda$ measures the strength of the two-body
interactions  in $ S_\lambda .$\\
Given $ S_\lambda ,$ one further reduces the effective momentum space around 
the Fermi
surface  to $ \Omega_{\lambda'} ,$ with $ \lambda' = M \lambda\,  ,\, M>1\, ,$ 
and calculates the corresponding effective action $ S_{\lambda'} .$ This calculation is organized
as a perturbation expansion in the parameters $ g_\lambda $ and $ \lambda^{-1} ,$
assuming that they are small. Rescaling the resulting system by a factor $M ,$ we obtain a
system that is similar to the initial one, and we can compare the relevant and marginal
parameters characterizing the corresponding effective actions. The RG method consists in
executing these transformations iteratively, with the aim of deriving the scaling limit of the system,
as $\lambda \rightarrow \infty .$\\
The underlying calculations can be interpreted in a transparent way by decomposing the
shell $ \Omega_\lambda $ into $ N \sim \lambda^{d-1} $ boxes $\, B_{\om_i}(\lambda) \, ,
\, i=1, ... ,N \, ,$ with sides of length $ \sim \frac{1}{\lambda} .$ The d-dimensional
unit vectors $ \,\vom_i  \, , \, i=1, ... ,N \, ,$ point to the centers of the corresponding
boxes. This decomposition implies a decomposition of the electron field $ \Psi^\sharp $
(where  $ \Psi^\sharp $ stands for  $ \Psi^\ast $ or $ \Psi $) into $N$ components
$ \;\psi^\sharp_{\om_i} \, , \, i=1, ... ,N \, ,$ whose Fourier modes have support in the
corresponding boxes. A component field $ \psi^\sharp_{\om_i} $ describes quasi
$1+1$-dimensional, "relativistic" electrons moving along the direction $ \vom_i $ with
velocity $ v_F .$ The  decomposition of the electron field into such component fields reproduces
the electron propagator to leading order in an expansion in $\frac{1}{\lambda} .$
In the non-interacting system,  the  subsystems along the different directions are independent.
Two-body interactions couple the subsystems. After rescaling, each
interaction process is suppressed by a factor $ \frac{1}{N} .$ Hence,  the
 interacting system can be analysed by using  "large-$N$" expansion techniques, \cite{20,7,12}.
\\
For systems with a strictly convex Fermi surface, superconductivity is shown to be the only
symmetry-breaking instability that can develop in the system. We propose to study electron
systems which do not undergo such a symmetry breakdown -- i.e. whose Cooper channel is
turned off -- by means of "bosonization".  The framework of the RG will serve  to characterize
the approximations involved by "bosonizing" the system.

In Sect.3, we introduce the bosonization technique by calculating the scaling limit of
the effective gauge field action (the generating
function of connected  current Green functions) for a non-interacting electron system.  
The effective
gauge field action is obtained by coupling an external gauge field to the electron system
and  by integrating out the fermionic degrees of freedom. \\
The calculation
of the scaling limit of the effective gauge field action is reduced to the calculation of the
gauge field action of a family of independent Schwinger models, 
because the $d$-dimensional, non-interacting
electron system decomposes -- to leading order in an expansion in the inverse scale 
parameter $ \lambda^{-1} $ -- into independent subsystems of quasi $1+1$ dimensional,
"relativistic" fermions, one along each direction $ [\om] = \{\vom,-\vom\} .$ \\
Gauge invariance must hold for each subsystem -- i.e.,  in each direction --
separately, and implies   local conservation laws for the associated quasi $1+1$ dimensional
current densities $ j^{A}_{[\om]}\, , A=0,1\, .$ These conservation laws imply that each
current density $ j^{A}_{[\om]} $ can be expressed as a derivative of a bosonic field $ \bo .$
It turns out that, for the non-interacting system, these fields are massless and gaussian, 
propagating only along the direction $  \rom .$\\
By taking the effective action $ S_\lambda $ determined in the context of an RG analysis as an 
input, we can apply the bosonization technique to interacting systems, too. For systems whose
Cooper channel is turned off, the electron-electron interactions are described -- to leading
order in the inverse scale parameter $ \lambda^{-1} $ -- by an expression quadratic in the 
currents $ j^{A}_{[\om]} .$ By replacing the fermionic currents $ j^{A}_{[\om]} $ by the
corresponding bosonized expressions, we obtain a gaussian bosonic theory.
Because the theory is gaussian, it can be studied more easily than the original fermionic theory
where the interaction term is quartic in the electron fields.\\
In the calculation of the effective gauge field action, this technique is shown to reproduce the leading
order of a (fermionic) perturbation expansion in powers of $ \lambda^{-1} .$ The leading contributions
coincide with the ones of an RPA approximation. We  determine the
 scaling limit of
the effective gauge field action for systems with short range interactions and discuss the
extension of the bosonization method to systems with long range interactions in three examples :
systems with longitudinal density-density or transversal current-current interactions, and
 "Tomographic Luttinger Liquids".

In Sect.4, we apply the bosonization technique to the calculation of the electron propagator at 
large
distance and time scales. We start from an effective action $ S_\lambda $ at energy scales
$ \sim \frac{v_F k_F}{\lambda} $ and  the corresponding decomposition of the electron fields
$ \Psi^\sharp $ into $N$  quasi $1+1$ dimensional components $\psi^\sharp_\om .$ 
We replace the action $ S_\lambda $ by its bosonized version,  introduced in Sect.3, and , 
for each ray
$ \rom = \{\vom,-\vom\} $ , we express the pair $ \psi_\om^\sharp, \psi_{-\om}
^\sharp $ in terms of a bose field $ \bo .$   This is accomplished by applying the well-known 
bosonization formalism for $1+1$ dimensional relativistic fermions . 
However, one has to cope with a subtlety arising from
the dependence of the  quasi  $1+1$ dimensional electron fields $ \psi^\sharp
_\om $  on the components of the momentum  perpendicular to the direction $ \rom $ . 
Again, the approximations involved can be characterized in the context of a (formal)
perturbation expansion in powers of the inverse scale parameter $\lambda^{-1}.$
Special care is taken to discuss the implications of linearizing the Fermi surface inside the
boxes $ B_\om(\lambda) .$\\
We verify that the electron propagator of a system with short-range interactions tends, for
large arguments, to the standard Landau Liquid form. Because of screening -- which is reproduced
by bosonization -- the same result holds for systems with
long-range density-density interactions.\\
For systems of electrons interacting via  long-range, transverse
current-current interactions, we find the possibility for a deviation from LFL behaviour,
depending on the exponent $\alpha$ which characterizes the singularity of the interaction
potential in momentum space  $\, \hat{V}(\vec{p}) \sim \frac{1}{|\vec{p}|^\alpha}\; .$
The critical value for $\alpha$ is $d-1 .$ For $0 < \alpha < d-1 $ and $ \lambda \rightarrow 
\infty ,$ the electron propagator has the standard LFL form, whereas, for
 $d-1 \leq \alpha \leq 2 $ and $ \lambda \rightarrow \infty ,$we argue that it shows
 MFL behaviour.
In the second case, in order to obtain a theory which is form-invariant under scale 
transformations, the parameters
of the effective action $S_\lambda ,$ used as an input of bosonization, must be functions of the
scale parameter $\lambda .$ A resulting consistency condition determines the expected
flow of the parameters of $S_\lambda$ under RG transformations.\\
Our analysis yields a  view somewhat complementary to those arrived at in \cite{2,3,4,5}.\\
\vspace{0.5cm}\\
\noindent {\bf Acknowledgements.} We thank P.-A. Marchetti and M. Seifert for many
helpful discussions.\\
\vspace{.5cm}

\section{Effective Action on Large Scales and Fermionic Perturbation Theory}

The action of a system of non-interacting, non-relativistic electrons in
$d+1$ space-time dimensions is given by
$$
S^{0}\left(\Psi^{\ast},\Psi;\mu\right) \;=\; \int dt \, d^dx\;\bigg[\; 
\Psi^{\ast}(\vec{x},t)
\left( i \partial_0 + \mu\right) \Psi(\vec{x},t) \hspace{4cm}
$$
\be
\hspace{7cm}- \; \frac{1}{2m}
 \sum_{\ell = 1}
^{d} \left( \partial_\ell\Psi(\vec{x},t)\right)^\ast \partial_\ell\Psi(\vec{x},t)\; \bigg]
\quad ,
\label{1}
\ee
where $m$ denotes the bare electron mass and the chemical potential, $\mu$,
specifies the mean electron density $n^0_\mu$. We choose units such that $ \hbar 
 = 1  .$ \\
In a functional integral quantization, the electrons are described by two independent,
two-component Grassmann fields $\;\Psi(\vec{x},t) \,=\,
\Big(
\begin{array}{c}
\psi_{\uparrow}(\vec{x},t) \\
\psi_{\downarrow}(\vec{x},t)
\end{array}
\Big)\;$ and 
$\;\Psi^\ast(\vec{x},t) \,=\,
\Big(
\begin{array}{cc}
\psi^\ast_{\uparrow}(\vec{x},t) &
\psi^\ast_{\downarrow}(\vec{x},t)
\end{array}
\Big)\;.$
The arrows $\;\uparrow,\downarrow\;$ stand for "spin up" and "spin down".

An ionic background and interactions between  electrons are described
 by additional terms in the action  of the form
$$
 S^{U} \quad   = \quad -\int dt\,d^dx\, U(x) \Psi^\ast(x) \Psi(x) 
$$
and
$$
 S^{V} \quad = \quad -  
     \frac{1}{2}\int dt\,d^dx \int ds\,d^dy\, \Psi^\ast(x) \Psi(x) 
       V(\vec{x}-\vec{y})\;\delta(t-s)\; \Psi^\ast(y) \Psi(y)\quad,
$$
resp.,  where $\,U\,$ is a one-body potential and $\,V\,$ a two-body potential.
The total action is given by 
\be
  S \;= \;S^0 \;+\; S^{U}\; +\; S^{V}\quad.
\label{2}
\ee
Expectation values of functionals, $\,{\cal F}(\Psi^\ast,\Psi)\,,$ of the $\Psi$
and $\Psi^\ast$
fields are calculated by evaluating  Berezin-Grassmann integrals :
\bea
\langle\;{\cal F}(\Psi^\ast,\Psi)\;\rangle_{\mu} 
&= &
\frac{\int {\cal{D}}(\Psi^\ast,\Psi)\;e^{ i  S(\Psi^\ast,\Psi\, ;\, \mu)}\;
{\cal F}(\Psi^\ast,\Psi)}{ \int {\cal{D}}(\Psi^\ast,\Psi)\;e^{  i  S(\Psi^\ast,\Psi ; \mu)}}
\quad.
\label{3}
\eea
 We work in
the Grand Canonical Ensemble, where the chemical potential $\mu$ is held fixed. This 
implies that, in general, the mean electron density $ n_{\mu} $ of the interacting
 system does not coincide with the mean electron density $ n^0
_{\mu} $  of the free system. \\
In order to calculate expectation values of the form in \eq{3}, the total action $S$
is split into a quadratic part, $ S^{2} = S^0 + S^{U} $, and a quartic part $S^{V}$.
One expands the exponential $e^{ i S^{V}}$ in a power series and calculates
the expectation values of the resulting polynomials in $\Psi^\ast$ and $\Psi ,$ with
 respect to the Gaussian Berezin integration  determined by $ S^{2} ,$  by using 
Wick's theorem.
 \\In this paper  we consider only systems  invariant under translations
and rotations of space, i.e., the background is described by a constant potential $
U(x) \equiv U $  that can be absorbed in a redefinition of the chemical potential
("jellium model"). 

For technical convenience, we analyse the system in the euclidean region reached
by analytic continuation in the time $t$ to the halfplane $ \Im m[t] > 0$ and setting 
$\, x_0 = i t \,$ (Wick rotation). For a system at a finite temperature $T$,
the $x_0$-variable in the euclidean action is integrated over the interval 
$ \,[-\frac{\beta}{2},\frac{\beta}{2}]\,,$ 
where $ \beta $ is proportional to the inverse temperature $T^{-1} ,$
and anti-periodic boundary conditions are imposed at $ x_0 = \pm \frac{\beta}{2}.$ 
 In this paper,
however,  we only consider systems at zero temperature, 
($ \beta \rightarrow \infty $). \\
Then the covariance of the Gaussian integration  --  i.e., 
the unperturbed electron propagator --  is given by
\bea
G^0_{\alpha,\beta} (x-y) & := & 
-  \langle \psi_\alpha(x) \psi^\ast_\beta(y) \rangle^0_{\mu}\;
= -  \; \int \frac{{\cal{D}}(\Psi^\ast,\Psi)}{\Xi^0_{\mu}}\; e^{ - S^0(\Psi^\ast,\Psi\, ;\, \mu)}
\;\psi_\alpha(x) \psi^\ast_\beta(y) \nonumber \\
& = & \delta_{\alpha,\beta}\, \int_{\bf{R}} \dbar p_0 \int_{\bf{R^d}} \dbar^{\,d}p\;
\frac{e^{-i p_0(x_0-y_0) + i\vec{p}(\vec{x}-\vec{y}) }}
{ip_0 -  \varepsilon^0_{\mu}(|\vec{p}|)} \quad ,
\label{4}
\eea
where $\varepsilon^0_{\mu}(|\vec{p}|) = \frac{\vec{p}^{\,2}}{2m} - \mu$ 
 is the energy of a free electron, and the indices 
$\alpha,\beta$
label the spin orientations. We use the short-hand notation $\int \dbar p (\cdot)  := 
\int \frac{dp}{2\pi} (\cdot) $.

In the ground state of the unperturbed system, all one-particle states with wave
vectors $\;\vec{p} \; \mbox{satisfying} \; |\vec{p}| \leq k_F = \frac{1}{2} \sqrt{2m\mu}
 $ are occupied, 
and the electron density is given by
\be
 n^0_{\mu} = 2 \int \dbar ^{\,d}p\; \theta(k_F -|\vec{p}|) \;.
\label{Z4}
\ee
The  Fermi surface , $ {\cal S}^{d-1}_{k_F} $, is defined as the surface of this sphere and the
Fermi  wave number, $ k_F $ ,  sets the fundamental momentum scale of the 
system. \\
For the main results and conclusions of this paper to hold, the spherical
symmetry is not essential. Our analysis applies as long as the Fermi surface is strictly
convex. The situation changes, however, radically for systems of electrons hopping
on a square lattice, at half filling, where the Fermi surface contains  flat, parallel faces
 giving rise to "nesting phenomena". In addition to the superconducting
instability, other instabilities like charge- or spin- density-wave instabilities
can occur,  and their interplay can lead to rather complicated phenomena. Such 
systems can be analyzed by using methods similar to those described in this section,
but their properties are not yet fully understood.

We are interested in universal large-scale and low-energy properties of electron
systems, i.e.,  in the so called {\sl scaling limit} of such systems.
The scaling limit can be constructed by using renormalization group (RG) techniques
which are based on  successively integrating out the modes of the electron
fields $ \Psi^\ast, \Psi $ corresponding to wave vectors far from the Fermi surface,
with the aim of deriving an effective action for the modes close to the Fermi surface;
(the energy vanishes for modes whose momenta lie on the Fermi surface) .\\
An alternative way to obtain some "non-perturbative", large-scale and low-energy
information about an electron system consists in using the  so called 
"Luther-Haldane" (LH) bosonization technique \cite{9,10,11} . However, for electron systems
in more than one space dimension,  LH bosonization is not an exact method,
i.e., it does not exactly resum  the perturbation expansion in the 
 coupling constant,  $g$, of the quartic electron-electron interaction . {\sl The aim of
 this paper is to incorporate the bosonization technique into
the  systematic framework of the RG, in order to estimate the effects of the 
approximations
involved by "bosonizing" an electron system in more than one space dimension.}

In general,  RG calculations for interacting electron systems require assuming that
the coupling constant $ g $  of the quartic electron-electron 
interaction is small. For one-dimensional systems, the
calculations have been carried out  a long time ago (for a review, see \cite{6}).
In more than one space dimension, the situation is more complicated. However,
during the last few years, there has been substantial progress accomplished by
 introducing
the inverse scale parameter , $ \lambda^{-1} $,   proportional to
the width of the effective wave vector space around the Fermi surface,  as a 
supplementary expansion parameter, \cite{7}. In this way,  the dominant 
contributions to the scaling  limit are obtained in a natural manner  . 

Here we   sketch a version of the  RG method for electron systems in more
 than one space dimension based on a double expansion in $\lambda^{-1} $ and the
running coupling constant, $g_\lambda,$ which measures the strength
of electron interactions at energy scales $\sim \frac{v_F k_F}{\lambda} ;$
 for  original results,  see \cite{1,19,7,8} , and for   reviews,  \cite{12,15} . 
(An alternative way would consist in artificially introducing 
$S$  different species of electrons and to organize the perturbation expansion in
powers of $\lambda^{-1}$ and $S^{-1}$ or $S$,  for $\; S \rightarrow \infty $
or $ S \rightarrow 0 $, resp. ; see \cite{2}.)

We  adopt a Wilson-type formulation of the RG and accomplish the mode reduction
in momentum space. 
The electron fields $ \Psi ^\ast(x), \Psi(x) $ are expressed in terms of
their Fourier modes
$$
\hat{\Psi}(p)\;=\;\int d^{d+1}x\;e^{ipx}\,\Psi(x) \qquad \qquad
 \hat{\Psi}^\ast(p)\;=\;\int d^{d+1}x\;e^{-ipx}\,\Psi^\ast(x) \quad,
$$
with $\; p x := p_0 x_0 - \vec{p} \vec{x}\;.$\\
We assume that,  for a sufficiently small value of the dimensionless coupling constant
 $g$, there
exists a large scale factor $\,\lambda_0 \gg 1\,$ (with $ \lambda_0 \rightarrow \infty,\,
\mbox{as}\; g \rightarrow 0 $), such that, in a first step, the integration over
the electron modes $\;\hat{\Psi}(p), \hat{\Psi}^\ast(p)\; $ with momenta $\vec{p}$  outside the
shell
$$
 \Omega_{\lambda_0} \; := \; \left\{ \vec{p} \in {\bf R}^d \; , \; \left| \vec{p} - 
k_F \frac{\vec{p}}{|\vec{p}|} \right| \; \leq \; \frac{k_F}{2\lambda_0} \right\} 
$$
of  width $ \frac{k_F}{\lambda_0}  $ around the Fermi surface leads to an
{\sl effective action} that has essentially the same form as the
original action $S$. More precisely, the effective action for the remaining modes 
with momenta inside the shell $ \Omega_{\lambda_0} $ is given by\\
\vspace{.2cm}
 \bea
S^{\rm{eff}}_{\lambda_0}\left(\hat{\Psi}^\ast(p),\hat{\Psi}(p)\right) & =&
- \log \int_{\vec{p} ' \in {\bf R}^d \setminus \Omega_{\lambda_0}}
{\cal D}\left(\hat{\Psi}^\ast(p'),\hat{\Psi}(p')\right)\, e^{ - \left[
S^0(\hat{\Psi}^\ast,\hat{\Psi} ; \mu)\; +\; g\, S^V(\hat{\Psi}^\ast,\hat{\Psi}) \right] }
\nonumber \\
 && 
 \hspace{6cm} + \qquad cte. \qquad.
\label{5}
\eea
We assume that
$$ S^{\rm{eff}}_{\lambda_0}\left(\hat{\Psi}^\ast(p),\hat{\Psi}(p)\right)\quad \sim
\quad S\left(\hat{\Psi}^\ast(p),\hat{\Psi}(p)\right) \quad.
$$
In other words,
for sufficiently small  coupling constants $ ( < \frac{1}{\lambda_0^2} ) $ 
, possible instabilities develop only at energy scales smaller
 than $\; v_F \frac
{k_F}{\lambda_0} $  ; in  \cite{1}, it has been proven  rigorously that
this is true for a two-dimensional system with short-range interactions. 
  
Given $ S^{\rm{eff}}_{\lambda_0} $ , we wish  to determine the effective action, 
$  S^{\rm{eff}}_{\lambda_1} $,  on a lower energy scale $ \sim v_F \frac{k_F}{\lambda_1}\;
, \mbox{where}\; \lambda_1 = M \lambda_0 \;,\;M>1 \;  $ , 
by integrating out the modes in the shells
$ \; \Omega_{\lambda_0} \setminus \Omega_{\lambda_1} $ . Functional
integration leads to a perturbation expansion for $  S^{\rm{eff}}_{\lambda_1} $ 
with a fermion propagator determined by the quadratic part of 
$  S^{\rm{eff}}_{\lambda_0} $ .  

In order to illustrate characteristic features of the
resulting perturbation theory, we determine the unperturbed electron propagator
$ G^0 (x) $ , defined in  eq.(\ref{4}), for large values $ x = \lambda
\xi $ of its arguments, where $ \xi \sim \frac{1}{k_F} $ and $ \lambda 
\rightarrow \infty $ . One finds
\bea
G^0(\lambda \xi) &= & (\frac{k_F}{2\pi})^{d-1} \int_{S_1^{d-1}} d^{d-1}\omega\;
e^{i k_F \vec{\omega} \lambda \vec{\xi}}  \cdot \hspace{5cm}
\nonumber\\
&&\hspace{2.5cm}\cdot \int_{\bf R} \dbar p_0 
\int_{-\frac{k_F}{2 \lambda}}^{\frac{k_F}{2 \lambda}}  \dbar p_\| \; 
\frac{e^{-i (p_0\lambda \xi_0 - p_\| \lambda \vec{\omega}
\vec{\xi})}}{i p_0 - v_F p_\| } \left[ 1 + O(\frac{1}{\lambda}) \right]\; ,
\label{6}
\eea
 cf. \cite{13}, where $ \vec{\omega} $ denotes a $d-$dimensional unit vector ($\vom \in
 {\cal S}^{d-1}_{1} $),  and the Fermi
velocity $ v_F = \frac{k_F}{m} $ is given by the linearization of the energy function
$\; \varepsilon^0_\mu (k_F\vec{\omega} + \vec{p}) = v_F | \vec{p}\cdot\vec{\omega} |
\, + \,O(\vec{p}^{\,2}) \; $ around the Fermi sphere . For large arguments $ (\sim\lambda) $ of the electron 
propagator, only momenta inside the shell $ \Omega_\lambda $ are important.

One can subdivide the shell $ \Omega_\lambda $  into $\; N = (\frac{\lambda}
{k_F})^{d-1}\mbox{Vol}({\cal S}_{k_F}^{d-1})\; $ disjoint, congruent blocks $ B_{\omega^{(i)}}(
\lambda) $ with sides of length $ \sim \frac{k_F}{\lambda} $ , centered at
the points $ k_F\vec{\omega}^{(i)} \;,\; i = 1, ... , N \;$, on the Fermi surface, as
indicated in Fig.1 .The endpoints of the unit vectors $ \vec{\omega}^{(i)}\;
,\; i = 1, ... , N\;$, form a regular grid on $ {\cal S}^{d-1}_1 .$ 

\begin{center}
\begin{picture}(200,190)(0,0)

\CArc(16,16)(180,5,71)
\DashCArc(16,16)(160,5,63){2}
\CArc(16,16)(140,5,71)

\Vertex(144.1,112.1){1.5} \Vertex(116.3,141.4){1.5} \Vertex(163,79.4){1.5}
\LongArrow(16,16)(143.8,111.8)
\Line(89.7,135.8)(110.3,169.7)  \Line(116.2,114.5)(143.5,142.0)
\Line(137.4,86.8)(171.4,107.2)  \Line(151.1,55.5)(189,66.6)
\LongArrow(163,79.4)(145.6,71.8) \LongArrow(163,79.4)(180.7,87.1)
\Text(176,74)[]{$\frac{k_F}{\lambda}$}  \Text(180,130)[]{$B_{\omega_i}(\lambda)$}
\Text(76,166)[]{${\cal S}^{d-1}_{k_F}$}  \Text(72,74)[]{$k_F \vec{\omega}_i$}
\Text(176,8)[]{$\Omega_\lambda$}
\end{picture}\nopagebreak
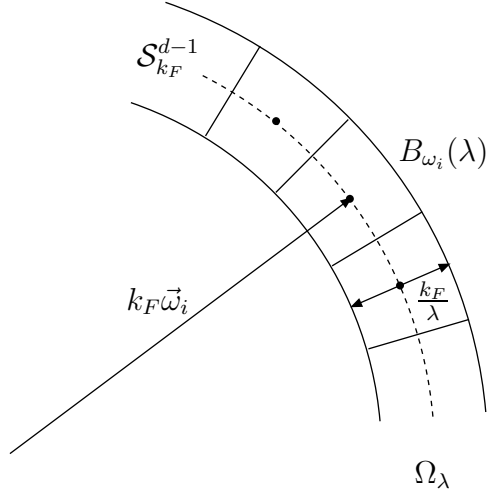
\\ {\sl Fig.1}
\end{center}

 This discrete decomposition
of the shell $ \Omega_\lambda $   into $ N $ boxes $ B_{\omega^{(i)}}(
\lambda) $  yields a formula analoguous to eq.(\ref{6}) for the large distance 
behaviour of the electron propagator :
\bea
G^0(\lambda\xi) &= & \sum_{i = 1}^N
e^{i k_F \vec{\omega}^{(i)} \lambda \vec{\xi}}\int_{\bf R} \dbar p_0 
\int_{\overline{B}_{\omega^{(i)}}(\lambda)}  \dbar^{\,d} p \; 
  \cdot \hspace{3cm}
\nonumber\\
&&\hspace{3cm}\cdot\quad \frac{e^{-i (p_0\lambda \xi_0 -\vec{p} \lambda 
\vec{\xi})}}{i p_0 - \varepsilon^0_{\mu}(k_F\vec{\omega}^{(i)} + \vec{p}) } 
 \;\left[ 1 + O(\frac{1}{\lambda}) \right]\quad ,
\label{7}
\eea
with $ \;\overline{B}_{\omega^{(i)}}(\lambda) := \left\{ \vec{q} - k_F\vec{\omega}^{(i)}
\; , \; \vec{q} \in B_{\omega^{(i)}}(\lambda) \right\} \; $ .\\
The energy function
$ \varepsilon^0_{\mu} $  of the unperturbed system can be written as
\be
\varepsilon^0_{\mu}(k_F\vec{\omega}^{(i)} + \vec{p})  \; = \; v_F  p_\| \;+\;
\left[ (\vec{p}_\perp)^2 + (\vec{p}_\|)^2 \right] \quad ,
\label{Z7}
\ee
where the momenta $ \vec{p} $ are supposed to lie in the box $\overline{B}_{
\omega^{(i)}}(\lambda) $ , with $\; \vec{p}_\| := p_\| \,\vec{\omega}\;, \;p_\| = (\vec{
\omega}\cdot\vec{p}) \;, \quad\mbox{and} \quad \vec{p}_\perp := \vec{p} - \vec{p}
_\parallel \; $ . Compared to the first, linear term, the quadratic
contributions are of higher order in $ \frac{1}{\lambda} $ , and one is tempted to
neglect them. Neglecting the dependence on the perpendicular momenta, $ \vec{p}
_{\perp} ,$  amounts to replacing the piece of the Fermi surface, $ {\cal S}_{k_F}^{d-1}
\cap B_{\omega^{(i)}}(\lambda) $ , in the box $ B_{\omega^{(i)}}(\lambda) $ by a plane,
 cf. Fig 2 .\\
 This approximation is harmless  as long as
$\; p_\| \gg (\frac{k_F}{\lambda})^2 \;.$ 
 I.e., {\sl given the decomposition of the shell
 $ \Omega_\lambda $ into boxes $ B_{\omega^{(i)}}(\lambda) $ , one can neglect
the corrections to the linear part of the energy function $ \varepsilon^0_\mu $ 
 in the propagator, eq.(\ref{7}), 
as long as one only integrates out  modes with momenta 
$\; p_\| \gg (\frac{k_F}{\lambda})^2 \;.$ }

\begin{center}
\begin{picture}(400,300)
\put(0,0){
\begin{picture}(400,300)(0,25)
\CArc(200,-448)(768,75,105)
\CArc(200,-448)(656,75,103)
\CArc(200,-448)(544,75,105)
\Line(114.4,88.6)(70.6,308) \Line(285,88.6)(329.4,308)
\DashLine(80,208)(320,208){2} 
\DashLine(80,202.8)(320,202.8){5} 
\DashLine(80,198.6)(320,198.6){2}
\LongArrow(200,200)(200,320) \LongArrow(200,200)(200,96)
\LongArrow(330,202.8)(330,208) \LongArrow(330,202.8)(330,198.6)
\Text(34.1,185.3)[]{${\cal S}^{d-1}_{k_F}$}
\Text(210,235)[l]{$\frac{k_F}{\lambda}$}
\Text(120,270)[]{$B_{\omega_i}(\lambda)$}
\Text(335,202.8)[l]{$\sim \frac{k_F}{\lambda^2}$}  
\Text(200,30)[]{\sl Fig.2}
\end{picture}
}
\end{picture}
\end{center}

After having linearized the energy function $\varepsilon^0_\mu  ,$ \eq{7} can be 
reproduced by decomposing the electron fields $ \Psi^\ast, \Psi $ 
into $ N $ independent components $ \psi^\ast_{\omega^{(i)}}, \psi_{\omega^{(i)}}\; $
\be
 \Psi^\ast(\lambda\xi)\;\approx\;\sum_{i = 1}^N\;e^{-i k_F \vec{\omega}^{(i)} \lambda
\vec{\xi}}\;\lambda^{-\frac{d}{2}}\;\psi^{\ast}_{\omega^{(i)}}(\xi) \qquad
\Psi(\lambda\xi)\;\approx\;\sum_{i = 1}^N\;e^{i k_F \vec{\omega}^{(i)} \lambda 
\vec{\xi}}\;\lambda^{-\frac{d}{2}}\;\psi_{\omega^{(i)}}(\xi) 
 \quad,
\label{8}
\ee
with propagators
$$
-  \left< \psi_{\omega^{(i)}}(\xi)\;
                \psi^{\ast}_{\omega^{(i)}}(\eta) \right>
\; = \; \delta_{\omega^{(i)},\omega^{(j)}}\;
\int_{\bf R} \dbar k_0 
\int_{\overline{B}_{\omega^{(i)}}}  \dbar^{\,d} k \; 
  \frac{e^{-i (k_0 (\xi_0 - \eta_0) -\vec{k}  
(\vec{\xi} - \vec{\eta}))}}{i k_0 -  v_F \vec{\om}^{(i)}\vec{k} } \quad. 
$$
Here, "$ \approx $" stands for "equal to leading order in  $ \frac{1}{\lambda} $ " and
$ \;\overline{B}_{\omega^{(i)}}\;:=\;\overline{B}_{\omega^{(i)}}(1) \;.$ \\
The Fourier modes of the  component fields $ \psi^{\sharp}_{\omega^{(i)}}(\xi) $ , i.e. 
$\;\hat{\psi}^{\sharp}_{\omega^{(i)}}(k_0,\,\cdot\,) \quad, \mbox{with} 
\;\Psi^\sharp = \Psi \,\mbox{or}\,\Psi^\ast  \;$, have support in the boxes 
 $ \overline{B}_{\omega^{(i)}} $ with sides of length $\sim k_F$  , and their propagators are given by
$$  
-  \left< \hat{\psi}_{\omega^{(i)}}(k_0,\vec{k})\;
                \hat{\psi}^{\ast}_{\omega^{(i)}}(k_0',\vec{k}\, ') \right> \; =
\hspace{8cm}
$$
\be
\hspace{2.5cm} = \; \delta_{\omega^{(i)},\omega^{(j)}}\;(2\pi)^{d+1}\;\delta^{(d+1)}
(k - k') \; \frac{1}{i k_0 -  v_F \vec{\omega}^{(i)}\cdot  
\vec{k} } \; \mbox{\bf \Large 1}_{\overline{B}_{\omega^{(i)}}}(\vec{k}) \quad ,
 \label{Z10}   
\ee
where
$$
\mbox{\bf \Large 1}_{\overline{B}_{\omega^{(i)}}}(\vec{k}) \quad := \quad
\left\{ \quad 
\begin{array}{ll}
1 & ,\quad \vec{k} \in \overline{B}_{\omega^{(i)}} \\
0 & ,\quad \mbox{else}
\end{array}
\right.
\qquad.
$$
Note that the momenta $ k $ in \eq{Z10} are related to the momenta $ p $ in \eq{7}
by the scale transformation$\; p = \frac{k}{\lambda} .$ 
Whereas the  electron fields $\Psi^\sharp(x=\lambda\xi) $ are functions of
the physical coordinates $(x)$, the component fields $\psi^\sharp_{\om^{(i)}}(\xi) $ are 
functions of the rescaled coordinates $(\xi) .$ In the rescaled system, the Fermi surface
has the radius $\lambda k_F ,$ and the boxes  $ \overline{B}_{\omega^{(i)}} $ cover
a shell $\, \Omega\, $ of thickness $ k_F $ around the Fermi surface. \\
{\sl Eq.(\ref{8})  shows that, in order to describe the large scale physics of the
 unperturbed system,
the electron fields $ \Psi^\ast, \Psi $ can be decomposed into 
$N \sim \lambda^{d-1}$  components  $ \psi^\ast_\omega , \psi_\omega $ 
which  propagate only in the direction perpendicular to the
 Fermi surface.}

We now return to the interacting system. After having integrated out the modes with
momenta in $ \;{\bf R}^d \setminus \Omega_{\lambda_0}\;,$ we can apply the 
decomposition (\ref{8}) to the remaining electron modes in the effective action
$\; S^{\rm{eff}}_{\lambda_0}\left(\hat{\Psi}^\ast(p),\hat{\Psi}(p)\right)\;,$
the error being of order $\frac{1}{\lambda_0} .$ According to the assumptions stated 
above, the effective action $\; S^{\rm{eff}}_0 \equiv S^{\rm{eff}}_{\lambda_0}\;$
at an energy scale $\sim v_F \frac{k_F}{\lambda_0} $ then has the form 
\be
S^{\rm{eff}}_{0} \quad = \quad S^2_{0} \; + \; \delta S^2_{0} \; + \; 
S^4_{0} \; + \; \mbox{" higher-order terms "}\quad ,
\label{11}
\ee
with
\bea
S^2_{0} & \approx &
\sum_\omega \sum_{\sigma = \uparrow,\downarrow} \int_{I_\omega}
\dbar^{\,d+1}k\; \frac{-1}{Z_0} \; \hat{\psi}^\ast_{\omega,\sigma}(k)\,
\left( i k_0 - v_{F0} \vec{\omega}\vec{k} \right) \,\hat{\psi}_{\omega,\sigma}(k)
\label{12}\\
\delta S^2_{0} & \approx &
\sum_\omega \sum_{\sigma = \uparrow,\downarrow} \int_{I_\omega}
\dbar^{\,d+1}k\; \lambda_0 \frac{\delta \mu_0}{Z_0} \; 
\hat{\psi}^\ast_{\omega,\sigma}(k)\;
\hat{\psi}_{\omega,\sigma}(k)
\label{Z12}
\eea
and
\bea
S^4_{0} & \approx &
\frac{1}{2 }\frac{k_F^{1-d}}{\lambda_0^{d-1} \,Z^2_0}
\sum_{\omega_1, ... ,\omega_4} \sum_{\sigma,\sigma' } 
\int_{I_{\omega_1}}\dbar^{\,d+1}k^{(1)}\, \cdots\,\int_{I_{\omega_4}}\dbar^{\,d+1}
k^{(4)} \; \cdot \hspace{0.1cm}
\nonumber \\
 & & \hspace{1cm} \cdot \;\delta_{\omega_1+\omega_2 , \omega_3 + \omega_4} \; 
\delta^{(d+1)}\left(k^{(1)} + k^{(2)} - k^{(3)} - k^{(4)} \right) \;\cdot
\nonumber\\
& & \hspace{0.8cm} g_0^{\sigma,\sigma'}( \,\underline{\omega} \,; {\scriptstyle\frac{1}{\lambda_0}}
\underline{k} )  \quad \hat{\psi}^\ast_{\omega_4 \sigma}(k^{(4)}) \,
\hat{\psi}^\ast_{\omega_3 \sigma'}(k^{(3)}) \,
\hat{\psi}_{\omega_2 \sigma'}(k^{(2)}) \,
\hat{\psi}_{\omega_1 \sigma}(k^{(1)}) \;  ,
\label{13}
\eea
where $\;\underline{\om} := \{\vom_1,\vom_2,\vom_3,\vom_4\}\; ,\quad \underline{
k} := \{k_1,k_2,k_3,k_4\}\;,$ and  
 $ I_\omega $ stands for the integration domain
 $ {\bf R} \times \overline{B}_{\omega} $  .

For small values of $g \lambda_0^2 ,$ the parameters   $\; v_{F0} , Z_0 ,
g_0^{\sigma,\sigma'}
( \,\underline{\omega} \,; {\scriptstyle\frac{1}{\lambda_0}}
\underline{k} ) \; $ 
are  renormalized only weakly with respect to
 the parameters $\; v_F ,\, Z = 1,\, g \hat{V}\; $ in the original action. 
The set of dimensionless
 coupling functions  $\; g_0^{\sigma,\sigma'}
( \,\underline{\omega} \,; {\scriptstyle\frac{1}{\lambda_0}}
\underline{k} )\;$ is related to the Fourier transform, $ g \hat{V}(\frac{\vec{k}}
{\lambda_0}) ,$ of the original interaction potential by
\be
k_F^{1-d} g_0^{\sigma,\sigma'}( \,\underline{\omega} \,; {\scriptstyle\frac{1}{\lambda_0}}
\underline{k} ) \quad \approx \quad 
  g_0\, \delta^{\sigma,\sigma'}\, \hat{V}\left( k_{F0} 
(\vec{\omega}_4-\vec{\omega_1}) \,+\,\frac{1}{\lambda_0} 
(\vec{k}^{(4)}-\vec{k}^{(1)}) \right) \; . 
\label{Z13}
\ee 
The Fourier transform, $ \hat{V}
 ,$  of a {\sl short-range} two-body potential $V$  is smooth in momentum space,
 so that -- to leading order in $ 
\frac{1}{\lambda_0} $ -- we can neglect the dependence on the small momenta.
Thus, the coupling functions $\; g_0^{\sigma,\sigma'}
( \,\underline{\omega} \,; {\scriptstyle\frac{1}{\lambda_0}}
\underline{k} )\;$  in eq.(\ref{13}) can be replaced by 
a set of coupling constants $ g_0^{\sigma,\sigma'}( \underline{\omega} ) $ .\\
For a {\sl long-range} potential, whose  Fourier transform $ \hat{V}(p) $  is singular 
in momentum space, i.e., $ \; \hat{V}(p) \sim \frac{1}{|p|^\alpha} \;$ , with $ \alpha
> 0 $ , we set
\be
k_F^{1-d} g_0^{\sigma,\sigma'}( \,\underline{\omega} \,; {\scriptstyle\frac{1}{\lambda_0}}
\underline{k} ) \; \approx \; 
\left\{ \quad 
\begin{array}{ll}
  g_0\, \delta^{\sigma,\sigma'}\, \hat{V}\left( k_{F0} 
(\vec{\omega}_4-\vec{\omega_1})  \right) & ,\quad
\mbox{for} \; \vec{\omega}_4 \neq \vec{\omega}_1\; , \\
  g_0\, \delta^{\sigma,\sigma'}\, \hat{V}\left( \frac{1}{\lambda_0} 
(\vec{k}^{(4)}-\vec{k}^{(1)}) \right) & ,\quad
\mbox{for} \; \vec{\omega}_4 = \vec{\omega}_1\; .
\end{array}
\right.
\label{ZZ13}
\ee 

We first restrict our analysis to short-range potentials; 
 comments about long-range potentials will be made at the end of this section.

In  general, a quadratic term $ \delta S^2_0 $ of the form given in eq.(\ref{Z12})
is generated. It displaces the origin of the energy spectrum -- i.e.,  the Fermi wave
number -- to $ \; \lambda_0 k_{F0} \approx \lambda_0 ( k_F - \frac{\delta \mu_0}
{v_{F0}} ) \; $ . Under the condition that $ \; \lambda_0 \frac{\delta \mu_0}
{v_{F0}} \ll k_F \;, $ we can absorb this term in a change of the parallel momentum
, $ \; k_\| \rightarrow k'_\| =  k_\| + \lambda_0 \frac{\delta \mu_0}{v_{F0}} \; ,$ 
obtaining
\be
\overline{S}^2_0 \; := \; S^2_{0} \; + \; \delta S^2_{0} \; = \;
\sum_\omega \sum_{\sigma = \uparrow,\downarrow} \int_{I_\omega'}
\dbar^{\,d+1}k'\; \frac{-1}{Z_0} \; \hat{\psi}^\ast_{\omega,\sigma}(k')\,
\left( i k'_0 - v_{F0} \vec{\omega}\vec{k'} \right) \,\hat{\psi}_{\omega,\sigma}(k')
\quad .
\label{ZZ12}
\ee
Thus the propagators obtained from $ \overline{S}^2_0 $ are equal to the propagators 
determined by $  S^2_{0} $ , except for a small displacement of their support, $
\;\overline{B}_\omega \rightarrow \overline{B}'_\omega \;$ .

By "higher-order terms",  in \eq{11},  we mean contributions corresponding to
higher orders in the Taylor expansion of the coefficient functions of the quadratic
and quartic terms in the momentum variables,  or contributions involving more than
4 electron fields.
Engineering scaling suggests that both types of contributions are irrelevant, but we shall
analyze these terms more carefully later in this section.

As usual, one divides the terms in the action into relevant, marginal and
irrelevant ones,  in accordance with their scaling dimension. The exponent , $ \nu $ ,
characterizing the (leading) behaviour of an expression $ {\cal F}\left(\hat{\Psi}
^\ast(k) , \hat{\Psi}(k) , 
k \right) $  under scale transformations, i.e. ,   
$$
 {\cal F}\left(\hat{\Psi}^\ast_{
(\lambda)}({\scriptstyle \frac{k}{\lambda}}) , \hat{\Psi}_{(\lambda)}({\scriptstyle 
\frac{k}{\lambda}}) \,;\, 
{\scriptstyle \frac{k}{\lambda}} \right) \quad \approx \quad
\lambda^\nu\;{\cal F}\left(\hat{\Psi}^\ast
(k) , \hat{\Psi}(k) \,;\, 
k \right)
\quad , \;\mbox{ for }  \lambda \rightarrow \infty \;,
$$
is called the scaling dimension of $ {\cal F} $ .  The scaling dimension of the electron modes
is $ \frac{d}{2}+1 ,$ i.e.,
\be
\hat{\Psi}^\sharp_{
(\lambda)}({\scriptstyle \frac{k}{\lambda}})\; = \; \lambda^{\frac{d}{2}+1}\;
\hat{\Psi}^\sharp(k)\quad.
\label{10}
\ee
This scaling dimension is fixed by the requirement that the scaling dimension of 
the quadratic action $ S^2_0 $ is zero. \\
Terms with scaling dimension greater than / equal to 0 are called relevant / 
marginal, and terms with negative scaling dimension are called irrelevant, as they 
are suppressed by inverse powers of $ \lambda $ , as $ \lambda \rightarrow
\infty $ .\\
The two terms of the quadratic action $ S^2_{0} $ , given by \eq{12}, are marginal.
Contributions to $ S^2_{0} $  which arise from higher orders in the Taylor expansion 
of the energy function, $ \varepsilon $ , in the momentum variables are not displayed, as they are irrelevant.
The quadratic term $ \delta S^2_{0} $ , cf. \eq{Z12} , which causes a displacement
 of the Fermi wave number has scaling dimension 1 , i.e. , it is relevant.
By holding the chemical potential  $ \mu $  fixed, the average electron density
of the system -- related to the Fermi wave number -- changes during the RG
iterations, and this requires a continual readaptation of the 
linearization point of the
energy function.\\
The quartic interaction , \eq{13} , has scaling dimension $ 1- d $ , i.e. , it is irrelevant
in dimension $ d \geq 2 $ . However, there are of order $ N \sim
\lambda_0^{d-1} $ different interaction terms, cf. eqs. (\ref{13}) and (\ref{Z13}). 
It can happen that -- for special
exterior momentum configurations --  the sum over the different interaction terms
compensates the scaling factor $ \frac{1}{\lambda_0^{d-1}} $ (as
discussed below ) .  Moreover,
engineering scaling arguments are only valid as long as the dimensionless 
running coupling
constants    remain small during the RG iterations. 
If an instability is developing, in the sense that some couplings  diverge as $ \lambda $ grows,
our weak coupling analysis breaks
down. The RG method will  permit us to identify those processes that
lead to instabilities.\\
Corrections due to higher orders in the Taylor expansion of the interaction potential
around the points  $ \;k_{F0} (\vec{\omega}^{(4)} - \vec{\omega}^{(1)})\; $  are 
sub-leading, cf. \eq{Z13} .  (Of course, this is only true for short-range
 interactions).\\
{\sl Local} terms in the action  involving $ 2\ell $ electron fields ( $\ell > 2$ ) have
scaling dimension $ -\ell d + (d+1) $ ; they are irrelevant and can be neglected.
However, {\sl non-local} terms of the type shown in Fig. 3a (straight lines stand for electron
propagators $G_\om$ and wiggly lines for local 4-vertices)  can appear which have scaling 
dimension $ (\ell-1)(1-d) $ . 
These are tree-level contributions  constructed
exclusively out of $\ell-1 $ local 4-vertices;  (the non-locality arises from the
inner propagator lines). Although they have a negative scaling dimension, they can
contribute in leading order to the renormalization of some special, lower
order vertices.
This happens in  cases where momentum conservation allows one to sum over
the $\omega-$labels of contracted pairs of incoming and outgoing propagator lines,
leading to a contribution of $O(\lambda_0^{d-1})$ per contracted pair ( cf. Fig. 3b) .\\
\begin{center}
\begin{picture}(220,100)(0,0)
\ArrowLine(10,10)(10,30)  \ArrowLine(10,30)(10,90)  
\ArrowLine(60,10)(60,30)  \ArrowLine(60,30)(60,70)  \ArrowLine(60,70)(60,90)
\ArrowLine(110,10)(110,70)  \ArrowLine(110,70)(110,90)  
\ArrowLine(210,10)(210,50)  \ArrowLine(210,50)(210,90)
\Photon(10,30)(60,30){1.5}{8}
\Photon(60,70)(110,70){1.5}{8}
\Photon(170,50)(210,50){1.5}{6}
\Text(140,50)[]{$\cdots\cdots$}
\Text(8,80)[r]{$\omega_1$} \Text(8,20)[r]{$\omega_1$} 
\Text(58,80)[r]{$\omega_2$}  \Text(58,20)[r]{$\omega_2$}
\Text(58,50)[r]{$\omega_2$}
\Text(108,80)[r]{$\omega_3$}  \Text(108,20)[r]{$\omega_3$}
\Text(208,80)[r]{$\omega_\ell$}  \Text(208,20)[r]{$\omega_\ell$}
\end{picture}
\\
{\sl Fig.3a}
\end{center}
\begin{center}
\begin{picture}(220,100)(0,0)
\ArrowLine(10,10)(10,50)  \ArrowLine(10,50)(10,90)  
\ArrowLine(210,10)(210,50)  \ArrowLine(210,50)(210,90)
\Photon(10,50)(35,50){1.5}{4}
\Photon(85,50)(110,50){1.5}{4}
\Photon(185,50)(210,50){1.5}{4}
\Text(172,50)[]{$\cdots$}
\ArrowArcn(60,50)(25,0,180)  \ArrowArcn(60,50)(25,180,360)
\ArrowArcn(135,50)(25,0,180)  \ArrowArcn(135,50)(25,180,360)
\Text(8,80)[r]{$\omega_1$} \Text(8,20)[r]{$\omega_1$}
\Text(60,77)[b]{$\omega_2$}  \Text(60,23)[t]{$\omega_2$}
\Text(135,77)[b]{$\omega_3$}  \Text(135,23)[t]{$\omega_3$}
\Text(208,80)[r]{$\omega_\ell$}  \Text(208,20)[r]{$\omega_\ell$}
\Text(60,50)[]{$\sum\limits_{\omega_2}$}	\Text(135,50)[]{$\sum\limits_{\omega_3}$}
 \end{picture}
\\
{\sl Fig.3b}
\end{center}
The influence of such higher order vertices to the RG flow of the parameters 
$\;v_F, Z, \delta \mu, g(\underline{\omega})\;$  is analyzed 
in \cite{14} ;
they contribute  in leading order in the inverse scale parameter $\lambda ,$  but 
do not cause any 
{\sl qualitative}
change of the flow. {\sl For  the sake of simplicity,  we shall ignore them in the 
following, because we assume that $g$ is small.}
(However, in the RG teatment of systems with long-range interactions, they play
 an important role in obtaining the correct form of screening) .

Given $ S^{\rm{eff}}_{0} $ , we attempt to calculate the effective action $ S^{\rm{eff}}_
{1} $ at a larger scale $ \lambda_1 = M \lambda_0 > \lambda_0 $ .
Assuming that the dimensionless quartic coupling constants $ g_0(\underline
{\omega}) $  are
  small, i.e. $ \; |g_0(\underline{\omega})| \leq g\; $ , we organise the perturbation theory in powers of 
$ g $ . The appearance of  the small  factor $ \frac{1}{\lambda_0^{d-1}} $ in 
front of the quartic interaction term, \eq{13}, suggests to introduce the inverse
scale parameter $ \frac{1}{\lambda_0} $ as a supplementary expansion parameter.
This is also the parameter that controls the approximation of replacing the
electron fields $ \Psi^\ast, \Psi $ by the component fields
$ \psi^\ast_\omega, \psi_\omega ,$ cf. \eq{8} .  
In order to obtain the effective action, $ S^{\rm{eff}}_{n} $ ,  on successively larger 
scales $ \; \lambda_n = M^n \lambda_0 \;,\; n \rightarrow \infty\;$ , we shall
 execute this 
procedure iteratively. This leads to the following formulation of the RG.

In the iteration step $ n \rightarrow n+1 $ , we  lower the energy scale in
the effective action $ S^{\rm{eff}}_{n} $ by a factor $ M $ which we choose to be 
an integer greater than 1 . The action  $ S^{\rm{eff}}_{n} $  is supposed to have
the form specified in eqs. (\ref{11}) - (\ref{13}), with the scale parameter 
$ \lambda_0 $ 
replaced by $ \;\lambda_n = M^n \lambda_0\; $,  and the parameters $ \;Z_0, v_{F0},
\delta \mu_0, g_0(\underline{\omega})\; $ replaced by $ \;Z_n, v_{Fn},
\delta \mu_n, g_n(\underline{\omega})\; $. 
Then, the symbol "$\approx$" stands for "equal to leading order in $ \frac{1}
{\lambda_n} $" .
There are $ \; N_n = (\frac{\lambda_n}
{k_F})^{d-1} \mbox{Vol}({\cal S}_n) \; $ boxes $ B^{(n)}_\omega $ which cover the shell 
$ \Omega_n $ of standard width $ k_F $ around the spherical surface 
$ {\cal S}_n $ of radius $ \; \lambda_n k_{F n-1} = \lambda_n ( k_F - \sum_{i=0}^
{n-1} \frac{\delta \mu_i}{v_{F i}} ) \;$.\\
The iteration can be continued as long as the running coupling constants
$ g_n(\underline{\omega}) $ remain small, i.e., $ \;|  g_n(\underline{\omega}) | \leq  g  
\ll 1\; $ and the displacement, $ \lambda_n \frac{\delta \mu_n}{v_{Fn}} $ 
, of the origin of the energy 
function from $\; \lambda_{n-1} k_{F n-1} \;\mbox{to}\; \lambda_{n} k_{F n} :=
\lambda_n ( k_{F n-1} - \frac{ \delta \mu_n}{v_{Fn}} ) \; $ is small compared to
$ \frac{k_F}{M} ,$ cf. Fig.4 . As in  step 0, cf. \eq{ZZ12}, this allows us to absorb
the displacement of the chemical potential -- described by $ \delta S^2_{n} $ --
in the term $ \overline{S}^2_{n} $ by  shifting the variables $\; k_\| \rightarrow
k_\|' = k_\| + \lambda_n \frac{\delta \mu_n}{v_{Fn}} \; $. 
This yields a displacement of the Fermi wave number from $\; \lambda_{n} 
k_{F n-1} \; $ to $\; \lambda_{n} k_{F n} \;( \approx \lambda_{n}  k_{F n-1} ) \; $ .\\
We denote the shell of width $ \frac{k_F}{M} $  around the spherical surface
$ \widetilde{\cal S}_n $ of radius $ \lambda_{n} k_{F n} $ by $ \widetilde{\Omega}_n $.
In the step $ n \rightarrow n+1 ,$  we eliminate modes with 
momenta $ \vec{k} $ lying in $ \;\Omega_n \setminus \widetilde{\Omega}_n\; $ . 
This yields an effective action
\vspace{.3cm}
$$
S^{\rm{eff}}_{n+1}\left(\{\hat{\psi}_\omega^\sharp(k)\}\right) 
\; \approx \hspace{11cm} 
$$
\be
 \hspace{2cm} \approx \;
- \log \;\int_{(k_0',\vec{k}') \in {\bf R}\times \Omega_n\setminus 
\widetilde{\Omega}_n}
{\cal D}{\scriptstyle \left\{\hat{\psi}_\omega^\sharp(k')\right\}}
\, e^{ - \left[
S^{\rm{eff}}_{n}(\{\hat{\psi}_\omega^\sharp(k')\}) \right] }\quad + 
\quad cte. \quad,
\label{B14}
\ee
where $ \{ \psi_\om^\sharp \} $ denotes the set of component fields with
$\, \psi_\om^\sharp \, = \, \psi_\om^\ast \;\mbox{or}\;\psi_\om\;.$ 
In order to compare $ S^{\rm{eff}}_{n+1} $ to $  S^{\rm{eff}}_{n} $ , we subdivide
each box $ \; B_{\omega_i}^{(n)} \cap \widetilde{\Omega}_n \; $ into $ M $ 
 congruent boxes, $\;\widetilde{ B}_{\omega_{i,\ell}}^{(n)}\;,\; \ell = 1, ... ,
M \; $, with sides of length $ \sim \frac{k_F}{M} $  (cf. Fig 4) .
\\
\begin{center}
\begin{picture}(400,280)(0,0)

\CArc(32,-128)(400,32,89)
\DashCArc(32,-128)(352,30,85){5}
\CArc(32,-128)(304,32,89)
\CArc(32,-128)(347,30,85)
\Line(82.2,172.2)(97.2,266.6) \Line(160.3,147.3)(201.7,234)
\Line(229.7,102.8)(292.6,175.4) \Line(285.3,41.4)(364.9,94.3)
\Vertex(136.5,207.9){2} \Vertex(223.5,167.7){2} \Vertex(295.8,105.1){2}

\CArc(32,-128)(362.8,32,87.5)
\CArc(32,-128)(331.2,32,87.5)
\Line(198.3,158.8)(213.7,185.3) \Line(224,142.2)(241.8,167.3)
\Vertex(193.2,179.2){1.5} \Vertex(220.3,163.8){1.5} \Vertex(246,145.4){1.5}
\LongArrow(30,224)(30,272) \LongArrow(30,224)(30,176)
\LongArrow(38,219)(38,234.8) \LongArrow(38,219)(38,203.2)
\Text(28,240)[r]{$k_F$} \Text(40,219)[l]{$\frac{k_F}{M}$}
\Text(342,45.3)[l]{$\lambda_n k_{Fn-1}$} \Text(331.2,34)[l]{$\lambda_n k_{Fn}$}
\Text(330,230)[]{\large M = 3} \Text(100,20)[r]{\sl Fig.4}
\Text(240,218.5)[lb]{$B_{\omega_i}^n$} 
\Text(195.8,195.5)[lb]{$\tilde{B}_{\omega_{i,1}}^n$}
\Text(252.5,167.2)[l]{$\tilde{B}_{\omega_{i,M}}^n$}

\end{picture}
\end{center}
The resulting number of boxes is $ \; N_{n+1} = M^{d-1} N_n \; $ . This requires a
refinement of the decomposition (\ref{8}) of the electron field.
{\sl Note that, in this way, we implicitly take  into account the curvature
of the Fermi surface.} Using \eq{10},
we then rescale the momentum variables, $ \; \vec{k} \rightarrow \vec{k}' = M \vec{k} \;,$
with $\; \vec{k}' \in B^{(n+1)}_{\omega_{i,\ell}} \;, $ where the boxes
$ B^{(n+1)}_{\omega_{i,\ell}} $ have standard size.\\
We end up with an effective action $ S^{\rm{eff}}_{n+1} $  of the same form as
$  S^{\rm{eff}}_{n} $. The new effective parameters $\; Z_{n+1}, v_{Fn+1}, \delta
\mu_{n+1}, g_{n+1}(\underline{\omega}) \; $ can be expressed as functions
of the set of parameters $\; {\cal P}_n := \{ \delta \mu_n , v_{Fn} , g_n(\underline{\omega})
 , \lambda_n \} \; $   :
\bea
\lambda_{n+1} \frac{\delta \mu_{n+1}}{Z_{n+1}} & \approx &
\frac{M}{Z_n} \left. \Sigma_\omega ( k ; {\cal P}_n ) \right|_{k=0}
\nonumber\\
\frac{1}{Z_{n+1}} & \approx &
\frac{1}{Z_n} \left[ 1 + i \frac{\partial}{\partial k_0} \left. \Sigma_
\omega ( k ; {\cal P}_n ) \right|_{k=0} \right]
\label{14} \\
\frac{ v_{F n+1}}{Z_{n+1}} & \approx &
\frac{1}{Z_n} \left[ v_{Fn} +  \frac{\partial}{\partial k_\|} \left. \Sigma_
\omega ( k ; {\cal P}_n ) \right|_{k=0} \right]
\nonumber
 \eea
and
\be
k_F^{1-d} \frac{g_{n+1}(\underline{\omega})}{Z_{n+1}^2 }
\quad \approx \quad
\frac{1}{Z_n^2} \left. \Gamma ( \underline{\omega} , \underline{k} ;
{\cal P}_n ) \right|_{\underline{k}=0} \quad .
\label{Z14}
\ee 
 The functionals $ \Sigma_
\omega $ and $ \Gamma ,$ which appear in  these flow equations,  turn out
not to depend on the iteration step  (to leading order in  $ \frac{1}{\lambda_n} $). 

The functional   $  \Sigma_\omega $ is the self-energy and is obtained from the
amputated one-particle irreducible
(1PI) connected graphs renormalizing the propagator line $ G_{\omega,n} ,$  and
$ \Gamma(\underline{\omega}) ,$ the 4-vertex function,  contains all 1PI connected 
graphs that renormalize the  dimensionless coupling constant $ g_n(\underline{\omega}) $ .
We restrict our attention to determining the functionals $  \Sigma_\omega $  and 
$ \Gamma(\underline{\omega}) $ to leading order in a double
expansion in $ g $ and $ \frac{1}{\lambda_n} $ . \\
\begin{center}
\begin{picture}(400,75)(0,0)
\ArrowLine(20,10)(50,10)  \ArrowLine(50,10)(80,10)
\ArrowLine(120,10)(150,10)  \ArrowLine(150,10)(180,10)
\ArrowLine(220,10)(250,10)  \ArrowLine(250,10)(280,10)
\ArrowLine(320,10)(350,10)  \ArrowLine(350,10)(380,10)
\GCirc(50,11){2}{0} \ArrowArc(50,24)(12,271,269)
\GCirc(150,11){2}{0}  \GCirc(150,37){2}{0}
\GCirc(250,11){2}{0}  \GCirc(250,37){2}{0}
\GCirc(250,63){2}{0}  
\GCirc(350,11){2}{0}  \GCirc(350,37){2}{0}  
\GCirc(350,63){2}{0}  \GCirc(363,50){2}{0}
\ArrowArc(150,24)(12,270,90) \ArrowArc(150,24)(12,90,270)
\ArrowArc(150,50)(12,271,269)
\ArrowArc(250,24)(12,270,90) \ArrowArc(250,24)(12,90,270) 
\ArrowArc(250,50)(12,270,90) \ArrowArc(250,50)(12,90,270)
\ArrowArc(250,76)(12,271,269)
\ArrowArc(350,24)(12,270,90) \ArrowArc(350,24)(12,90,270) 
\ArrowArc(350,50)(12,270,0) \ArrowArc(350,50)(12,0,90) \ArrowArc(350,50)(12,90,270)
\ArrowArc(350,76)(12,271,269) \ArrowArc(376,50)(12,181,179)
\Text(50,24)[]{$\omega_1$} \Text(150,24)[]{$\omega_1$} 
\Text(250,24)[]{$\omega_1$} \Text(350,24)[]{$\omega_1$}
 \Text(150,50)[]{$\omega_2$} 
\Text(250,50)[]{$\omega_2$} \Text(350,50)[]{$\omega_2$}
\Text(250,76)[]{$\omega_3$} \Text(350,76)[]{$\omega_3$}
\Text(376,50)[]{$\omega_4$}
\Text(24,8)[t]{$\omega,\sigma$} \Text(76,8)[t]{$\omega,\sigma$}
\Text(124,8)[t]{$\omega,\sigma$} \Text(176,8)[t]{$\omega,\sigma$}
\Text(224,8)[t]{$\omega,\sigma$} \Text(276,8)[t]{$\omega,\sigma$}
\Text(324,8)[t]{$\omega,\sigma$} \Text(376,8)[t]{$\omega,\sigma$}
\end{picture}
\\
\vspace{.4cm}
{\sl Fig.5}
\end{center}

In Fig.5 , we display   diagrams   contributing
to the self energy $ \Sigma_\omega $ in leading order in $  \frac{1}{\lambda_n} $ ("cactus diagrams") .
To leading order in $ g $ , only the first one is retained. Straight lines 
stand for  electron propagators  $ G_{\omega,n} $   given by
\be
G_{\omega,n}(k) \quad= \quad \frac{1}{i k_0 - v_{Fn} k_\|} \cdot \mbox{\bf \Large 1}_
{\overline{B}_\om^{(n)}} ( \vec{k} ) \quad,
\label{ZZ14}
\ee 
 and  dots stand for the following combinations of  interaction vertices
$\; - k_F^{1-d} \frac{g_n(\underline{\omega})}{\lambda_n^{d-1}}\; ,$ represented
by wiggly lines :\\
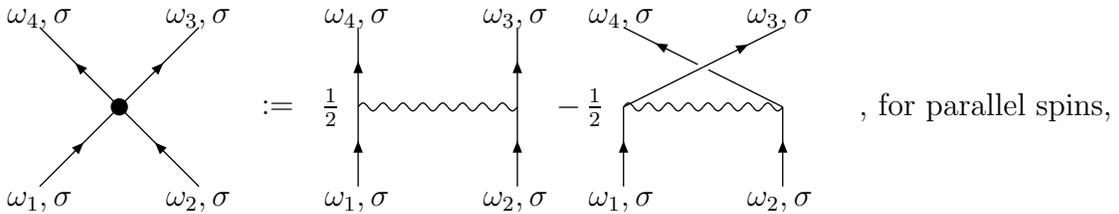
\vspace{-.5cm}
\begin{center}
\begin{picture}(400,80)(0,0)

\ArrowLine(10,10)(40,40) \ArrowLine(40,40)(10,70)
\ArrowLine(70,10)(40,40) \ArrowLine(40,40)(70,70)
\ArrowLine(130,10)(130,40) \ArrowLine(130,40)(130,70)
\ArrowLine(190,10)(190,40) \ArrowLine(190,40)(190,70)
\ArrowLine(230,10)(230,40) \ArrowLine(290,10)(290,40)
\Line(230,40)(260,55) \Line(290,40)(262,54)
\ArrowLine(260,55)(290,70) \ArrowLine(258,56)(230,70)
\Text(100,40)[]{$ := $} \Text(210,40)[]{$ - $}
\Text(10,8)[t]{$\omega_1 , \sigma$} \Text(70,8)[t]{$\omega_2 , \sigma$}
\Text(10,72)[b]{$\omega_4 , \sigma$} \Text(70,72)[b]{$\omega_3 , \sigma$}
\Text(130,8)[t]{$\omega_1 , \sigma$} \Text(190,8)[t]{$\omega_2 , \sigma$}
\Text(130,72)[b]{$\omega_4 , \sigma$} \Text(190,72)[b]{$\omega_3 , \sigma$}
\Text(230,8)[t]{$\omega_1 , \sigma$} \Text(290,8)[t]{$\omega_2 , \sigma$}
\Text(230,72)[b]{$\omega_4 , \sigma$} \Text(290,72)[b]{$\omega_3 , \sigma$}
\GCirc(40,40){3}{0}
\Photon(130,40)(190,40){1.5}{8}     \Photon(230,40)(290,40){1.5}{8}
\Text(320,40)[l]{, for parallel spins, }
\Text(120,40)[]{$\frac{1}{2}$} \Text(220,40)[]{$\frac{1}{2}$}
\end{picture}
\end{center}
and
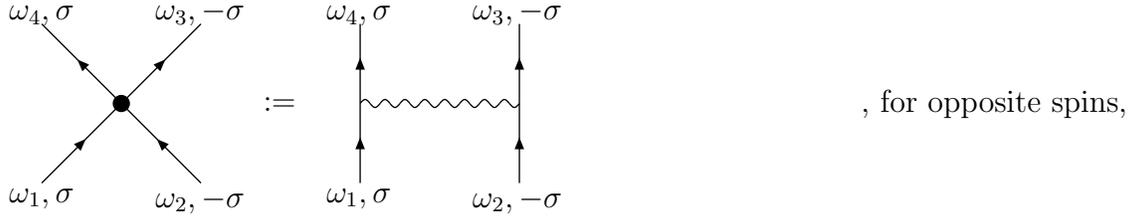
\begin{center}
\begin{picture}(400,80)(0,0)

\ArrowLine(10,10)(40,40) \ArrowLine(40,40)(10,70)
\ArrowLine(70,10)(40,40) \ArrowLine(40,40)(70,70)
\ArrowLine(130,10)(130,40) \ArrowLine(130,40)(130,70)
\ArrowLine(190,10)(190,40) \ArrowLine(190,40)(190,70)
\Text(100,40)[]{$ := $} 
\Text(10,8)[t]{$\omega_1 , \sigma$} \Text(70,8)[t]{$\omega_2 , -\sigma$}
\Text(10,72)[b]{$\omega_4 , \sigma$} \Text(70,72)[b]{$\omega_3 , -\sigma$}
\Text(130,8)[t]{$\omega_1 , \sigma$} \Text(190,8)[t]{$\omega_2 , -\sigma$}
\Text(130,72)[b]{$\omega_4 , \sigma$} \Text(190,72)[b]{$\omega_3 , -\sigma$}
\GCirc(40,40){3}{0}
\Photon(130,40)(190,40){1.5}{8}    
\Text(320,40)[l]{, for opposite spins, }
\end{picture}
\end{center}
where momentum conservation requires that $\; \vom_1+\vom_2 = \vom_3+\vom_4\;.$\\
In Fig.5,  a summation, $ 
\sum\limits
_{\omega}( \cdot ) $ , over $ \vom \in {\cal S}^{d-1} $  is associated with every electron loop and 
compensates the factor $ \frac{1}{\lambda_n
^{d-1}} $ per interaction vertex
( for graphical reasons, the spin indices are omitted in the loops; we sum over all compatible
spin configurations).
 One verifies easily  that  -- for the case of 
short-range interactions  -- the contribution of the first diagram in Fig. 5
 is a {\sl constant} of order $ g $ . 
Using \eq{14}, this implies that, to leading order, the parameters of the quadratic part of the action
do not flow, i.e. ,
 $$ Z_{n+1}\quad \approx \quad Z_n \approx \quad Z_0 $$
\be v_{Fn+1} \quad \approx \quad v_{Fn} \approx \quad v_{F0}
\label{15} \ee
$$ \lambda_{n+1} \, \delta \mu_{n+1} \quad \approx \quad O( g ) \quad. $$

In order to analyze the flow of the quartic coupling constants $ g_n^{\sigma,\sigma'}(\underline{
\omega}) ,$  we  classify them in terms of 
qualitatively different channels. Besides the spin indices, $ \sigma,\sigma' $ , they
 depend on the 4 discrete momenta $ \;\underline{\omega} = \{ \vec{\omega_1}, 
... , \vec{\omega_4} \}\; $. At the $ n^{th} $ iteration step, each of these unit vectors
 can take $ N_n \sim \lambda_n^{d-1}
$ different values $\; \vec{\omega}^{(i)} \;, \; i = 1, 
... , N_n \; $ . However, the four momenta $ \underline{\omega} $ are not all 
independent, but must satisfy the momentum conservation   $ \; \vec{\omega_1} + 
\vec{\omega}_2 = \vec{\omega}_3 + \vec{\omega}_4 \;  $ , as required by translation
invariance. Studying the geometry of momentum conservation, we can subdivide the 
set of coupling constants into three different channels. There is a qualitative
difference between two and more than two dimensions.

In $ d=2 $, cf. Fig. 6a, given the two incoming momenta $ \vec{\omega}_1,
\vec{\omega}_2 $, with $ \vec{\omega}_1 + \vec{\omega}_2 \neq \vec{0} ,$  there are 
exactly 2 possibilities to choose the outgoing momenta, either $ \vec{\omega}_ 3 =
\vec{\omega}_2 $ and $ \vec{\omega}_4 = \vec{\omega}_1 $ or $ \vec{\omega}_ 3 =
\vec{\omega}_1 $ and $ \vec{\omega}_4 = \vec{\omega}_2 $ . In the exceptional case
where $ \vec{\omega}_1 + \vec{\omega}_2  =  \vec{0}  ,$  one is free to choose arbitrarily
one of the $ \lambda_n \sim N_n $ discrete values for $ \vec{\omega}_4 $ ; 
$ \vec{\omega}_3 $ is then determined as its antipode, $ \vec{\omega}_3 = -\vec{
\omega}_4 $ (i.e. , in this case, not $ \vec{\omega}_1 $ and $ \vec{\omega}_2 $ are
independent momenta, but $ \vec{\omega}_1 $ and $ \vec{\omega}_4 $ ) .\\
\begin{center}
\begin{picture}(400,200)(0,0)
\CArc(100,90)(70,0,360)
\LongArrow(100,90)(65.5,150.3) \LongArrow(100,90)(159.6,126)
\LongArrow(100,90)(126,186.8)
\Text(80,56)[]{\bf\large  d = 2} \Text(124,167)[l]{$\vec{\omega}_1+\vec{\omega}_2$}
\Text(82,118)[rt]{$\vec{\omega}_1$} \Text(136,108)[lt]{$\vec{\omega}_2$}
\Text(100,0)[b]{\sl Fig.6a}

\begin{picture}(200,200)(-200,0)
\CArc(100,90)(70,0,360)
\LongArrow(100,90)(66,149) \LongArrow(100,90)(158,125)
\Text(63,152)[rb]{${\cal S}^{d-2}(\vom_1,\vom_2)$}
\LongArrow(100,90)(126,186.8)
\Text(80,56)[]{\bf\large  d = 3} \Text(124,167)[l]{$\vec{\omega}_1+\vec{\omega}_2$}
\Text(74,135)[rt]{$\vec{\omega}_1$} \Text(136,108)[lt]{$\vec{\omega}_2$}
\Text(95,124)[r]{$\vec{\omega}_3$} \Text(126,118)[]{$\vec{\omega}_4$}
\Oval(112.3,138.1)(8,47.7)(-15)
\LongArrow(100,90)(95.2,133.8) \LongArrow(100,90)(128.8,140.8)
\DashLine(95.3,133.8)(128.8,140.8){2}
\DashLine(65.5,150.3)(159.6,126){2}
\Text(100,0)[b]{\sl Fig.6b}
\end{picture}
\end{picture}

\end{center}

In higher dimensions (cf. Fig.  6b for d=3 ), given the two incoming momenta
$  \vec{\omega}_1, \vec{\omega}_2 $ ,  with $ \vec{\omega}_1 + \vec{\omega}_2 
\neq \vec{0} $ ,  one has $ O(\lambda^{d-2}_n) $ choices for the outgoing momenta.
One of the outgoing momenta, e.g. , $ \vec{\omega}_3 $ can be chosen arbitrarily
on the $ d-2 $ dimensional sphere $ {\cal S}^{d-2}( \vec{\omega}_1,\vec{\omega}_2)
$ , $ \vec{\omega}_4 $ being determined as its antipode on $ {\cal S}^{d-2}( \vec
{\omega}_1,\vec{\omega}_2) .$  Again,  in the exceptional case, where $ \vec{\omega}_1 + 
\vec{\omega}_2  =  \vec{0}  $ , there are $ O(\lambda^{d-1}_n) $ choices, as in 2
dimensions ( i.e.,  the number of choices of outgoing momenta  is larger than for a generic 
configuration by a factor of $ \sim \lambda_n )  $ . 

\noindent Thus, in two dimensions, the couplings can be classified as follows :
\bea
g^d( \vec{\omega}_1 \cdot \vec{\omega}_2 , \sigma \cdot \sigma' ) &
:= &  g^{\sigma \sigma'}( \vec{\omega}_1, \vec{\omega}_2, \vec{\omega}_3=
\vec{\omega}_2, \vec{\omega}_4=\vec{\omega}_1 )
\nonumber \\
 g^e( \vec{\omega}_1 \cdot \vec{\omega}_2 , \sigma \cdot \sigma' ) &
:= &  g^{\sigma \sigma'}( \vec{\omega}_1, \vec{\omega}_2, \vec{\omega}_3=
\vec{\omega}_1, \vec{\omega}_4=\vec{\omega}_2 ) 
\label{17} 
\\
 g^c( \vec{\omega}_1 \cdot \vec{\omega}_4 , \sigma \cdot \sigma' ) &
:= &  g^{\sigma \sigma'}( \vec{\omega}_1, \vec{\omega}_2=-\vom_1, \vec{\omega}_3=
-\vec{\omega}_4, \vec{\omega}_4 ) \quad ,
\nonumber
\eea
cf. Fig.7 .  Here, "d" stands for "direct", "e" for "exchange" and "c" for "Cooper";
we use the convention that $ \sigma = \pm 1$ for "spin up" or "spin
down", resp. . 
\\
\vspace{-.6cm}
\begin{center}
\begin{picture}(400,100)(0,0)
\ArrowLine(10,10)(10,50)  \ArrowLine(10,50)(10,90)
\ArrowLine(80,10)(80,50)  \ArrowLine(80,50)(80,90)
\ArrowLine(165,10)(165,50)  \ArrowLine(165,50)(165,90)
\ArrowLine(235,10)(235,50)  \ArrowLine(235,50)(235,90)
\ArrowLine(320,10)(320,50)  \ArrowLine(320,50)(320,90)
\ArrowLine(390,10)(390,50)  \ArrowLine(390,50)(390,90)
\Photon(10,50)(80,50){1.5}{10}
\Photon(165,50)(235,50){1.5}{10}
\Photon(320,50)(390,50){1.5}{10}

\Text(8,10)[r]{$\omega_1$}  \Text(8,90)[r]{$\omega_1$}
\Text(14,10)[l]{$\sigma$}  \Text(14,90)[l]{$\sigma$}
\Text(78,10)[r]{$\sigma'$}  \Text(78,90)[r]{$\sigma'$}
\Text(84,10)[l]{$\omega_2$}  \Text(84,90)[l]{$\omega_2$}
\Text(45,45)[t]{$ \displaystyle \frac{g^d(\sigma\cdot\sigma')}{N}$} 

\Text(163,10)[r]{$\omega_1$}  \Text(163,90)[r]{$\omega_2$}
\Text(169,10)[l]{$\sigma$}  \Text(169,90)[l]{$\sigma$}
\Text(233,10)[r]{$\sigma'$}  \Text(233,90)[r]{$\sigma'$}
\Text(239,10)[l]{$\omega_2$}  \Text(239,90)[l]{$\omega_1$}
\Text(200,45)[t]{$ \displaystyle \frac{g^e(\sigma\cdot\sigma')}{N}$} 

\Text(318,10)[r]{$\omega_1$}  \Text(318,90)[r]{$\omega_4$}
\Text(324,10)[l]{$\sigma$}  \Text(324,90)[l]{$\sigma$}
\Text(388,10)[r]{$\sigma'$}  \Text(388,90)[r]{$\sigma'$}
\Text(394,10)[l]{$-\omega_1$}  \Text(394,90)[l]{$-\omega_4$}
\Text(355,45)[t]{$ \displaystyle \frac{g^c(\sigma\cdot\sigma')}{N}$} 
 \end{picture}
\\ \nopagebreak
{\sl Fig.7}
\end{center}
Note that because of rotational invariance the coupling constants, 
$ g(\underline{\om}) ,$ 
really only depend on the scalar product, $ \vec{\omega} \cdot \vec{\omega}' $ ,
of two vectors on the unit sphere, i.e., on the angle, $ \angle(\vec{\omega},\vec
{\omega}') ,$  between them. Analogously, they only depend on the relative
orientations, $ \sigma \cdot \sigma' ,$  (i.e. parallel or antiparallel) of the spin degrees
of freedom.\\
For an electron system in more than 2 dimensions, momentum conservation
allows more independent coupling constants than the ones listed in \eq{17}.
However, the analysis of their renormalization flow in leading order in 
$ \frac{1}{\lambda_n} $ is similar to the one for two-dimensional systems.
 To leading order in $ \frac{1}{\lambda_n} ,$  they neither flow nor do they influence
the flow of the ones listed in \eq{17}.

The contributions  to the functional $ \Gamma(\underline{\omega} , {\cal P}_n ) $
 of leading order in $ \frac{1}{\lambda_n} $  can  be determined
by applying similar geometric considerations. However, now one has to study the geometry
of momentum conservation for vectors lying {\sl in thin shells around} the
Fermi surface . \\
In Fig.8 ,   the  contributions of leading order in
$ \frac{1}{\lambda_n} $ are displayed (except that self-energy contributions to the inner 
propagator lines are omitted).
 A summation, $ 
\sum\limits
_{\omega}( \cdot ) $ , over $ \vom \in {\cal S}^{d-1} $  is associated with every electron loop 
( for graphical reasons, the spin indices are omitted in the loops; we sum over all compatible
spin configurations).\\
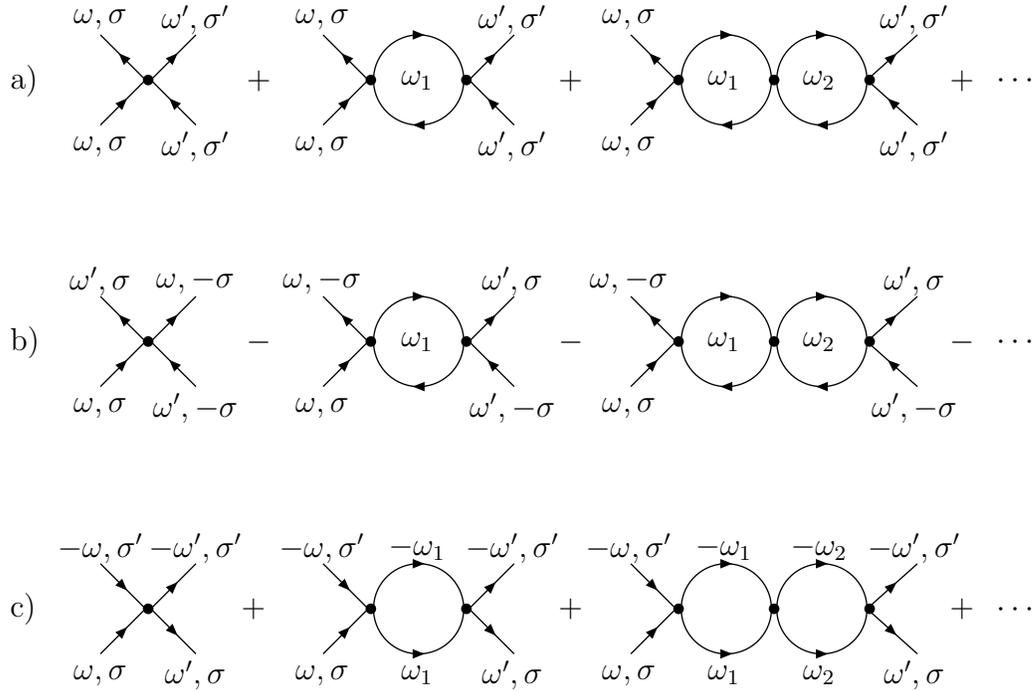
\vspace{-1cm}
\begin{center}
\begin{picture}(400,80)(0,0)
\Text(0,40)[l]{a)}
\ArrowLine(34,23)(51,40) \ArrowLine(51,40)(34,57)
\ArrowLine(70,23)(53,40) \ArrowLine(53,40)(70,57)
\Vertex(52,40){2}
\Text(34,15)[]{$\omega,\sigma$}  \Text(70,15)[]{$\omega',\sigma'$} 
\Text(34,63)[]{$\omega,\sigma$}  \Text(70,63)[]{$\omega',\sigma'$}

\Text(94,40)[]{+}

\put(0,0){
\begin{picture}(400,80)(-10,0)
\ArrowLine(104,23)(121,40) \ArrowLine(121,40)(104,57)
\ArrowArcn(140,40)(17,0,180) \ArrowArcn(140,40)(17,180,0)
\ArrowLine(176,23)(159,40) \ArrowLine(159,40)(176,57)
\Vertex(122,40){2} \Vertex(158,40){2}
\Text(104,15)[]{$\omega,\sigma$}  \Text(176,15)[]{$\omega',\sigma'$} 
\Text(104,63)[]{$\omega,\sigma$}  \Text(176,63)[]{$\omega',\sigma'$}
\Text(140,40)[]{$\omega_1$} 
\end{picture}
}

\Text(212,40)[]{+}

\put(0,0){
\begin{picture}(400,80)(-20,0)
\ArrowLine(210,23)(227,40) \ArrowLine(227,40)(210,57)
\ArrowArcn(246,40)(17,0,180) \ArrowArcn(246,40)(17,180,0)
\ArrowArcn(282,40)(17,0,180) \ArrowArcn(282,40)(17,180,0)
\ArrowLine(318,23)(299,40) \ArrowLine(299,40)(318,57)
\Vertex(228,40){2} \Vertex(264,40){2} \Vertex(300,40){2}
\Text(210,15)[]{$\omega,\sigma$}  \Text(318,15)[]{$\omega',\sigma'$} 
\Text(210,63)[]{$\omega,\sigma$}  \Text(318,63)[]{$\omega',\sigma'$}
\Text(246,40)[]{$\omega_1$} \Text(282,40)[]{$\omega_2$}
\end{picture}
}

\Text(360,40)[]{+}
\Text(380,40)[]{$\cdots$}

\end{picture}
\end{center}
  \vspace{-.45cm}
\begin{center}
\begin{picture}(400,80)(0,0)
\Text(0,40)[l]{b)}
\ArrowLine(34,23)(51,40) \ArrowLine(51,40)(34,57)
\ArrowLine(70,23)(53,40) \ArrowLine(53,40)(70,57)
\Vertex(52,40){2}
\Text(34,15)[]{$\omega,\sigma$}  \Text(70,15)[]{$\omega',-\sigma$} 
\Text(34,63)[]{$\omega',\sigma$}  \Text(70,63)[]{$\omega,-\sigma$}

\Text(94,40)[]{$-$}

\put(0,0){
\begin{picture}(400,80)(-10,0)
\ArrowLine(104,23)(121,40) \ArrowLine(121,40)(104,57)
\ArrowArcn(140,40)(17,0,180) \ArrowArcn(140,40)(17,180,0)
\ArrowLine(176,23)(159,40) \ArrowLine(159,40)(176,57)
\Vertex(122,40){2} \Vertex(158,40){2}
\Text(104,15)[]{$\omega,\sigma$}  \Text(176,15)[]{$\omega',-\sigma$} 
\Text(104,63)[]{$\omega,-\sigma$}  \Text(176,63)[]{$\omega',\sigma$}
\Text(140,40)[]{$\omega_1$}
\end{picture}
}

\Text(212,40)[]{$-$}

\put(0,0){
\begin{picture}(400,80)(-20,0)
\ArrowLine(210,23)(227,40) \ArrowLine(227,40)(210,57)
\ArrowArcn(246,40)(17,0,180) \ArrowArcn(246,40)(17,180,0)
\ArrowArcn(282,40)(17,0,180) \ArrowArcn(282,40)(17,180,0)
\ArrowLine(318,23)(299,40) \ArrowLine(299,40)(318,57)
\Vertex(228,40){2} \Vertex(264,40){2} \Vertex(300,40){2}
\Text(210,15)[]{$\omega,\sigma$}  \Text(318,15)[]{$\omega',-\sigma$} 
\Text(210,63)[]{$\omega,-\sigma$}  \Text(318,63)[]{$\omega',\sigma$}
\Text(246,40)[]{$\omega_1$} \Text(282,40)[]{$\omega_2$}
\end{picture}
}

\Text(360,40)[]{$-$}
\Text(380,40)[]{$\cdots$}

\end{picture}
\end{center} 
\vspace{-.45cm}
\nopagebreak
\begin{center}
\begin{picture}(400,80)(0,0)
\Text(0,40)[l]{c)}
\ArrowLine(34,23)(51,40) \ArrowLine(34,57)(51,40)
\ArrowLine(53,40)(70,23) \ArrowLine(53,40)(70,57)
\Vertex(52,40){2}
\Text(34,15)[]{$\omega,\sigma$}  \Text(70,15)[]{$\omega',\sigma$} 
\Text(34,63)[]{$-\omega,\sigma'$}  \Text(70,63)[]{$-\omega',\sigma'$}

\Text(92,40)[]{+}

\put(0,0){
\begin{picture}(400,80)(-10,0)
\ArrowLine(104,23)(121,40) \ArrowLine(104,57)(121,40)
\ArrowArc(140,40)(17,180,360) \ArrowArcn(140,40)(17,180,0)
\ArrowLine(159,40)(176,23) \ArrowLine(159,40)(176,57)
\Vertex(122,40){2} \Vertex(158,40){2}
\Text(104,15)[]{$\omega,\sigma$}  \Text(176,15)[]{$\omega',\sigma$} 
\Text(104,63)[]{$-\omega,\sigma'$}  \Text(176,63)[]{$-\omega',\sigma'$}
\Text(140,15)[]{$\omega_1$} \Text(140,63)[]{$-\omega_1$}
\end{picture}
}

\Text(212,40)[]{+}

\put(0,0){
\begin{picture}(400,80)(-20,0)
\ArrowLine(210,23)(227,40) \ArrowLine(210,57)(227,40)
\ArrowArc(246,40)(17,180,360) \ArrowArcn(246,40)(17,180,0)
\ArrowArc(282,40)(17,180,360) \ArrowArcn(282,40)(17,180,0)
\ArrowLine(299,40)(318,23) \ArrowLine(299,40)(318,57)
\Vertex(228,40){2} \Vertex(264,40){2} \Vertex(300,40){2}
\Text(210,15)[]{$\omega,\sigma$}  \Text(318,15)[]{$\omega',\sigma$} 
\Text(210,63)[]{$-\omega,\sigma'$}  \Text(318,63)[]{$-\omega',\sigma'$}

 \Text(246,15)[]{$\omega_1$} \Text(282,15)[]{$\omega_2$}
 \Text(246,63)[]{$-\omega_1$} \Text(282,63)[]{$-\omega_2$}

\end{picture}
}

\Text(360,40)[]{+}
\Text(380,40)[]{$\cdots$}

\end{picture}
\\ \vspace{-.1cm}
\nopagebreak
{\sl Fig.8}
\end{center}

The renormalization of the coupling constants to
leading order in $ g $ is determined -- for each channel -- by the first two
diagrams. Diagrams a) renormalize the direct and exchange channel for electrons 
with parallel spins and the direct channel for electrons with opposite spins, diagrams
b) renormalize the exchange channel for electrons with opposite spins and diagrams
c) the Cooper channel.\\
The contributions corresponding to diagrams a) and b) only lead to a weak renormalization of
the direct and exchange channel and, solving \eq{Z14}, one can show that -- for
sufficiently small initial conditions -- the running coupling constants corresponding 
to these channels stay bounded. 
(This result remains true if one includes the leading order
contributions of the higher vertices shown in Fig. 3 , cf \cite{14}) . \\
The renormalization of the Cooper channel is -- to leading order in $ \frac{1}
{\lambda_n} $ and omitting  self energy renormalizations of  inner propagator lines --
independent of the direct and exchange channel  :
\be
g_{n+1}^C( \ell ; \sigma \cdot \sigma' ) \; \approx \; 
\frac{g_{n}^C( \ell ; \sigma \cdot \sigma' )}{1 + \beta 
g_{n}^C( \ell ; \sigma \cdot \sigma' )} \left[ 1 + 0(\frac{1}{\lambda_n}) \right]
\quad,
\label{Z17}
\ee
where  $ \beta $ is a positive constant of order unity,  and $ \ell = 0,1,2, ... $ refers to
the angular decomposition of the Cooper channel : e.g., in $ d=2 $,   the
trigonometric decomposition
$$
g_n^C( \vec{\omega}_1\cdot\vec{\omega}_2 = cos\vartheta ) \; = \;
\sum_{\ell = 0}^\infty g_n^C(\ell)\, cos(\ell \vartheta) 
$$
is used ( in $ d=3 $, the Cooper channel coupling function is expanded in a sum of Legendre functions,
rather than  trigonometric functions ).\\
The solution of \eq{Z17} is
\be
g_{n+1}^C( \ell ; \sigma \cdot \sigma' ) \; \approx \; 
\frac{g_{0}^C( \ell ; \sigma \cdot \sigma' )}{1 + (n+1)\beta 
g_{0}^C( \ell ; \sigma \cdot \sigma' )} 
\quad.
\label{ZZ17}
\ee
Suppose that  one of the Cooper channel couplings, e.g. $ \;g_0^C(\ell^\ast)\;,$
 is negative, ($\ell^\ast = 0,1,2, ... $) .  Then we encounter a
singularity after $\; n^\ast = \frac{1}{\alpha | g_0^C(\ell^\ast) |} \; $ iteration
steps,  corresponding to an energy scale $\; v_F \frac{k_F}{\lambda_0}
(\frac{1}{M})^{n^\ast} \;$. Thus, perturbation theory breaks down when $ 
n \rightarrow n^\ast $. This is the celebrated superconducting instability
(for a more detailed discussion, see \cite{15}).\\
However, one should note that this instability only appears if the Fermi surface
is symmetric under reflections at the origin, i.e., $\, \vec{k} \in {\cal S}^{d-1}_{k_F} \,$
implies that $\,  -\vec{k} \in {\cal S}^{d-1}_{k_F} \,.$  If the reflected Fermi surface
intersects the original Fermi surface transversally, the superconducting channel is
not renormalized to leading order in $ \frac{1}{\lambda_n} $.\\
If all initial values of the superconducting couplings are positive, i.e., $\; g_0^C(\ell)
\ge 0\;,\;\ell = 0,1,2, ... \;$, then, following \eq{ZZ17}, they all appear to flow to zero. However,
in general, higher order corrections (in $g$ and $ \frac{1}{\lambda_n}$)
to \eq{Z17} invalidate this conclusion; this is the "Kohn-Luttinger" - effect \cite{16}.
An analysis of this effect can be found in \cite{1,12}.

{\sl In the following, we assume that the Cooper channel is turned off,}
as e.g. in systems of electrons coupled to magnetic impurities with
broken parity- and time reversal invariance \cite{1} . Then, for
short-range interactions, all remaining coupling constants are marginal, i.e.,
$ g_{n+1} \approx g_n .$  (This result turns out to be stable against adding higher order
corrections). In this case, as the coupling constants in the effective action
$  S^{\rm{eff}}_{n} $ are suppressed by a factor $ \frac{1}{\lambda_n^{d-1}} $ ,
our calculations indicate
that the system flows to a Landau Liquid Fixed Point, as $ n \rightarrow \infty $.
In the following sections we confirm this expectation by calculating the electron 
propagator using {\sl bosonization}.  The effective
action $  S^{\rm{eff}}_{n} $ obtained by the RG analysis serves as an input for
this calculation.

For long-range interactions, the situation is more complicated. For long-range
density-density interactions of the form displayed in \eq{2},  the effective action
$  S^{\rm{eff}}_{0} $ at a large, initial scale $ \lambda_0 $ has the form given
in eqs. (\ref{11}) - (\ref{13}) . The effective interaction potential 
can be replaced by a set of coupling functions,  as described in \eq{ZZ13}. 
Due to the singularity of the interaction potential  $ \hat{V}(|\vec{p}|) $ at
$ |\vec{p}| = 0 $, the interaction processes with $\; \vec{\omega}_4 \ne 
\vec{\omega}_1 \; $ are less important than the direct scattering
processes,  where $\; \vec{\omega}_4 =  \vec{\omega}_1 \; $ and $\; \vec{\omega}_3 
=  \vec{\omega}_2 \; $. In this channel, the singularity contributes a 
supplemantary factor $ \lambda_0^\alpha $  to the quartic term (\ref{13}) in the 
action which can partially or completely compensate 
the factor $ \frac{1}{\lambda_0^{d-1}} $ .
However, resumming all  diagrams of leading order in $ \frac{1}{\lambda_0 } $ leads to
"screening"  which renders the long-range interaction effectively short-ranged.
Therefore similar results as for the short-range case are expected to hold.
We shall confirm this expectation in the approximation obtained by bosonizing the 
system. \\
For long-range transverse current-current interactions, as they occur in
Quantum Hall fluids at filling factors $\; \nu = \frac{1}{2},\frac{1}{4}, ... \;$, the
screening  mechanism is ineffective. Calculating the electron propagator by the
bosonization technique, we shall observe  the possibility for
 a deviation from Landau
Liquid behaviour,  depending on the exponent $\alpha$ which characterizes the singularity
of the interaction potential in  momentum space : for $\alpha \geq d-1 ,$ we argue that 
the system is a MFL
(similar predictions have previously been made in \cite{2,3,4,5}).

\section{Effective Gauge Field Action and Bosonization}

It is easiest to understand the meaning and accuracy of "bosonization"  by 
calculating the scaling limit of the effective gauge field action, $  {\cal W}( A ) $ ,
where $ A $ is an external electromagnetic vector potential.
 From the effective gauge
field action one can determine the (connected) Green functions of currents by 
differentiating with respect to the gauge field $ A $ . Calculating the Green functions
for the electron fields $ \Psi^\ast, \Psi $ is more complicated and is accomplished in
the next section.

The external gauge field $ \;A_\rho, \; \rho = 0,1, ... ,d \; , $ is coupled to the electron 
system by replacing derivatives in the free action $ S^0\left(\Psi^{\ast},\Psi;
\mu\right) $ , \eq{1}, by covariant ones $ \;  D_\rho(A) := \partial_\rho - i e A_\rho
\;$ ("minimal coupling"). Here $ e $ is the elementary electric charge  (we choose units
such that $ \hbar = c = 1 $) .  Eq. (\ref{1})  is then replaced by
\be
 S^0\left(\Psi^{\ast},\Psi;\mu , A\right) 
\; =  \; S^0\left(\Psi^{\ast},\Psi;\mu\right)  \; + \;
 S^{J}\left(\Psi^{\ast},\Psi;A\right) \quad ,
\label{21}
\ee
where
$$
 S^{J}\left(\Psi^{\ast},\Psi;A\right) \; = \; \int d^{d+1}x \sum_{\rho = 0}^d
\, A_\rho(x)\, j^\rho\left(\Psi^{\ast},\Psi;A\right)(x) \quad,
$$
and the current density, $ j^\rho $ , is defined (in euclidean space-time) by
\be
\begin{array}{lcc}
 j^0\left(\Psi^{\ast},\Psi \right)(x) & = & - i e \Psi^{\ast}(x) \Psi(x)
\\
&&\\
j^\ell\left(\Psi^{\ast},\Psi;A\right)(x) & = & \frac{ i e}{2 m} 
\left[ \Psi^{\ast}(x) D_\ell(A) \Psi(x)  -  \left(\Psi(x) D_\ell(A)\right)^\ast \Psi(x) \right] \;,
\end{array}
\label{22}
\ee
for $ \ell = 1, ... ,d $ .  \\
The effective gauge field action $ {\cal W}( A )  $  is obtained by integrating out
the degrees of freedom of the electrons,
\be
 {\cal W}( A )  \; := \; - log\;\left\{(\Xi^V_{\mu})^{-1} \int 
{\cal D}(\Psi^\ast,\Psi) 
\; e^{- \left[  S^0\left(\Psi^{\ast},\Psi ; \mu,A\right)  + S^V\left(\Psi^{\ast},
\Psi\right) \right]}\right\} \quad ,
\label{23}
\ee
with $  S^V\left(\Psi^{\ast},\Psi\right) $ given by \eq{2} . It is the generating 
functional for the connected Green functions. At non-coinciding arguments, one has
that
\begin{eqnarray}
\left.
\prod^n_{i=1} \frac{\delta}{\delta A_{\rho _i} (x_i)}\,\, {\cal W}
(A)\right|_{A = 0 } & = & (-1)^n  \left\langle
\prod^n_{i=1}\, j^{\rho _i} \left( \psi^\ast ,\psi ; A;
x_i\right)\right\rangle^{\mbox{\small con}}_{A=0}   \;. 
\label{24}
\end{eqnarray}

First, we consider a system of non-interacting electrons. The scaling limit,
$ {\cal W}^0_\ast( A ) $ , of the effective gauge field action $ {\cal W}^0( A ) $  
has been calculated in \cite{13}. Here,  we just sketch the essential ideas and recall the 
main results. \\
We expand the effective action  $ {\cal W}^0 (A) $\quad  -- \eq{23}, with $ S^V
\equiv 0 $ --\quad in powers of the field $ A $ :
\be
 {\cal W}^0 ( A)  = \sum^\infty_{n=1}
\frac{1}{n!} \int \prod^n_{i=1} dx^{d+1}_i\; C^{\rho_{1},...,\rho_n} (x_{1},...,
x_n)\; A_{\rho_1} (x_1) \cdots A_{\rho_n} (x_n)\;.
\label{25}
\ee
The expansion coefficients, $ C $ , are given -- at non-coinciding arguments --
by the current Green functions, cf. \eq{24}.\\
Next, we map the physical system in a space-time region $ \Lambda^{(\lambda)} $
to a reference system in the region $ \Lambda^{(1)} \equiv \Lambda  $ ,
 where $ \lambda > 1 $
is a scale parameter,  and $ \mu $ is kept fixed. We shall be interested in the
asymptotics when $ \lambda \rightarrow \infty .$ 
Under the rescaling map, points in $ \Lambda^{(\lambda)} $ transform as
$$
x = \lambda \xi \;\in \; \Lambda^{(\lambda)} \; \longmapsto \;
 \xi \; \in \;  \Lambda\quad .
$$
The gauge field $ A_\rho^{(\lambda)}(x) $   -- which probes the
response of the electron system to small external electromagnetic fields -- is 
chosen as follows :
\be
A_\rho^{(\lambda)}(x) \; = \; \frac{1}{\lambda}\, a_\rho(\xi) \quad ,
\label{26}
\ee
where $ a_\rho(\xi)  $ is an arbitrary, but fixed function on $ \Lambda  .$  Thus, the
gauge field scales like the momentum operator.
In the following calculations we consider the formal thermodynamic limit
$ \Lambda \rightarrow {\bf R}^{d+1} .$ To construct the scaling limit
 $ {\cal W}^0_\ast (a)  $   of $ {\cal W}^0 (A^{(\lambda)}) $ , we
study the asymptotic form of the current Green functions, \eq{24}, in the limit
$ \lambda \rightarrow \infty $ , using \eq{7}, and plug the result into \eq{25}.
 We define the scaling limit  $ {\cal W}^0_\ast (a)  $ as the {\sl coefficient} of
 the most divergent term in an expansion  of  $ {\cal W}^0 (A^{(\lambda)}) $
in powers of $ \lambda $ and $ \lambda^{-1} $ . \\
In the scaling limit, one encounters ultraviolet divergencies in the perturbation
series of $ {\cal W}^0_\ast (a) $ . To fix a resulting ambiguity, we use
standard Ward identities implied by gauge invariance, i.e., 
$$
\frac{\partial}{\partial x_i^{\rho_i}} C^{\rho_1,...,\rho_i,...,\rho_n}
\left( x_1,..., x_i,..., x_n\right) = 0    \quad .
$$

The result of our analysis can be summarized as follows.
In a calculation of electron Green functions at distance- and time scales of order
$ \lambda ,$ the free  fermion action 
 $  S^0\left(\Psi^{\ast},\Psi;\mu\right) $  can be replaced by the following
approximate action :
\bea
 S^0(\Psi^{\ast},\Psi) & \approx & 
\sum_{\omega \in {\cal S}_1^{d-1}} \int_{I_{\omega}} \dbar^{\,d+1}k \;
\hat{\psi}^\ast_\omega(k) \, \left( - i k_0 + v_F \vec{\omega}\vec{k} \right) \,
\hat{\psi}_\omega(k)
\nonumber \\
& \approx &
\sum_{[\omega] \in {\cal S}^+} \int d^{d+1}\xi \; \overline{\psi}_{[\omega]}(\xi) \,
\left( \gamma^0  \frac{ \partial}{\partial \xi_0} + v_F \gamma^1
\vec{\omega} \frac{ \partial}{\partial \vec{\xi}}
\right) \, \psi_{[\omega]}(\xi) 
\nonumber\\
& =:& S^0\left(\{\psi_\om^\sharp\}\right) \quad.
\label{27}
\eea
The symbol "$\approx$" indicates that the approximate action reproduces
the large distance- and time asymptotics of electron Green functions to
leading order in $ \frac{1}{\lambda} .$  The sum, $ \sum\limits_
{\om \in{\cal S}_1^{d-1}} $ , extends over the discrete set of unit vectors, 
$ \vec{\omega}_i
 , \;  i = 1, ... , N \sim \lambda^{d-1} \; , $ whose endpoints lie on the surface,
$ {\cal S}_1^{d-1} ,$  of the $d-$dimensional unit sphere (cf. Section 2).
The Fourier modes $ \hat{\psi}^\sharp_\omega(k) $ have support on
$ I_\om := {\bf R}\times\overline{B}_\om $ .
In the second line of \eq{27}, we introduce an (arbitrary) partitioning of  
$ {\cal S}_1^{d-1} $ into a positive, $ {\cal S}^+ $ , and a negative, $ {\cal S}^- $ ,
hemisphere and denote the ray through $ \vec{\omega} $ , for $ \;\vec{\omega}  \in
{\cal S}^+  ,\; $ by    $ \; [ \omega ] := \{ \vec{\omega}, -\vec{\omega} \} \; $ .
This allows us to use the "relativistic" notation
\be
\Psi_{[\omega]} \;:=\;
\left(
\begin{array}{c}
\psi_{-\omega} \\
\psi_{\omega}
\end{array}
\right)
\qquad
\overline{\Psi}_{[\omega]} \;:=\; \Psi^\ast_{[\omega]} \gamma^0 \; = \; 
\left(
\begin{array}{cc}
\psi^\ast_{\omega} &
\psi^\ast_{-\omega}
\end{array}
\right) \quad , 
\label{28}
\ee
with $ \gamma^0 = \sigma_1 $ and $ \gamma^1 = \sigma_2 $ . In the following
, we assume that electron spins are frozen in a fixed direction (we shall omit 
 spin indices) . 

In determining the effective gauge field action ${\cal W}^0_\ast ,$  the term
 $ S^{J}\left(\Psi^{\ast},\Psi;A\right) ,$  coupling 
 the gauge field  $ A_\rho^{(\lambda)}(x) \; = \; \frac{1}{\lambda}\,
 a_\rho(\xi)  $  to the electron fields, can be replaced by
\bea
S^J\left(\Psi^{\ast},\Psi;A^{(\lambda)}\right) & \approx &
\int d^{d+1}\xi \sum_{\omega \in {\cal S}^{d-1}_1} \left[ - i a_0(\xi) -
v_F \vec{\omega}\vec{a}(\xi) \right] \;\cdot \qquad \nonumber \\
& & \qquad \qquad \cdot \; \sum_{\omega' \in {\cal S}^{d-1}_1}\;
e^{ i \lambda k_F (\vec{\omega} - \vec{\omega}\,') \vec{\xi}}\;
:\psi^\ast_{\omega'}(\xi) \psi_\omega(\xi): 
\nonumber\\
& =: &  S^J\left(\{\psi_\om^\sharp\} ; a \right) \quad.
\label{29}
\eea
Again, the symbol "$\approx$" indicates that, in a calculation of
 electron Green functions at distance- and time scales of order $\lambda, $ this
approximation yields the leading contribution in   
 $ \frac{1}{\lambda} $ .
As mentioned above, in the expansion of the effective gauge field action 
 ${\cal W}^0_\ast $ in powers of $A, $ we evaluate the expansion coefficients
-- i.e. the current Green functions -- only at non-coinciding points.  The
omission of self contractions is indicated by the normal ordering of the product
of electron fields in \eq{29}. The local terms of the expansion coefficients are
determined by requiring gauge invariance.
 \\
Given a ray $ [\omega] $ , we define a current density
\be
j_\rom(\xi) \; := \; \sum_{\vom' \in {\cal S}^+} \left\{
e^{ i k_F (\vom' - \vom) \lambda\vec{\xi}}\; :\psi^{ \ast}_{-\om'}(\xi) 
\psi_{-\om}(\xi): \; + \;
e^{ i k_F (\vom - \vom') \lambda\vec{\xi}}\; :\psi^{ \ast}_{-\om}(\xi) 
\psi_{-\om'}(\xi):  \right\} \quad,
\label{30}
\ee
corresponding to electron motion in the direction of $ -\vom $ , and a current
density
\be
\overline{\jmath}_\rom(\xi) \; := \; \sum_{\vom' \in {\cal S}^+ } \left\{
e^{ -i k_F (\vom' - \vom) \lambda\vec{\xi}}\; :\psi^{ \ast}_{\om'}(\xi) 
\psi_{\om}(\xi): \; + \; 
e^{ -i k_F (\vom - \vom') \lambda\vec{\xi}}\; :\psi^{ \ast}_{\om}(\xi) 
\psi_{\om'}(\xi): \right\} 
\label{Z30}
\ee
corresponding to electron motion in the direction of $ \vom $ .\\
Using \eq{8}, one can establish that the  quasi $1+1$
dimensional current densities
\be
\begin{array}{ccc}
j^0_\rom(\xi) & := &  \frac{i}{2}\,\left( j_\rom(\xi) + \overline{\jmath}_\rom(\xi) \right) 
\quad\\
&&\\
j^1_\rom(\xi) & := &  \frac{1}{2}\,\left( j_\rom(\xi) - \overline{\jmath}_\rom(\xi) \right) 
\quad ,  \end{array}
\label{31}
\ee
for $ \; \rom \in [{\cal S}^{d-1}_1] \; $ ,  are related to the  $d+1$ dimensional current
 density, $\;
j^\rho, \; \rho = 0, ... , d \; $ , defined in \eq{22}, by 
\be
\begin{array}{lcccl}
j^0(x) & := & j^0\left(\Psi^\ast,\Psi\right)(x) & \approx &
\lambda^{-d} \sum\limits_{\om \in {\cal S}^+} j^0_\rom(\xi) \quad
\\
&&&&\\
j^\ell(x) & := & \left. j^\ell\left(\Psi^\ast,\Psi;A\right)\right|_{A = 0}(x) & \approx &
 \lambda^{-d} \sum\limits_{\om \in {\cal S}^+} v_F\, \om^\ell j^1_\rom(\xi) \quad ,
\end{array}
\label{32}
\ee
where $\; \ell = 1, ... , d \; $ .
 Eq. (\ref{32}) holds  for
current Green functions  to leading order in $ \frac{1}
{\lambda} ,$ at non-coinciding arguments . For simplicity, the electron charge, $ -e $ , has been set to $ -1$.\\
Eq. (\ref{29}) can be written in the suggestive form
\be
S^J\left(\Psi^{\ast},\Psi;A^{(\lambda)}\right) \; \approx \;
S^J\left(\{\psi_\om^\sharp\} ; a \right) \; := \;
\sum_{\omega \in {\cal S}^+}\sum_{B = 0,1}\int d^{d+1}\xi \;
a_B^\om(\xi) \; j^B_\rom(\xi) \quad ,
\label{33}
\ee
with
$$
a_0^\om(\xi) := a_0(\xi) \quad , \quad a_1^\om(\xi) := v_F \vom \vec{a}(\xi) \quad.
$$

{\sl One observes that -- to leading order in the inverse scale parameter
 $ \lambda^{-1} $ --  the original $d+1$ dimensional system decomposes into
 independent, quasi $1+1$ dimensional subsystems, one along each direction
$ \rom $ of $ {\bf R}^d $ . The $1+1$ dimensional subsystems describe
relativistic fermions moving along the direction $ \rom $ with velocity $ \pm v_F $.
Hence, the calculation of $  {\cal W}^0_\ast(a) $ is reduced to the calculation of the 
effective gauge field action of a family of independent Schwinger models .}\\
One obtains
\bea
{\cal W}^0_\ast(a) & = & \frac{1}{\lambda^{d-1}}\frac{1}{\pi} (\frac{k_F}{2\pi})^{d-1} \frac{1}{v_F} \sum_{\om \in
{\cal S}^+} \int d^{d+1}\xi \; \sum_{B=0,1}\; a_B^{\om, T}(\xi) a_B^{\om, T}
(\xi) \nonumber \\
 & = & \frac{1}{2} \int_{{\bf R} \times \overline{B}} \dbar^{\,d+1}k\;
 \sum_{\rho,\sigma = 0}^d\; \hat{a}_\rho(-k)\,{\bf \Pi}^{\rho\sigma}_{\ast} (k)\,
\hat{a}_\sigma (k) \quad ,
\label{34}
\eea
where we used
$$
 a_B^{\om, T}(\xi) \; := \;\sum_{C =0,1} \left( \delta_{B,C} - 
\frac{\partial_B^\om \partial_C^\om}{(\partial_0^\om)^2 + 
(\partial_1^\om)^2} \right) \; a_C^\om(\xi)
$$
\be
\mbox{and} \hspace{15cm} \label{35}
\ee
$$
\partial_0^\om := \frac{\partial}{\partial \xi_0} \qquad
\partial_1^\om :=  v_F \vom \frac{\partial}{\partial \vec{\xi}} \quad.
$$
Note the non-trivial (and important) fact that $ {\cal W}^0_\ast $ is only quadratic in
 the gauge field.\\
Invariance under  gauge transformations, space rotations and parity - or time -
reversal implies that the (euclidean) polarization tensor $ {\bf \Pi}^{\rho\sigma}_{\ast} $
has the general form
 $$
{\bf \Pi}^{00}_{\ast} (k) \; = \; \frac{\vec{k}^2}{k^2_0} {\bf \Pi}^\ell_{\ast} (k)
\qquad
\qquad {\bf \Pi}^{0i}_{\ast} (k) \; = \;{\bf \Pi}^{i0}_{\ast} (k) \; = \; - \frac{k^{i}}
{k_0} {\bf \Pi}^\ell_{\ast}(k) 
$$
\be
\hfill
\label{36}
\ee
$$
{\bf \Pi}^{ij}_{\ast} (k) \; = \; {\bf \Pi}^t_{\ast} (k) \left(
\delta^{ij} - \frac{k^ik^j}{\vec{k}^2}\right) + {\bf \Pi}^\ell_{\ast} (k)
\frac{k^ik^j}{\vec{k}^2}\quad\,\,\,\, i,j = 1,...,d \; .
$$
The calculation of the two independent functions $ {\bf \Pi}^\ell_{\ast} $ and  
$ {\bf \Pi}^t_{\ast}   ,$  for non-interacting electrons,  yields the result :
\bea
{\bf \Pi}^\ell_{\ast} (k) &=& \left\{
\begin{array}{ll}
\frac{e^2}{\pi} k_F v_F \left[
\left(
1 + \sqrt{1 + \left( \frac{v_Fk}{k_0}\right)^2}\right)
\sqrt{ 1+ \left( \frac{v_F k}{k_0}\right)^2}\right]^{-1} & d= 2\\
\frac{e^2}{\pi^2} \frac{(k_F)^2}{v_F} \frac{k^2_0}{\vec{k}^2} \left(
1 - \left|\frac{k_0}{v_Fk}\right|\,\, \mbox{arctg}\,\,
\left|\frac{v_Fk}{k_0}\right|\right) & d = 3
\end{array}
\right.
\nonumber \\
&& \label{Z36} \\
{\bf \Pi}^t_{\ast} (k)& =& \left\{
\begin{array}{ll}
\frac{e^2}{\pi} k_F v_F \frac{|k_0|}{\sqrt{k^2_0 + (v_F k)^2}} -
{\bf \Pi}^\ell_{\ast} (k) \; = \frac{e^2}{\pi} k_F v_F\left[ 
1 + \sqrt{1 + \left( \frac{v_Fk}{k_0}\right)^2} \right]^{-1}
 & d = 2\\
\frac{1}{2} \left( \frac{e^2}{\pi^2} (k_F)^2 \left|\frac{k_0}{k}\right| \,\,
\mbox{arctg}\,\, \left|\frac{v_Fk}{k_0}\right| - {\bf \Pi}^\ell_{\ast} (k)\right) & d
= 3 \quad .
\end{array}
\right.
\nonumber
\eea
Comparing this result to the small-$k$ asymptotics  of formulas
found in standard textbooks, one finds agreement -- up to the "diamagnetic term"
of the transversal part which is proportional to $ |\vec{k}|^2 $ 
and yields  lower order corrections in $ \frac{1}{\lambda} .$

Eq.(\ref{34}) is the result of resumming the  leading contributions in
the (formal) expansion of the effective gauge field action $ {\cal W}^0(A) $ in
powers of $\frac{1}{\lambda} .$\\
By power counting,  the  possible diagrams contributing to leading order in
$ \frac{1}{\lambda} $ to the effective gauge field action are the following ones :\\

\begin{center}
\begin{picture}(400,100)(0,0)
\Text(25,50)[]{$ \textstyle \sum\limits_{\omega \in {\cal S}^{d-1}}$}
\Line(50,15)(50,85)  \Line(50,15)(55,15) \Line(50,85)(55,85)
\Line(379,15)(379,85)  \Line(379,15)(374,15) \Line(374,85)(379,85)
\ArrowArcn(87,50)(15,180,0) \ArrowArcn(87,50)(15,360,180)
\crossoncircle(87,50,15,0,6,1.8)  \crossoncircle(87,50,15,180,6,1.8)
\ArrowArcn(174,50)(22,135,45) \ArrowArcn(174,50)(22,45,315)
\ArrowArcn(174,50)(22,315,225) \ArrowArcn(174,50)(22,225,135)
\crossoncircle(174,50,22,45,6,1.8)  \crossoncircle(174,50,22,135,6,1.8)
\crossoncircle(174,50,22,225,6,1.8)  \crossoncircle(174,50,22,315,6,1.8)
\ArrowArcn(276,50)(30,90,30) \ArrowArcn(276,50)(30,30,330)
\ArrowArcn(276,50)(30,330,270) \ArrowArcn(276,50)(30,270,210)
\ArrowArcn(276,50)(30,210,150) \ArrowArcn(276,50)(30,150,90)
\crossoncircle(276,50,30,30,6,1.8)  \crossoncircle(276,50,30,90,6,1.8)
\crossoncircle(276,50,30,150,6,1.8)  \crossoncircle(276,50,30,210,6,1.8)
\crossoncircle(276,50,30,270,6,1.8)  \crossoncircle(276,50,30,330,6,1.8)
\Text(127,50)[]{$+$} \Text(221,50)[]{$+$}
\Text(321,50)[]{$+$} \Text(346,50)[]{$\cdots$}
\Text(87,50)[]{$\omega$}  \Text(174,50)[]{$\omega$}  \Text(276,50)[]{$\omega$}
\end{picture}\\
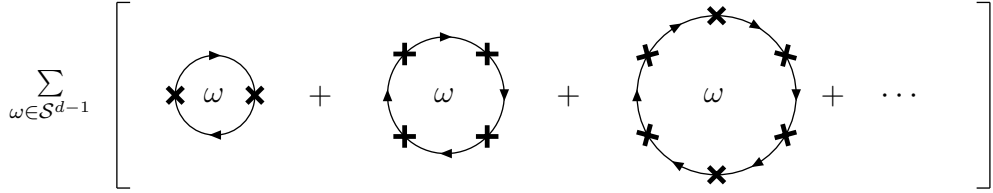
{\sl Fig.9}
\end{center}

\noindent Straight lines stand for the electron
propagators 
$$ G^0_{\om}(k)\; = \; \frac{1}{i k_0 - v_F \vom\vec{k}} \cdot {\bf 1}_{
\overline{B}_\om}(\vec{k})
$$ 
and crosses for the gauge field insertions $ a_\mu^{\om} $
(note that, in contrast to the diagrammar introduced in the context of the RG,
the electron propagators are understood to be integrated over the entire domain 
 $ I_\om = {\bf R}\times
\overline{B}_\om $ ) .\\
 However, \eq{34} states that, because of the special form of the propagators 
$ G^0_{\om} ,$  only the first term in Fig.9 with two gauge field insertions contribute to
$ {\cal W}^0_\ast(a) $ .\\
In addition, gauge invariance requires the introduction of local Schwinger terms
which couple, along a ray $\rom ,$ right- and left-movers. 

Remembering that, in the $1+1$ dimensional Schwinger model,  the calculation of current
Green functions in the fermionic theory can be reproduced by expressing the current densities
in terms of a free, massless bose field (see, e.g., \cite{13,17}, and references therein), we introduce
following identifications \nolinebreak :
\bea
\;j_\rom(\xi)\; & \longleftrightarrow & \frac{2}{\sqrt{\pi}} (\frac{k_F}
{2\pi})^{\frac{d-1}{2}}  \partial^\om \varphi_\rom
(\xi)
\nonumber \\
&& \label{37}\\
\;\overline{\jmath}_\rom(\xi)\; & \longleftrightarrow &  - \frac{2}
{\sqrt{\pi}} (\frac{k_F}
{2\pi})^{\frac{d-1}{2}}  \overline{\partial}^\om \varphi_\rom(\xi)\quad,
\nonumber
\eea
for $ \;\vom \in {\cal S}^+ \; $ , with $\; \rom = \{\vom,-\vom\}\; $ denoting the
 corresponding ray, and
$$
\partial^\om := \frac{1}{2} \left( - i \frac{\partial}{\partial\xi_0} + v_F \vom \frac{ 
\partial}{\partial\vec{\xi}} \right)
\qquad
\overline{\partial}^\om := \frac{1}{2} \left( i \frac{\partial}{\partial\xi_0} + v_F \vom
 \frac{\partial}{\partial\vec{\xi}} \right) \quad .
$$
Eq.(\ref{37}) is equivalent to
$$
\;j^B_\rom(\xi)\; \; \longleftrightarrow \; \frac{i}{\sqrt{\pi}} 
(\frac{k_F}{2\pi})^{\frac{d-1}{2}} \varepsilon^{BC} 
 \partial^\om_C \varphi_\rom(\xi)
\quad ,
$$
for $ \;B,C = 0,1 \; $ and  $ \;\vom \in {\cal S}^+ \; $  .\\
The action  of the set $ \{\bo(\xi)\} $  of bose fields  is given by
\be
S^0(\{\bo\}) \; = \; - \frac{1}{2} \sum_\rom  \int_{{\bf R} \times \overline{B}_\om}
\dbar^{\,d+1}k\; \hbo(-k)\, \left[ k^2_0 + (v_F \vom \vec{k})^2 \right] \;\hbo(k) 
\quad , \label{39}
\ee
where the Fourier modes, $ \hbo(k) ,$ of the bose field $ \bo(\xi) $
  have support on $ {\bf R} \times \overline{B}_\om .$ ( For a given value
$k_0 ,$ the modes $ \hbo(k_0, \cdot ) $ have a compact support. We choose this
support to be $ \overline{B}_\om ,$ but, in principle, there is no need to choose
a support identical to the one of the fermion modes $ \hat{\psi}^\sharp_\om(k_0, \cdot ) .$ ) 

\noindent One can verify that 
$$
\Xi^{-1}_\varphi \int  {\cal D} \{\bo\} \; e^{-  S^0(\{\bo\})}
\; \left\{\; \prod_{j=1}^n\, \left[ \frac{i}{\sqrt{\pi}} 
(\frac{k_F}{2\pi})^{\frac{d-1}{2}} \varepsilon^{B_jC_j}\; \partial^\om_{C_j} \varphi_
{[\om_j]}(\xi_j) \right] \right\} \; \approx \hspace{6cm}
$$
\be 
\hspace{6cm} \approx \; \Xi^{-1}_{\psi} \int   {\cal D} \{\psi^\sharp_\om\} \; e^
{-  S^0(\{\psi^\sharp_\om\})} \;\left\{\; \prod_{j=1}^n\, j^{B_j}_{[\om_j]}(\xi_j)\,
 \right\}\;,
\label{Z39}
\ee
in the sense of distributions.

The representation (\ref{37}) of the one-dimensional current densities, $ 
j^B_\rom $ , in terms of the bose fields, $ \varphi_\rom $ , is implied by the 
local conservation of the electron current, \eq{32} , along each ray $ \rom $ 
seperately.  These formulas are equivalent to Luther-Haldane bosonization \cite{9,10,11}.\\
{\sl The identification (\ref{37}) reproduces only the leading term
of the fermionic perturbation theory, i.e.} :
\be
{\cal W}^0_\ast(a) \; = \;"\lim_{\lambda \rightarrow \infty}" - log \left\{Z_\varphi^{-1}
\int {\cal D}\{\bo\}\;
e^{-\left[ S^0(\{\bo\}) + S^J(\{\bo\};a) \right]} \right\} \quad ,
\label{40}
\ee
with
$$
S^J(\{\bo\} ; a) \; = \; \sum_{\om \in {\cal S}^+ } \int d^{d+1}\xi \;
a_B^\om(\xi) \, \varepsilon^{BC}  \,\frac{i}{\sqrt{\pi}} 
(\frac{k_F}{2\pi})^{\frac{d-1}{2}} 
 \partial^\om_C \varphi_\rom(\xi)  \quad .
$$
The symbol $ "\lim_{\lambda \rightarrow \infty}"$ stands for determining
the {\sl coefficient} of the leading term of the expansion in $ \lambda $ and
$ \lambda^{-1}.$\\
The derivatives $ \partial^\om \bo $ and $ \overline{\partial}^\om \bo $ of
the bosonic fields $ \bo $  describe the current density fluctuations, $ j_\rom $ and
$ \overline{\jmath}_\rom , $  determining the scaling limit of the system. For a fixed
direction $ \rom $  they are composed of electron modes with momenta near the
points $ k_F \vom $ and $ -k_F \vom , $  resp., on the Fermi surface. In the scaling
limit, they probe the Fermi surface only locally around the points $ k_F \vom $ and
 $ -k_F \vom $ .

This  implementation of bosonization is a special
realization of a more general formalism, presented in \cite{18}.\\
Abelian gauge invariance implies the local conservation law
\be
\partial_\rho j^\rho (x) \; = \; 0 \qquad. \label{54}
\ee
Instead of expressing the current density in terms of the elementary fields 
$ \Psi^\ast,\Psi $ and imposing the constraint (\ref{54}), one can introduce
new field variables which guarantee \eq{54} by construction. For $d+1$
dimensional currents, $ j^\rho $ , one needs, in general, antisymmetric
gauge forms of rank $d-1$ . However, considering only the scaling limit
of the system, we can take advantage of a substantial simplification : 
the fermionic theory decomposes into a family of quasi $1+1$ dimensional
 subsystems ("dimensional reduction"), one along each direction $ \rom $ of
$ {\bf R} $ . Each subsystem describes quasi $1+1$ dimensional, "relativistic"
electrons. Gauge inveriance has to be fullfilled for each subsystem
-- i.e. in each direction -- seperately, and implies the conservation laws for the 
associated currents, $ j_\rom^B \;,\; B = 0,1 \;$ , i.e.
\be
\frac{\partial}{\partial x_0} j^0_\rom(x) \;+\; v_F \vom \frac{\partial}{\partial\vec{x}}
 j^1_\rom(x) \; = \; 0
\qquad. \label{55}
\ee
Expressing the currents $ j^B_\rom $  in terms of the free, massless bose fields
 $ \bo $  guarantees that \eq{55} holds.

\noindent Next, we  study the effective gauge field action for interacting systems. \\
In section 2, by using the RG group method, we successively eliminated the electron
modes with momenta outside the shell
$ \; \Omega_{\lambda_n} := \left\{\; \vec{p} \in {\bf R}^d \;;\;
\left| \vec{p} - k_F\frac{\vec{p}}{|\vec{p}|} \right| \; \leq \; \frac{k_F}
{2 \lambda_n}\;\right\} \; $ of thickness $\; \frac{k_F}{\lambda_n} \ll k_F\; $
around the Fermi surface,  in order to determine the effective action $ S_n $
for the remaining modes with momenta inside the shell $ \Omega_{\lambda_n} .$
The resulting effective action $ S_n $ describes the physics at energy scales smaller
than $ v_F \frac{k_F}{\lambda_n} .$
Under the assumptions specified in Sect.2,  one can reach very small energy scales
 $\; v_F \frac{k_F}{\lambda_n} \ll
v_F k_F \;$ before the form of the corresponding effective action $ S_n $
differs considerably from the original action $ S .$ \\
In this section, we consider
"spinless" electrons and suppose that, in addition, the Cooper channel is
turned off. Then the terms $ S_n^2 $ and $ S_n^4\, ,$  quadratic and 
quartic in the fields $\hat{\psi}^\sharp_\om(k) ,$ are given by
\be
S^2_n\; \approx \; 
\sum_{\omega \in {\cal S}_1^{d-1}} \int_{I_{\omega}} \dbar^{\,d+1}k \;
\hat{\psi}^\ast_\omega(k) \, \frac{-1}{Z_n} \,  \left( i k_0 - v_{Fn} \vec{\omega}\vec{k} \right) \,
\hat{\psi}_\omega(k)
\label{B42}
\ee
and
$$
S^4_n \;  \approx \;
\frac{k_F^{1-d}}{2\lambda^{d-1}_n} \frac{1}{Z_n^2}
\sum_{\omega_1,\omega_2 \in {\cal S}^{d-1}_1}  
\int_{I_{\omega_1}}\dbar^{\,d+1}k^{(1)}\, \cdots\,\int_{I_{\omega_1}}\dbar^{\,d+1}
k^{(4)} \; \cdot
\hspace{2cm}
$$
$$
\hspace{3cm} \cdot \;
\delta^{(d+1)}\left(k^{(1)}\!+\!k^{(2)}\!-\!k^{(3)}\!-\!k^{(4)} \right) \; 
g_n(\vom_1\vom_2 ; {\scriptstyle \frac{k^{(1)}-k^{(4)}}{\lambda_n}}) \;\cdot
$$
\be
\hspace{3.5cm} \cdot\; 
\hat{\psi}^\ast_{\omega_1}(k^{(4)}) \,
\hat{\psi}^\ast_{\omega_2 }(k^{(3)}) \,
\hat{\psi}_{\omega_2 }(k^{(2)}) \,
\hat{\psi}_{\omega_1}(k^{(1)}) \quad , \qquad \quad
\label{42}
\ee
with
\be
g_n(\vom_1\vom_2 ; {\scriptstyle \frac{k}{\lambda_n}} )\quad=\quad
\left\{g_n^d(\vom_1\vom_2 ; {\scriptstyle \frac{k}{\lambda_n}})\;-
\;g_n^e(\vom_1\vom_2 ; {\scriptstyle \frac{k}{\lambda_n}})
\right\}\quad .
\label{43}
\ee

The "direct"- and "exchange"-coupling functions $ g^d $ and $ g^e $ have been
 defined in \eq{17}. The symbol "$\approx$" stands for "equal to leading order in
 $ \frac{1}{\lambda_n} $".\\
Contributions to $ S_n $ corresponding to higher orders in the Taylor expansion
of the coefficient functions of the quadratic and quartic terms in the momentum
variables,  or contributions involving more than 4 electron fields are neglected.
By engineering scaling they are irrelevant, cf. Sect.2 . For systems
with a short range two-body interaction potential $ V ,$ the coupling functions
$\;g_n(\vom_1\vom_2 ; {\scriptstyle \frac{k}{\lambda_n}} )\;$
can be replaced -- to leading order in $ \frac{1}{\lambda_n} $ -- by the coupling
constants $ \; g_n(\vom_1\vom_2 )\;.$ We first restrict our attention to such systems.\\
The term $ S^J_n(\{\psi_\om^\sharp\} ; a) $ coupling the electron fields
to the gauge field a has the form\\
\vspace{.2cm}
\be
S^J_n(\{\psi_\om^\sharp\} ; a) \; \approx \; 
\sum_{\omega \in {\cal S}^+}\sum_{B = 0,1}\int d^{d+1}\xi \;
a_B^\om(\xi) \; j^B_\rom(\xi) \quad ,
\label{Z43}
\ee

where $\;( a_0^\om , a_1^\om ) := ( a_0 , v_{Fn} \vom \cdot \vec{a} ) \;,$
and the quasi $1+1$ dimensional currents $  j^B_\rom(\xi) $ are defined as in the
non-interacting system, cf. eqs. (\ref{30}) - (\ref{32}), but with the products
$\;\psi^\ast_{\om'}(\xi) \psi_\om(\xi)\;$ of the electron fields replaced by
$\;\frac{1}{Z_n}\, \psi^\ast_{\om'}(\xi) \psi_\om(\xi)\;$ ( in the non-interacting
system $Z_n \equiv 1$). A Ward identity relates the renormalization of the vertex (\ref{Z43}) to
the renormalization of the electron propagator, preventing the appearence of a new, independent
renormalization factor.\\
The quartic interaction term $ S^4_n $ given by \eq{42} can be expressed in terms of 
the quasi $1+1$ dimensional currents :
$$
S^4_n\left(\{\psi_\om^\sharp\}\right)
\; = \; 
\frac{1}{8} \frac{k_F^{1-d}}{\lambda_n^{d-1}}  \sum_{\om_1,\om_2 \in {\cal S}^+} \int
 d^{d+1}\xi \int d^{d+1} \eta\; \delta_{k_F}^{d+1}(\xi-\eta) \; \cdot \hspace{6cm} 
$$
$$
\hspace{4cm}
\bigg\{
g_n(\vom_1\vom_2) \left[  j_\jom{1}(\xi)  j_\jom{2}(\eta)
+ \overline{\jmath}_\jom{1}(\xi)  \overline{\jmath}_\jom{2}(\eta) \right] \, -
\hspace{5cm}
$$
\be
\hspace{6cm} 
-  \,g_n(-\vom_1\vom_2) \left[ j_\jom{1}(\xi) 
\overline{\jmath}_\jom{2}(\eta)
+  \overline{\jmath}_\jom{1}(\xi)  j_\jom{2}(\eta) \right] \bigg\} \; ,
\label{ZZ43}
\ee
with
$$
\delta^{d+1}_{k_F}(\xi\!-\!\eta) \;=\; \int_{\bf R} \dbar k_0
\int_{|\vec{k}| \leq \frac{k_F}{2}} \dbar^{\,d}k\;
e^{i k (\xi-\eta)}\quad .
$$

Replacing the quadratic term $ S^2_n $  of the action by the bosonic action
$ S^0(\{\bo\}) $ given in \eq{39} and inserting the bosonization identities (\ref{37})
for the current densities in eqs. (\ref{Z43}) and (\ref{ZZ43}), we end up with
a {\sl gaussian } bosonic theory. The bosonized versions of eqs. (\ref{Z43}) and 
(\ref{ZZ43}) are given -- in momentum space -- by\\
\vspace{.2cm}
\be
S^J_n(\{\bo\} ; a) \; = \; \sum_{\om \in {\cal S}^+ } \sum_{B,C=0,1} \int_{I_\om}
\dbar^{\,d+1}k \;
\hat{a}_B^\om(-k) \, \varepsilon^{BC}  \,\frac{-1}{\sqrt{\pi}} 
(\frac{k_F}{2\pi})^{\frac{d-1}{2}} 
 k^\om_C \varphi_\rom(k)  
\label{B44}
\ee
and 
$$
S^V_n\left(\{\bo\}\right) \; = \;
\frac{1}{\lambda_n^{d-1}}(\frac{1}{2\pi})^{d}
\sum_{\om_1,\om_2 \in {\cal S}^+}  \int_{I_{\om_1}\cap I_{\om_2}} \dbar^{\,d+1}k
 \; \cdot \hspace{4cm}
$$
$$
\hspace{3.5cm} \cdot \; \bigg[\, g_n(-\vom_1\vom_2)\,
\varphi_\jom{1}(-k)\,\left( k^{\om_1}\overline{k}^{\om_2}
+ \overline{k}^{\om_1}k^{\om_2} \right)\,\varphi_\jom{2}(k)
\hspace{4cm}
$$
\be
\hspace{5.5cm}  +\;  g_n(\vom_1\vom_2)\;
\varphi_\jom{1}(-k)\,\left( k^{\om_1}k^{\om_2}
+ \overline{k}^{\om_1}\overline{k}^{\om_2} \right)\,\varphi_
\jom{2}(k)\, \bigg] \quad,
\label{44}
\ee
with
$$
k^\om\;=\;\frac{1}{2} ( -i k_0 + v_F \vom \vec{k} ) \qquad \mbox{and} \qquad
\overline{k}^\om\;=\;\frac{1}{2} ( i k_0 + v_F \vom \vec{k} )\quad.
$$

With the aim of calculating the scaling limit, $ {\cal W}^V_\ast(a) ,$ 
 of the effective gauge field action 
of the interacting electron system, we  replace the
fermionic action $ S_n ,$ given by eqs. (\ref{B42}) - (\ref{Z43}), by their bosonic
 version,  given by eqs. (\ref{39}), (\ref{B44}) and (\ref{44}), calculate the 
corresponding effective gauge field action, $ \widetilde{\cal W}^V(a ; {\cal P}_n) ,$
 and determine the {\sl coefficient}, $ \widetilde{\cal W}^V_\ast(a ; {\cal P}_n) ,$ of
the leading contribution in an expansion
of $ \widetilde{\cal W}^V(a ; {\cal P}_n) $ in powers of $ \lambda_n $ and $ \lambda_n^{-1} ,$
as $ \lambda_n \rightarrow \infty .$ $ {\cal P}_n $ stands for the set of parameters
characterizing the fermionic action $ S_n .$ As the bosonic theory is gaussian,
 it is easier to study than the fermionic theory where the
interaction term is quartic in the fields $ \hat{\psi}^\sharp_\om .$

But does the action $ \widetilde{\cal W}^V_\ast(a ; {\cal P}_n) $ derived by the
 bosonization 
procedure reproduce the correct result that one would obtain by iterating the
RG transformations and by determining the
resulting fixed point action ($ n \rightarrow \infty $)  ?\\
In order to answer this question, we have to clarify the approximations involved
by bosonizing the electron system described by the action $ S_n .$ The approximations
can be characterized in terms of the {\sl formal} perturbation expansion of the
fermionic theory in powers of $ \lambda_n^{-1} .$ {\sl In fact, by bosonizing an electron
system with  action $ S_n ,$ we resum all leading order contributions of 
the fermionic perturbation expansion of the effective gauge field action in powers
of $ \lambda_n^{-1} $ -- except self-energy renormalizations of inner propagator
lines.} The diagrams reproduced after bosonization are the following ones : \\
\vspace{-.4cm}
\begin{center}
\begin{picture}(390,80)(0,0)

\Text(25,40)[]{$\widetilde{\cal W}^V_\ast$}
\Text(65,40)[]{$\sim$}
\ArrowArcn(141,40)(18,180,0) \ArrowArcn(141,40)(18,360,180)
\ArrowArcn(241,40)(18,180,0) \ArrowArcn(241,40)(18,360,180)
\ArrowArcn(280,40)(18,180,0) \ArrowArcn(280,40)(18,360,180)
\Text(105,40)[]{$\sum\limits_{\omega_1}$}
\Text(205,40)[]{$\sum\limits_{\omega_1,\omega_2}$}
\Text(177,40)[]{$+$}
\Text(316,40)[]{$+$}
\Text(344,40)[]{$\cdots$}
\crossoncircle(141,40,18,0,6,1.8) \crossoncircle(141,40,18,180,6,1.8)
 \crossoncircle(241,40,18,180,6,1.8)
\crossoncircle(280,40,18,0,6,1.8) 
\Vertex(260.5,40){2.8}
\Text(141,40)[]{$\omega_1$} \Text(241,40)[]{$\omega_1$} \Text(280,40)[]{$\omega_2$}

\end{picture}\\
{\sl Fig.10}
\end{center}

As above, straight lines stand for electron propagators and crosses for gauge field
insertions. The interaction $ \frac{k_F^{1-d}}{\lambda_n^{d-1}}\; g_n(\vom_1\vom_2) \quad  $
 vertices are represented by dots. \\
In the diagrams in Fig.10 , each factor $ \frac{1}{\lambda_n^{d-1}} $ per interaction 
vertex is
compensated by a bubble-summation, $ \sum\limits_\om( \cdot ) $ .\\
{\sl These diagrams correspond to the so called $RPA$ approximation .}

In principle, to leading order in a (formal) expansion in $ \frac{1}{\lambda_n} ,$ 
 the propagator lines in
the polarization bubbles in Fig.10  are renormalized by  "cactus diagrams"
of the form :\\
\vspace{-.4cm}

\begin{center}
\begin{picture}(360,90)(0,0)

\Line(20,10)(50,10)  \Line(50,10)(80,10)
\Line(120,10)(150,10)  \Line(150,10)(180,10)
\Line(220,10)(250,10)  \Line(250,10)(280,10)
\Line(33,8)(37,12) \Line(63,8)(67,12)
\Line(133,8)(137,12) \Line(163,8)(167,12)
\Line(233,8)(237,12) \Line(263,8)(267,12)

\GCirc(50,11){2}{0} \ArrowArc(50,24)(12,271,269)
\GCirc(150,11){2}{0}  \GCirc(150,37){2}{0}

\GCirc(250,11){2}{0}  \GCirc(250,37){2}{0}  
\GCirc(250,63){2}{0}  \GCirc(263,50){2}{0}
\ArrowArc(150,24)(12,270,90) \ArrowArc(150,24)(12,90,270)
\ArrowArc(150,50)(12,271,269)
\ArrowArc(250,24)(12,270,90) \ArrowArc(250,24)(12,90,270) 
\ArrowArc(250,50)(12,270,0) \ArrowArc(250,50)(12,0,90) \ArrowArc(250,50)(12,90,270)
\ArrowArc(250,76)(12,271,269) \ArrowArc(276,50)(12,181,179)

\Text(50,24)[]{$\omega_1$} \Text(150,24)[]{$\omega_1$} 
\Text(250,24)[]{$\omega_1$} 
 \Text(150,50)[]{$\omega_2$} 
\Text(250,50)[]{$\omega_2$} 
\Text(250,76)[]{$\omega_3$} 
\Text(276,50)[]{$\omega_4$}
\Text(24,8)[t]{$\omega$} \Text(76,8)[t]{$\omega$}
\Text(124,8)[t]{$\omega$} \Text(176,8)[t]{$\omega$}
\Text(224,8)[t]{$\omega$} \Text(276,8)[t]{$\omega$}

\Text(100,10)[]{$+$} \Text(200,10)[]{$+$} \Text(300,10)[]{$+$}
\Text(330,10)[]{$\cdots$}
\end{picture}
\\
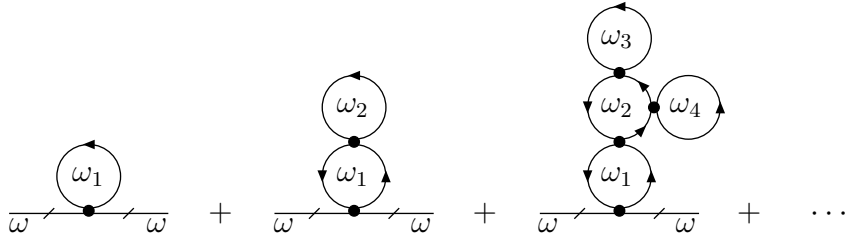
{\sl Fig.11}
\end{center}

\noindent For systems with a short-range two-body potential $ V ,$ their  contribution
 is -- to leading order in $ \frac{1}{\lambda_n} $ -- a {\sl constant},
i.e., they only lead to a displacement of the chemical potential. As the polarization
bubbles only depend on the difference of the momenta of the two
inner propagator lines, they are not modified by such a displacement of the
chemical potential.

Does this result imply that, by taking functional derivatives  of 
$ \widetilde{\cal W}^V_\ast(a ; {\cal P}_n) $ with respect to the gauge field $ a $ and 
by evaluating
the resulting current Green functions, cf. \eq{24}, for large arguments
$\;|x^{(i)}-x^{(j)}| = \lambda_n |\xi^{(i)}-\xi^{(j)}|\; ,$ with $ \lambda_n \rightarrow
\infty ,$ one obtains the leading contribution in $ \frac{1}{\lambda} $ of the
current Green functions of the interacting system ? One must remember
that we use the fermionic action $ S_n $ as an input of bosonization. Because of the
linearization of the pieces of the Fermi surface contained in the boxes $ \overline{B}_
\om ,$  this action describes the properties of the system
only correctly at momentum scales between $ \frac{k_F}{\lambda_n} $ and
$ b \frac{k_F}{\lambda_n^2} ,$ as argued in Sect.2 .
\\
It follows that the current Green functions derived from $ \widetilde
{\cal W}^V_\ast(a ; {\cal P}_n) $
reproduce the leading order contribution in $ \frac{1}{\lambda_n} $ of
the current Green functions of the interacting system for arguments
$\;|x^{(i)}-x^{(j)}| = \lambda_n |\xi^{(i)}-\xi^{(j)}| \; $ of  order 
between $ \lambda_n $ and $ \lambda_n^2 .$ In order to explore the current Green
functions of the interacting system at larger distance- and time scales, we first have
to  to iterate the RG transformations further,
with the aim of deriving  effective actions $ S_m\, ,\, m > n ,$ describing the properties of the system
at larger distance- and time scales. Such an effective action $ S_m $ can serve as a 
new  input for the bosonization procedure.

In Sect.2, we have analyzed the flow of the set $ {\cal P}_n $ of parameters
characterizing the effective action $ S_n $ for electron systems with a short-range
 two-body potential $ V .$ All the parameters of the set $ {\cal P}_n $ -- except the scale
factor $ \lambda_n $ -- tend to finite values $ {\cal P}_\ast $ in the limit
$ \lambda_n \rightarrow \infty .$ This implies that
\pagebreak
\be 
  {\cal W}^V_\ast(a)  \; = \;   "\lim_{n \rightarrow \infty}"\;\widetilde{\cal W}^V(a ; {\cal P}_n) \quad,
\label{B46}
\ee 
 i.e. the scaling limit of
the effective gauge field action is given by
\vspace{.5cm}
\be
 {\cal W}^V_\ast(a) \; = \;"\lim_{n \rightarrow \infty}"\; - log \left\{ \Xi_\varphi^{-1} 
\int {\cal D}\{\bo\}\; e^{- \left( S_\ast^0(\{\bo\}) + S_\ast^J(\{\bo\} ; a) \right)}\;
e^{-S^V_\ast(\{\bo\})} \right\} \quad,
\label{46}
\ee
where the action is defined by eqs. (\ref{39}), (\ref{B44}) and (\ref{44}). 
Again,  $ "{\lim_{\lambda_n \rightarrow \infty}}"$ stands for determining
the {\sl coefficient} of the leading term of the Laurent expansion in $ \lambda_n $ and
$ \lambda_n^{-1}.$\\
Carrying out the {\sl Gaussian}  functional integral, we find for the leading term proportional
to $ \lambda_n^{d-1} :$
\be
{\cal W}^V_\ast(a) \; = \;  \frac{1}{2}   \int\dbar^{\,d+1}k 
\sum_{\rho,\sigma = 0}^d\; \hat{a}_\rho(-k){\bf \Pi}_{V\ast}^{\rho\sigma}(k)
\hat{a}_\sigma(k) \quad. 
\label{Z46}
\ee
The polarization tensor $ {\bf \Pi}_{V\ast}^{\rho\sigma}(k) $ has the  general form
described in \eq{36}, where $ {\bf \Pi}_{V\ast}^t(k) $ is  given by
$ {\bf \Pi}_{\ast}^t(k)  ,$ and $ {\bf \Pi}_{V\ast}^\ell(k) $ can be expressed as a
Neumann series in $  g_\ast(\vom \vom') .$ The explicit expression for the Neumann
 series 
is rather complicated in the general case (but straightforward to determine).
We only write down the explicit expression for $ {\bf \Pi}_{V\ast}^\ell(k) $
in the (somewhat artificial) case when $ \tilde{g}_\ast(\vom_1\vom_2) = g_\ast .$
Then one obtains
$$
 {\bf \Pi}_{g\ast}^\ell(k)  \; = \;
  {\bf \Pi}^\ell_{\ast}(k)\;\frac{1}{1\;+\;\frac{\vec{k}^2}
{k_0^2}\,\frac{ g_\ast}{k_F^{d-1}}\, {\bf \Pi}^\ell_{\ast}(k)}\quad.
$$ 
(The functions $  {\bf \Pi}^{\ell,t}_{\ast} $  are defined in \eq{37}) .

For systems with a long-range two-body potential $ V ,$ the  renormalization 
flow of the
effective action  $ S_n $  is not yet fully understood. Given
the effective action $ S_n ,$ we can calculate calculate the corresponding effective
gauge field action $ \widetilde{W}(a , {\cal P}_n) $ by bosonizing the system.
This yields a result for the current Green functions in
the  domain of validity discussed above. In order to derive 
results  at larger distance- and time-scales, one is obliged to make assumptions on
the flow of the effective action $ S_n ,$ as $ n \rightarrow \infty .$ 

\noindent In the following, we discuss the
calculation of  $ \widetilde{W}(a , {\cal P}_n) $ for three examples of systems with
singular interactions :\\
\vspace{.3cm}
 
\noindent i) {\bf Long-range, density-density interactions}

We consider interactions of the type shown in \eq{2} with a long-range interaction
potential $ g V $, i.e., one whose Fourier transform, $ \;g \hat{V}(|\vec{p}|) =
\frac{g}{k_F^{d-1}} \left|\frac{k_F}{\vec{p}}\right|^\alpha \; $ , becomes singular at $ |\vec{p}| = 0 $ . 
The exponent $ \alpha $ and  the coupling constant $ g $ are supposed
to be positive. We assume that the effective action $ S_n $ at an energy scale
$ v_{Fn} \frac{k_F}{\lambda_n} $ has the form specified in eqs. (\ref{B42}) - (\ref{Z43})
and that all the parameters of the set $ {\cal P}_n $ characterizing the action
$ S_n $ -- except the scale parameter $ \lambda_n $ --
 tend to  finite values in the limit $ n \rightarrow \infty .$ In particular,
the set of coupling functions $\;g_n(\underline{\om}  ; {\scriptstyle \frac{1}
{\lambda_n}} \underline{k})\;$
is supposed to be related to the initial interaction potential by \eq{ZZ13},
for arbitrary $ n .$
 We can neglect the exchange
channel with respect to the direct channel.  In the direct channel -- contrary
to the case of  short-range interactions -- we have to retain the (singular)
dependence on the small momenta, $ {\scriptstyle \frac{1}{\lambda_n}} \underline{k} $ , 
as indicated in \eq{ZZ13}. Hence,  in \eq{42}, the dominant coupling
constants are given by
\be
g_n(\vom_1\vom_2 ; {\scriptstyle \frac{1}{\lambda_n}}\underline{k})
\; = \; k_F^{d-1} g_n \hat{V}( {\scriptstyle \frac{1}{\lambda_n}}|\vec{k}|) \; = \; 
g_n \left|\frac{k_F \lambda_n}{\vec{k}}\right|^\alpha   \quad.
\label{47}
\ee
By bosonizing the system, one obtains the result
\be
\widetilde{\cal W}^D(a ; {\cal P}_n)  \; = \;  \frac{1}{2} \lambda_n^{d-1}
  \int\dbar^{\,d+1}k 
\sum_{\rho,\sigma = 0}^d\; \hat{a}_\rho(-k){\bf \Pi}^{\rho\sigma}_{D}(k;{\cal P}_n)
\hat{a}_\sigma(k) \quad ,
\label{49}
\ee
where the polarization tensor $ {\bf \Pi}^{\rho\sigma}_{D}(k;{\cal P}_n) $ has the same form
as in \eq{36}, with the two independent functions $ {\bf \Pi}^\ell_{D} $ and
$ {\bf \Pi}^t_{D} $ given by
\begin{eqnarray*}
 {\bf \Pi}^\ell_{D} (k;{\cal P}_n) & = & {\bf \Pi}^\ell_{n}(k)\;\frac{1}{1\;+\;\frac{\vec{k}^2}
{k_0^2}     g_n \hat{V}(|\frac{\vec{k}}{\lambda_n}|) {\bf \Pi}^\ell_{n}(k) }
\\
 {\bf \Pi}^t_{D} (k;{\cal P}_n)  & = & \qquad\qquad{\bf \Pi}^t_{n}(k)\hspace{4cm}
\end{eqnarray*} 
(The functions $  {\bf \Pi}^{\ell,t}_{n} $ are defined in \eq{37}, with $ v_F \rightarrow
v_{Fn} .$)\\
As expected, this result coincides with the result of an $ RPA $ calculation

\noindent ii) {\bf Tomographic Luttinger Liquid}

We introduce interactions which describe  singular direct (or "forward")
scattering processes, where the quasi-electron with momentum near $ k_F \vom $
interacts only with one with momentum near  $- k_F \vom $ .
This leads to the coupling constants
\be
g_n(\vom_1\vom_2) \; = \; \delta_{\vom_1,-\vom_2}\;
\lambda_n^{d-1}\;(2\pi)^{d-1}\;g_n \quad .
\label{50}
\ee
By inserting \eq{50} into \eq{46}, one obtains :\\
\begin{eqnarray*}
\widetilde{\cal W}^{TL}(a ; {\cal P}_n) & = &  - log \bigg( \int \frac{{\cal D}\{\bo\}}{Z_\varphi}\;
e^{-\frac{1}{2} \sum\limits_{\scriptscriptstyle \om \in {\cal S}^+} 
 4 (1 + \frac{g_n}{2\pi} ) \int d^{d+1}\xi\;
\partial^\om\bo(\xi)\;\overline{\partial}^\om\bo(\xi)} \,\cdot\,e^{S_n^J(\{\bo\};a)}
\bigg) \\
& = & \frac{1}{2} \lambda_n^{d-1} \int\dbar^{\,d+1}k 
\sum_{\rho,\sigma = 0}^d\; \hat{a}_\rho(-k){\bf \Pi}^{\rho\sigma}_{TL}(k;{\cal P}_n)
\hat{a}_\sigma(k) \quad .
\end{eqnarray*}\\
The polarization tensor $ {\bf \Pi}^{\rho\sigma}_{TL} $ is renormalized by an 
overall factor $ \left( 1 + \frac{g_n}{2\pi} \right)^{-1}  $ with respect to the 
non-interacting system :
\be
{\bf \Pi}^{\rho\sigma}_{TL}(k;{\cal P}_n)\; = \;\left( 1 + \frac{g_n}{2\pi} \right)^{-1} 
 {\bf \Pi}^{\rho\sigma}_{n}(k) \quad .
\label{51}
\ee\\

\noindent iii) {\bf  Long-range, transverse, current-current interactions}

In the original system, interactions between transverse currents lead to an additional term
 in the action of the form
\be
S^T(\Psi^\ast,\Psi) \; = \; - \frac{1}{2} \int d^{d+1}x \int d^{d+1}y \sum_{k = 1}
^d j_T^k(x)\;g V(|\vec{x} - \vec{y}|) \delta(x_0-y_0)\;j_T^k(y) \quad, 
\label{52}
\ee
with
$$
j_T^k(\Psi^\ast,\Psi;x) \; = \; \sum_{l = 1}^d \left( \delta_{k,l} - \frac{\partial_k
\partial_l}{\sum_{i=0}^d \partial_i^2} \right)\,j^l(\Psi^\ast,\Psi;x)\quad.
$$\\
A physical realization of such a system is a QH system at filling
factors $\;\nu = \frac{1}{2},\frac{1}{4}, ...  ,$ which can be described as a system of free (composite)
fermions interacting via long-range, transverse, current-current interactions as diplayed in \eq{52},
 cf. \cite{2,3,5}  . \\
 One determines the effective
interaction for the low-energy modes whose momenta lie in a thin shell
$ \Omega_n $ around the Fermi surface following the same procedure
 as for longitudinal interactions. Using \eq{33},  one obtains
$$
S_n^T(\{\psi^\sharp_\om\}) \; = \; - \frac{g_n v_{Fn}^2}{2\lambda_n^{d-1}}
\sum_{\om_1,\om_2 \in {\cal S}^+}\int_{I_{\om_1}\cap I_{\om_2}}\dbar^{\,d+1}k
\;\hat{V}(|\frac{\vec{k}}{\lambda_n}|) \; \cdot \hspace{5cm}
$$
\be
\hspace{7cm} \cdot \; \sum_{i,j = 1}^d\, \om_1^i\,  j_\jom{1}^1(-k)\; {\cal P}^T_{ij}(k)
\; \om_2^j \,j_\jom{2}^1(k) \quad ,
\label{B53}
\ee
with
$$ {\cal P}^T_{ij}(k) = \delta_{ij} - \frac{k_i k_j}{\vec{k}^2}
\qquad \mbox{and} \qquad \hat{V}(|\vec{p}|) = \frac{1}{k_F^{d-1}} \left|\frac{k_F}{\vec{p}}\right|^
\alpha
\quad . 
$$
For a QH system with unscreened Coulomb interactions, the exponent $\alpha$ is equal
to 1 ; if one assumes the Coulomb interactions to be screened, the exponent $\alpha$ turns
out to be 2.\\
We propose to calculate the effective gauge field action of this system by replacing
the fermionic currents, $ j^\mu_\rom(\overline{\psi}_\rom,\psi_\rom) $ , 
by their bosonic versions, $  j^\mu_\rom(\bo) $ , given in \eq{37}. We have to
make assumptions analoguous to the ones stated in (i) .  One obtains
\be
\widetilde{\cal W}^T(a ; {\cal P}_n) \; = \;   \frac{1}{2} \lambda_n^{d-1}  \int\dbar^{\,d+1}k 
\sum_{\rho,\sigma = 0}^d\; \hat{a}_\rho(-k)\;{\bf \Pi}^{\rho\sigma}_{T}(k;{\cal P}_n)\;
\hat{a}_\sigma(k) \qquad ,
\label{53}
\ee
with
\begin{eqnarray*}
 {\bf \Pi}^\ell_{T} (k;{\cal P}_n) & = & 
\qquad\qquad{\bf \Pi}^\ell_{n}(k)\hspace{4cm}
\\
  {\bf \Pi}^t_{T} (k;{\cal P}_n)  & = & 
 {\bf \Pi}^t_{n}(k)\;\frac{1}{1\;+\;
 {\bf \Pi}^t_{n}(k) g_n \hat{V}(|\frac{\vec{k}}{\lambda_n}|)}\quad .
\end{eqnarray*}
Again, this reproduces the result of an $ RPA $ calculation.\\

\section{Electron Propagator}

In this section, we determine the bosonic expressions for the electron fields
$ \Psi^\ast , \Psi ,$  in order to calculate the electron propagator for  interacting
systems. We shall bosonize each one of the $ N \sim \lambda^{d-1} $ component-fields
 $ \psi^\sharp_\om  $   seperately ($ \psi^\sharp $ stands
for $ \psi^\ast $ or $ \psi $ ).  More precisely, for each ray
$ \rom = \{\vom,-\vom\} $ , we express the pair $ \psi_\om^\sharp, \psi_{-\om}
^\sharp $ in terms of a bose field $ \bo .$   This is accomplished by applying the well-known 
bosonization formalism for $1+1$ dimensional relativistic fermions  summarized
in appendix A . However, one has to cope with a subtlety arising from
the dependence of the  quasi  $1+1$ dimensional electron fields $ \psi^\sharp
_\om $  on the components of the momentum  perpendicular to the direction $ \rom $ .

We start our discussion with the  non-interacting system for which
\be 
-  \left< \hat{\psi}_{\omega^{(i)}}(k)\;
                \hat{\psi}^{\ast}_{\omega^{(i)}}(k') \right>^0 \; =
 \; \delta_{\omega^{(i)},\omega^{(j)}}\;(2\pi)^{d+1}\;\delta^{(d+1)}
(k - k') \; \frac{1}{i k_0 -  v_F \vec{\omega}^{(i)}
\vec{k} } \; \mbox{\bf \Large 1}_{\overline{Q}_{\omega^{(i)}}}(\vec{k}) \quad .
 \label{56}   
\ee 
In comparison to \eq{Z10}, we replaced the boxes $ \overline{B}_{\om^{(i)}} $ 
by cubes, $ \overline{Q}_{\omega^{(i)}} $ , with sides of length $  k_F $
( the exact shape of the integration domain is irrelevant) . Because of the isotropy
of the electron system, one observes a non-trivial  dependence on the components
of $ k $  only
in the $0-$ and $\vom-$direction. Propagation takes place   
 in the radial direction, $ \vom .$ This suggests a decomposition of the quasi particle
fields $ \psi_\om^\sharp(\xi) $  into  tensor products
\be
 \psi_\om^\sharp(\xi) \; = \chi_\rom(\xi_\perp) \otimes \psi^\sharp_{\om\parallel}
(\xi_0,\xi_\parallel) \quad,
\label{57}
\ee
for $ \vom \in {\cal S}^{d-1}_1 .$  \\
The "radial" fields, $ \psi^\sharp_{\om\parallel} $ , 
describe  $ 1+1$ dimensional, relativistic electrons, whereas the bosonic 
"angular" fields, $ \chi_\rom(\xi_\perp) $ , just guarantee  momentum conservation
in the perpendicular direction.\\ 
The angular fields  $ \chi_\rom(\xi_\perp) $
  are Gaussian, of mean $ 0 ,$ and their propagator (covariance)
is given , in momentum space,  by
\vspace{.4cm}
\be
\left\langle\, \hat{\chi}_\rom(\vec{k}_\perp)\;\hat{\chi}_{[\om']}(\vec{k}_\perp')
\right\rangle_\perp
\;= \;\delta_{[\om],[\om']}\,(2\pi)^{d-1} \, \delta^{(d-1)}(\vec{k}_\perp-\vec{k}_\perp')
\,\cdot\, {\bf  \Large 1}_{\overline{Q}^{(d-1)}_\om}(\vec{k}^\perp)\quad.
\label{58}
\ee
\vspace{.4cm}
Their Fourier modes, $ \hat{\chi}_\rom(\vec{k}_\perp) $ , have support in
$\overline{Q}^{d-1}_\om .$ \\
The radial fields $ \psi^\sharp_{\om\parallel} $ 
can be bosonized by applying the standard $ 1+1 $ dimensional formalism . For
each ray $ \rom $,
one uses the following identifications :
\vspace{.4cm}
\be
\begin{array}{lccclc}
\psi_{\om 1}(\xi_0,\xi_\|)& := &\psi_{-\om\|}(\xi_0,\xi_\|) & \longleftrightarrow & \frac{1}{(2\pi)^\frac{1}{4}}
\;
D_\rom(\xi_0,\xi_\| ; 1)\; :e^{ i \sqrt{\pi} \bo^\|(\xi_0,\xi_\|)}:&,
\nonumber\\
\psi^\ast_{\om 1}(\xi_0,\xi_\|) &:= &\psi^\ast_{-\om\|}(\xi_0,\xi_\|) & \longleftrightarrow & \frac{1}{(2\pi)^\frac{1}{4}}\;
D_\rom(\xi_0,\xi_\| ; -1)\; :e^{- i \sqrt{\pi} \bo^\|(\xi_0,\xi_\|)}:&,
\nonumber \\
\psi_{\om 2}(\xi_0,\xi_\|)& := & \psi_{\om\|}(\xi_0,\xi_\|)& \longleftrightarrow & \frac{1}{(2\pi)^\frac{1}{4}}\;
D_\rom(\xi_0,\xi_\| ; 1)\; :e^{- i \sqrt{\pi} \bo^\|(\xi_0,\xi_\|)}:&,
\nonumber\\
\psi^\ast_{\om 2}(\xi_0,\xi_\|) &:=&\psi^\ast_{\om\|}(\xi_0,\xi_\|) & \longleftrightarrow & \frac{1}{(2\pi)^\frac{1}{4}}\;
D_\rom(\xi_0,\xi_\| ; -1)\; :e^{ i \sqrt{\pi} \bo^\|(\xi_0,\xi_\|)}:&,
\end{array}
\label{Z60}
\ee
where $ D_\rom $ is a disorder operator and the normal ordered exponential
of $ \bo^\| $ is  a "vertex operator".
The precise definitions of the expressions on the r.s. of \eq{Z60} appear in appendix
A . \\
The fields $ \bo^\|(\xi_0,\xi_\|) $ are $1+1$ dimensional free, massless Bose fields
with an action given by
\be
S^0_\rom (\bo^\|) \; = \;  \frac{1}{2} \int_{{\bf R}\times[-\frac{k_F}{2},\frac{k_F}{2}]}
 \dbar k_0  
\dbar k_\| \; \hbo^\|(-k_0,-k_\|)\,
\left( k_0^2 + (v_F k_\|)^2  
\right) \,\hbo^\|(k_0,k_\|) \quad .
\label{61}
\ee
The Fourier modes $ \hbo^\|(k_0,k_\|) $  have support on $ \; {\bf R} \times
[-\frac{k_F}{2},\frac{k_F}{2}] \; $ . \\
One can verify that from the bosonization formulas, eqs. (\ref{57})-(\ref{Z60}),  for 
the electron fields the ones for the current densities, \eq{37}, follow. In appendix B,
we show this in an example.

It is important to understand the relation of the radial bose fields $  \bo^\|(\xi_0,\xi_\|) $  to
the fields $ \bo(\xi) $  introduced in eqs.(\ref{37})-(\ref{39}), which are related
 to the fermionic currents.\\
 Following \eq{39},  the action $ S^0(\bo) $  of the
 bose field $ \bo $   is given, in momentum space,  by\\
\be  
S^0 (\bo) \; = \; - \frac{1}{2}  \int_{{\bf R}\times\overline{Q}_\om} \dbar^{\,d+1} k  
 \; \hbo(-k)\,
\left( k_0^2 + (v_F k_\|)^2  
\right) \,\hbo(k)  \quad ,
\label{Z61}
\ee\\
i.e., the field $ \bo $  propagates only along the direction $ \rom $ . It follows that the
propagators of the $ \bo$-fields are related to the ones of the  $ \bo^\|$- fields 
by\\
$$
\left\langle\, \bo(\xi_0,\xi_\|,\vec{\xi}_\perp)\;\varphi_{[\om']}(\eta_0,\eta_\|,\vec{\eta}_\perp)\,
\right\rangle \; =\hspace{7cm}
$$
\be
 \hspace{4cm} \delta_{\rom,[\om']}\; \delta^{d-1}_{k_F}(\vec{\xi}_\perp-\vec{\eta}_
\perp) \, \left\langle\, \bo^\|(\xi_0,\xi_\|)\;\varphi_{[\om']}^\|(\eta_0,\eta_\|)\,
\right\rangle^\|\;,
\label{ZZ61}
\ee
with
$$
 \delta^{d-1}_{\om,k_F}(\vec{\xi}_\perp)\quad=\quad \int_{\overline{Q}_\om^{d-1}}
\dbar^{\,d-1}k_\perp\;e^{i \vec{k}_\perp \vec{\xi}_\perp}\quad. 
$$
\\

\noindent{\bf Remark :} 
\parbox[t]{12.5cm}
{
Comparing \eq{Z61} to \eq{61}, one can decompose the field $
\bo(\xi) $ into a tensor product $ \; \bo(\xi) = \bo^\|(\xi_0,\xi_\|)\otimes\chi
_\rom(\vec
{\xi}_\perp) \;,$ where $ \chi_\rom(\vec{\xi}_\perp) $ coincides with the bosonic
Gaussian field introduced in eqs. (\ref{57}) and (\ref{58}).
}\\

Next, we study the effect of interactions of the form displayed in \eq{42} .
As discussed in the preceeding sections, we use the effective actions $ S_n $
given in eqs. (\ref{B42}) and (\ref{42}) as an input for the bosonization calculations.
The action $ S_n $ is the result of an RG analysis where the electron modes with 
momenta $ \vec{p} $ outside the shell $ \Omega_n $ have been integrated
out. It describes the properties of the electron system
at energy scales smaller than $ v_{Fn} \frac{k_F}{\lambda_n} .$
Under the assumptions specified in Sect.2, one can reach very small energy scales
$\; v_{Fn} \frac{k_F}{\lambda_n} \ll  v_{Fn} k_F \;$ before the form of $ S_n $ deviates
from the one given in eqs. (\ref{B42}) and (\ref{42}). The (remaining) fermionic
degrees of freedom are described by the  fields $ \Psi^\sharp .$
These fields  $ \Psi^\sharp(\lambda_n\xi) $ can be decomposed into
$ \;N_n \sim \lambda_n^{d-1}\;$ independent components  $ \psi^\sharp_\om(\xi) : $
\be
 \Psi^\ast(\lambda_n\xi)\;=\;\sum_{\om}\;e^{-i k_F \vec{\omega} \lambda_n
\vec{\xi}}\;\lambda_n^{-\frac{d}{2}}\;\psi^{\ast}_{\omega}(\xi) \qquad
\Psi(\lambda_n\xi)\;=\;\sum_{\om}\;e^{i k_F \vec{\omega} \lambda_n 
\vec{\xi}}\;\lambda_n^{-\frac{d}{2}}\;\psi_{\omega}(\xi) 
 \quad,
\label{B62}
\ee
where the Fourier modes $ \hat{\psi}^\sharp_\om(k) $ have support in $ \overline
{Q}_\om $ (cf. \eq{8}) . Momentum conservation guarantees that
$ \; \langle \psi_{\om} \psi^\ast_{\om'} \rangle^V \sim \delta_
{\om,\om'}
 \; ,$ so that the interacting propagator, $ G $ , splits  into the contributions
 of the component fields $ \psi_\om^\sharp $ : 
\be
G(\lambda_n (\xi-\eta)) \; \approx \sum_{\om} \;e^{ i k_F \vom\lambda_n
 (\vec{\xi}-\vec{\eta})} \;\lambda_n^{-d}\;
G_{\om,n}(\xi-\eta) \quad,
\label{62}
\ee
where $ G_{\om,n}(\xi-\eta) $ are the propagators of the component fields calculated
by using the effective action $ S_n .$ The symbol "$\approx$" indicates that the 
equation holds to leading order in an expansion in $ \frac{1}{\lambda_n} .$\\
As in \eq{7}, this formula should be regarded as a discrete approximation to a
continuous angular decomposition of the interacting propagator,
$$
G(\lambda_n (\xi-\eta)) \; \approx \; \int_{{\cal S}_1^{d-1}} d^{d-1}\om \; 
e^{ i k_F \vom\lambda_n
 (\vec{\xi}-\vec{\eta})} \;\lambda_n^{-1}\;
 \; G^\|_{\om,n}\left(\xi_0-\eta_0,\vom(\vec{\xi}-\vec{\eta})\right)
\quad ,
$$
for large arguments $ \;  \lambda_n(\xi-\eta) \;,$ where the radial propagators
 $\, G^\|_{\om,n}\left(\xi_0-\eta_0, \vom
(\vec{\xi}-\vec{\eta})\right) \, $  only depend on the $0$-component and the 
component parallel to the direction $ \vom .$ 
Hence, for the interacting system, the decomposition (\ref{57}) of each component
 $ \psi_\om^
\sharp $  into a tensor product of a radial field $ \psi^\sharp_{\om\|}(\xi_0,\xi_\|) $
 and an angular field $ \chi_\rom(\vec{\xi}_\perp) $   -- with the propagator defined 
in \eq{58}  --
remains valid.\\
We calculate the propagator $ G^\|_{\om,n}(\xi_0,\xi_\|) $ in the radial direction
by bosonizing the system, i.e., we use the identifications (\ref{Z60}) for the
component fields and replace the fermionic action $ S_n(\{\psi^\sharp_\om\}) $
by its bosonic version $ S_n(\{\bo\}) $ given in eqs. (\ref{39}) and (\ref{44}).
Then, the calculation of the propagator $ G^\|_{\om,n}(\xi_0,\xi_\|) $ is reduced
to evaluating an expactation value in the interacting bosonic ground state of
a product
of a disorder- and a vertex-operator in $ \bo^\| $  (see \eq{Z60}).
 In order to calculate
this expectation value, we have to determine the action, $ S_n(\bo^\|) ,$ of
the radial bose field $ \bo^\| $ from the action $ S_n(\{\bo\}) $  of the family
$ \{\bo\,,\,\rom \in {\cal S}^+\} $ of $d+1$ dimensional bose fields.\\
Through the interaction (\ref{44}), the field $ \bo ,$ for a given ray $ \rom ,$
is coupled to all the other fields $ \varphi_{[\om']} ,$ for $ [\om'] \in {\cal S}^1 .$
Integrating out the fields $ \varphi_{[\om']} ,$ for $  [\om']  \neq \rom ,$
we obtain the ("effective") action  $ S_n(\bo) $ of the field $ \bo $ {\sl
which is still Gaussian \nolinebreak:} 
\be
S_n(\bo) \; = \; \frac{1}{2} \int_{{\bf R}\times\overline{Q}_\om} \dbar^{\,d+1}k
\;\hbo(-k) \left[ {\cal C}^n_\rom(k) \right]^{-1} \hbo(k) \quad,
\label{BB63}
\ee
where 
$$
{\cal C}^n_\rom(k) \; = \; \langle \hbo(-k) \hbo(k) \rangle_{n}
$$
is the propagator of the Fourier mode $ \hbo(k) $ in the interacting system.
 We determine the action $ S_n(\bo^\|) $ 
of the radial field $ \bo^\|(\xi_0,\xi_\|) $ by averaging the inverse propagator
$ \left[ {\cal C}^n_\rom(k) \right]^{-1} $ in \eq{BB63}
over the components $ \vec{k}_\perp $ of $ \vec{k} $ perpendicular to the direction
$ \rom :$
\be
 S_n(\bo^\|) \; = \; \frac{1}{2} \int_{\bf R}\dbar k_0 \int_{[\frac{-k_F}{2},\frac{-k_F}{2}]}
\dbar k_\| \;\hbo^\|(-k_0,-k_\|) \; h^n_\rom(k_0,k_\|)\; \hbo^\|(k_0,k_\|) \quad,
\label{B63}
\ee
with
$$
 h^n_\rom(k_0,k_\|) \;:= \; k_F^{1-d} \int_{\overline{Q}_\om^{d-1}} \dbar^{\,d-1}k_\perp
\; \left[ {\cal C}^n_\rom(k_0,k_\|;\vec{k}_\perp) \right]^{-1}\quad.
$$
Our procedure  to calculate the radial, fermionic 
propagator $ G^\|_{\om,n} $  is summarized in the following formula :
$$
G^\|_{\mp\om,n}(\xi_0-\eta_0,\xi_\|-\eta_\|) \; = \;
- \langle \, \psi_{\om\alpha}(\xi_0,\xi_\|) \, \psi^\ast_{\om\alpha}(\eta_0,\eta_\|) \,
\rangle^\|_n \; \approx 
\hspace{4cm}
$$
\be
\frac{-Z_n}{\sqrt{2\pi}}    
\left\langle 
D_\rom(\xi_0,\xi_\| ; 1) D_\rom(\eta_0,\eta_\| ; -1)
 :\!e^{ i (-1)^{\alpha-1} \sqrt{\pi} \bo^\|(\xi_0,\xi_\|)}\!:
 :\!e^{ i (-1)^{\alpha}  \sqrt{\pi} \bo^\|(\eta_0,\eta_\|)}\!:
 \right\rangle_{S_n(\bo^\|)} \;,
\label{63}
\ee
where $ \alpha = 1 ,$  for $ -\om ,$ and $ \alpha = 2 ,$ for $ +\om .$
The disorder field $ D_\rom $  in the bosonized expression of the electron field
 guarantees the correct anticommutation relations, regardless of the nature 
of interactions.

Before diving into explicit calculations, we have to clarify the meaning of the symbol
"$\approx$" in \eq{63}, i.e., we have to clarify the approximations involved
in calculating the electron propagator by means of bosonization.\\
As in Sect.3,  where we have analysed the calculation of the effective gauge field action, 
we characterize our approximations  in terms of the {\sl formal} perturbation
expansion of the fermionic theory in powers of $ \lambda_n^{-1} .$ By bosonizing
the electron system with  action $ S_n ,$ we resum all
leading order contributions of the (formal) fermionic perturbation expansion of
the electron propagator in powers of  $ \lambda_n^{-1}  $  --  except for
self-energy renormalizations of inner propagator lines (as in Sect.3) .\\
The diagrams contributing to the propagator $ G_{\om,n}(k) $ that are reproduced
by bosonization can be found in the following way  (we restrict our discussion
to 1PI self-energy contributions) :\\
First, draw all diagrams renormalizing the bare $\om$-propagator line that are 
composed
only of $\om$-propagators and contain loops of at most two propagator lines, i.e.
 "bubbles".
Then add all diagrams that are generated by replacing an $\om$-bubble by an $\om'$-bubble
formed by two $\om'$-propagator lines, for $\om' \neq \om . $ In the following Fig.12, we
display some examples :

\begin{center}
\begin{picture}(400,120)(0,0)
 
\ArrowLine(20,50)(40,50) \ArrowLine(40,50)(60,50)
\ArrowLine(100,50)(120,50) \ArrowLine(120,50)(140,50)
\ArrowLine(180,50)(198,50) \ArrowLine(198,50)(242,50) \ArrowLine(242,50)(260,50)
\ArrowLine(300,50)(320,50) \ArrowLine(320,50)(360,50) \ArrowLine(360,50)(380,50)

\ArrowArc(40,64)(12,271,269)
\ArrowArc(120,64)(12,270,90) \ArrowArc(120,64)(12,90,270)
\ArrowArc(120,90)(12,271,269)
\Vertex(40,51){2} \Vertex(120,51){2} \Vertex(120,77){2}

\Text(40,64)[]{$\omega_1$} \Text(120,64)[]{$\omega_1$} 
 \Text(120,90)[]{$\omega_2$} 
\Text(24,46)[t]{$\omega$} \Text(56,46)[t]{$\omega$}
\Text(104,46)[t]{$\omega$} \Text(136,46)[t]{$\omega$}

\ArrowArcn(220,50)(22,180,0) \ArrowArcn(220,50)(22,360,180)

\ArrowArcn(320,64)(12,270,38.7) \ArrowArcn(320,64)(12, 38.7,270)
\ArrowArcn(340,80)(12,218.7,321.3) \ArrowArcn(340,80)(12,321.3,218.7)
\ArrowArcn(360,64)(12,141.3,270) \ArrowArcn(360,64)(12,270,141.3)
\Vertex(198,50){2} \Vertex(242,50){2}
\Vertex(320,51){2} \Vertex(330,72){2} 
\Vertex(350,72){2}  \Vertex(360,51){2}

\Text(220,68)[t]{$\omega_1$} \Text(220,24)[t]{$\omega_1$}
\Text(320,64)[]{$\omega_1$} \Text(340,80)[]{$\omega_2$}
\Text(360,64)[]{$\omega_3$}
\Text(184,46)[t]{$\omega$} \Text(220,46)[t]{$\omega$}
\Text(256,46)[t]{$\omega$} \Text(304,46)[t]{$\omega$}
\Text(340,46)[t]{$\omega$} \Text(376,46)[t]{$\omega$}

\end{picture}
\\
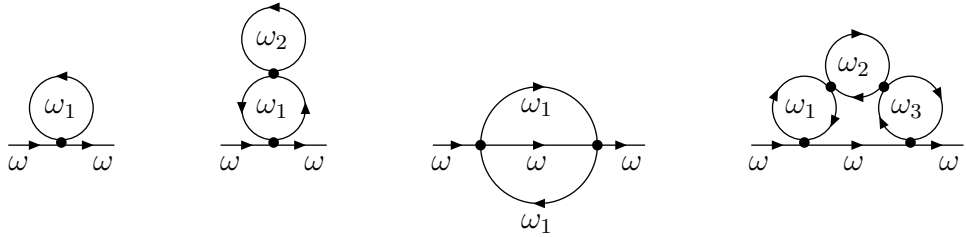
{\sl Fig.12}
\end{center}

\noindent (The corresponding
"diagrammar" has been defined in Sect.3 ; the summation, $ \sum_{\om_i} ,$ associated to each electron
loop, is not displayed explicitly.)\\
A characteristic feature of bosonization is that it reproduces only diagrams containing
loops of at most two propagator lines.

\noindent Introducing the effective interaction vertex

\begin{center}
\begin{picture}(370,60)(0,0)
\ArrowLine(10,17)(23,30) \ArrowLine(23,30)(10,43)
\ArrowLine(38,17)(25,30) \ArrowLine(25,30)(38,43) 
\ArrowLine(78,17)(91,30) \ArrowLine(91,30)(78,43)
\ArrowLine(106,17)(93,30) \ArrowLine(93,30)(106,43)
\BCirc(24,30){4}
\Vertex(92,30){2.5} 
\Text(58,30)[]{$:=$} \Text(124,30)[]{$+$}
\Text(10,13)[t]{$\omega$} \Text(10,46)[b]{$\omega$}
\Text(38,16)[t]{$\omega'$} \Text(38,46)[b]{$\omega'$}
\Text(78,13)[t]{$\omega$} \Text(78,46)[b]{$\omega$}
\Text(106,16)[t]{$\omega'$} \Text(106,46)[b]{$\omega'$}

\put(0,0)
{
\begin{picture}(360,60)(-10,0)
\ArrowLine(132,17)(145,30) \ArrowLine(145,30)(132,43)
\ArrowLine(188,17)(175,30) \ArrowLine(175,30)(188,43) 
\ArrowLine(224,17)(237,30) \ArrowLine(237,30)(224,43)
\ArrowLine(308,17)(295,30) \ArrowLine(295,30)(308,43)
\Vertex(146,30){2.5} \Vertex(174,30){2.5}
\Vertex(238,30){2.5} \Vertex(266,30){2.5} \Vertex(294,30){2.5}
\ArrowArcn(160,30)(13,180,0) \ArrowArcn(160,30)(13,360,180)
\ArrowArcn(252,30)(13,180,0) \ArrowArcn(252,30)(13,360,180)
\ArrowArcn(280,30)(13,180,0) \ArrowArcn(280,30)(13,360,180)
 \Text(206,30)[]{$+$}
\Text(326,30)[]{$+$} \Text(344,30)[]{$\cdots$}
\Text(160,30)[]{$\omega_1$} \Text(252,30)[]{$\omega_1$} \Text(280,30)[]{$\omega_2$}
\Text(132,13)[t]{$\omega$} \Text(132,46)[b]{$\omega$}
\Text(188,16)[t]{$\omega'$} \Text(188,46)[b]{$\omega'$}
\Text(224,13)[t]{$\omega$} \Text(224,46)[b]{$\omega$}
\Text(308,16)[t]{$\omega'$} \Text(308,46)[b]{$\omega'$}

\end{picture}
}

\end{picture}
\end{center}

\noindent we can represent the diagrams reproduced as follows
\vspace{-.5cm}
\begin{center}
\begin{picture}(400,140)(0,0)

\ArrowLine(20,70)(60,70) \ArrowLine(60,70)(100,70)
\ArrowArc(60,98)(24,271,269)
\BCirc(60,72){6}
\Text(60,98)[]{$\omega_1$}
\Text(24,67)[t]{$\omega$} \Text(96,67)[t]{$\omega$}
\Text(140,70)[]{$+$}

\put(0,0){
\begin{picture}(400,140)(40,0)
\ArrowLine(220,70)(256,70) \ArrowLine(256,70)(344,70) \ArrowLine(344,70)(380,70)
\ArrowArcn(300,70)(44,180,360) \ArrowArcn(300,70)(44,0,180)
 \BCirc(344,70){6}
\Vertex(256,70){4}
\Text(224,67)[t]{$\omega$} \Text(300,67)[t]{$\omega$}
\Text(376,67)[t]{$\omega$} \Text(300,33)[t]{$\omega_1$}
\Text(300,111)[t]{$\omega_1$}
\end{picture}
}

\end{picture}
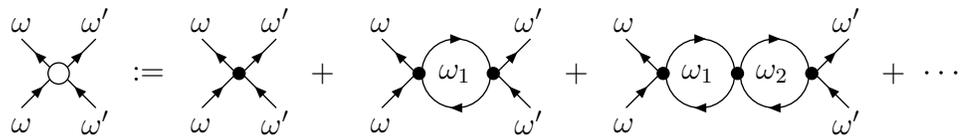
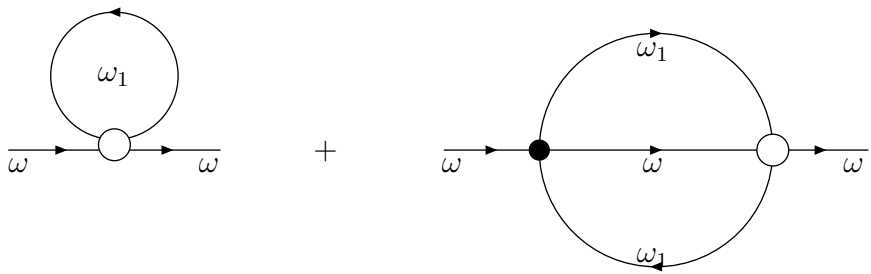
\\ {\sl Fig.13}
\end{center}

The first class of diagrams leads to a contribution of order 1, the second one to a
 contribution of order $ \frac{1}{\lambda_n} $ . One finds that all  leading order
 diagrams are reproduced by bosonization,
except for self-energy renormalizations of inner propagator lines.  For short-range
 interactions, one can show by explicit calculations
that the contribution of these self-energy renormalizations of inner propagator lines
 is {\sl zero} -- to the order considered in our calculation. 
For long range interactions, however, they could change the final result.\\
One should remember that we use the effective action $ S_n $ as an input of
bosonization. Because of the linearization of the pieces of the Fermi surface contained
in the boxes $ \overline{B}_\om ,$ this action describes the properties of the electron
system correctly at (unrescaled) momentum scales between $\frac{k_F}{\lambda_n} $
and  $  \frac{k_F}{\lambda^2_n} .$ Therefore, \eq{63} reproduces the electron 
propagator only correctly for arguments $ |\xi-\eta| $ of order between $ \frac{1}{k_F} $
and $ \frac{\lambda_n}{k_F} $ ( \eq{63} is written with respect to the rescaled 
system \nolinebreak ).
In this range, formula (\ref{63}) holds to leading order in a (formal) expansion in powers
of $ \frac{1}{\lambda_n} .$ \\
To determine the electron propagator for larger arguments,
we first have to calculate an effective (fermionic) action $ S_m ,\;m > n\;,$ describing
the properties of the system at larger distance- and time-scales. Such an effective action
$ S_m $ can serve as a new input for the bosonization procedure.

We now return to the calculation of the bosonic propagator of the interacting system.\\
First, we consider systems with a short-range two-body interaction potential.\\
For the non-interacting system, \eq{Z61} implies
\be
\langle \; \hbo(k) \; \hbo(k') \; \rangle^0 \; = \;
(2\pi)^{d+1} \delta^{(d+1)}(k+k') \frac{1}{k_0^2 + (v_F\vom\vec{k})^2} \; \mbox{\bf \Large 1}
_{\overline{Q}_{\omega^{(i)}}}(\vec{k})\quad.
\label{64}
\ee
If one turns on interactions given by \eq{44}, the propagator changes to become
$$
\langle \; \hbo(k) \; \hbo(k') \; \rangle_n \; = \;\hspace{11cm}
$$
\be\hspace{2.8cm}
(2\pi)^{d+1} \delta^{(d+1)}(k+k') \frac{1}{k_0^2 + (v_{Fn}\vom\vec{k})^2} 
\left[ \sum_{\ell=0}^\infty \left( -\frac{1}{\lambda_n^{d-1}} {\cal T}^n(k)
 \right)^\ell \right]_{\rom[\om']} \quad,
\label{65}
\ee
 where
$$
{\cal T}^n_{\rom[\om']}(k) \; = \;  (\frac{1}{2\pi})^{d} 
 \left[ g_n(\vom\cdot\vom') \left( k_0^2 + (v_{Fn}\vom\vec{k})
(v_{Fn}\vom'\vec{k}) \right) \right. +\hspace{6cm}
$$
$$
\left. \hspace{5cm}
 g_n(-\vom\cdot\vom') \left( -k_0^2 + (v_{Fn}\vom\vec{k})
(v_{Fn}\vom'\vec{k}) \right) \right] \frac{1}{k_0^2 + (\vom'\vec{k})^2} \quad.
$$
For sufficiently small coupling constants, $ \; | g_n(\vom\cdot\vom') |
\leq g_c \ll 1 \; $ , the Neumann series converges, and one obtains
\be
\langle \; \hbo(k) \; \hbo(k') \; \rangle_n \; = \;
(2\pi)^{d+1} \delta^{(d+1)}(k+k') \frac{1}{k_0^2 + (v_{Fn}\vom\vec{k})^2} 
\left[ 1 + \frac{1}{\lambda_n^{d-1}} f(k;g_n) \right] \quad,
\label{66}
\ee
where $ f(k;g_n) $ is a bounded function in $k$. 
From Sect.2, we know that -- for a system with short-range interactions --
 the coupling constants $  g_n(\vom\cdot\vom') $ tend to  finite limits,
as $ n \rightarrow \infty .$ Thus, in the limit
$ \lambda_n \rightarrow \infty $ , the effects of interactions disappear, and the system
is driven to the non-interacting (Landau liquid) fixed point with a propagator
given by (\ref{64}). For short-range interactions, the dependence on $ \vec{k}_\perp $
is irrelevant and suppressed by a factor $ \frac{1}{\lambda_n^{d-1}} .$ 

 Below, we shall see
that, for sufficiently long-range transverse current-current interactions, the
dependence on $ \vec{k}_\perp $ is significant ( i.e., singular, as $  |\vec{k}_\perp| \rightarrow
0 $ , with $ |k_0|, |k_\|| < |\vec{k}_\perp\| $\nolinebreak ). It is not suppressed by an 
inverse power of the
scale factor $ \lambda_n $. We have to average over the variable $ \vec{k}_\perp $ 
in order to determine the effective dynamics of the  bosonic degrees of freedom
$ \bo^\| $  in the direction along $ \rom $ . \\Using \eq{63}, we  then calculate the
propagator of the radial, quasi-electron components $ \psi^\sharp_{\om\|} $ .

We start by studying two technically easier classes of systems with singular interactions.
These systems correspond to examples (i)-(iii) of section 3 for which the polarization
tensor $ {\bf \Pi}^{\rho\sigma} $  has been calculated explicitly.
\vspace{.5cm}

\noindent i) {\bf Long-range, density-density interactions}
\\ 

The set of coupling constants is given by \eq{47}, with $ \;g_n 
\hat{V}(|\vec{p}|)
=\frac{ g_n}{k_F^{d-1}} |\frac{k_F}{\vec{p}}|^\alpha \;,\; \alpha >\nolinebreak 0 \;.$  One obtains
$$
\langle \; \hbo(k) \; \hbo(k') \; \rangle_n \; = \;
(2\pi)^{d+1} \delta^{(d+1)}(k+k') \frac{1}{k_0^2 + (v_{Fn}\vom\vec{k})^2}\; \cdot \hspace{5cm}
$$
\be
\hspace{3.5cm} \cdot \; 
\left\{ 1 + \frac{2}{\lambda^{d-1}} (\frac{1}{2\pi})^{d}
\frac{(v_{Fn}\vom\vec{k})^2}{k_0^2 + (v_{Fn}\vom\vec{k})^2}
 \left[ g_n^{-1} \hat{V}^{-1}(|\frac{\vec{k}}{\lambda_n}|) + {\bf \Pi}^{00}_{n}(k)
\right]^{-1} 
\right\} \quad,
\label{67}
\ee
where $ {\bf \Pi}^{00}_{n} $ is defined in \eq{36}, with $v_F \rightarrow v_{Fn} .$\\
Using that
$$
\lim_{|\frac{\vec{k}}{k_0}|\rightarrow 0}\; {\bf \Pi}^{00}_{n}(k)\; \longrightarrow
\; c\, (
\frac{\vec{k}}{k_0})^2 \qquad \mbox{and} \qquad
 \lim_{|\frac{k_0}{\vec{k}}| \rightarrow 0}\; {\bf \Pi}^{00}_{n}(k)\; \longrightarrow\; c' 
\quad ,
$$
\vspace{.3cm}
one verifies that
\vspace{.3cm}
$$
\frac{(v_{Fn}\vom\vec{k})^2}{k_0^2 + (v_{Fn}\vom\vec{k})^2}
 \left[ g_n^{-1} \hat{V}^{-1}(|\frac{\vec{k}}{\lambda_n}|) + {\bf \Pi}^{00}_{n}(k)
\right]^{-1} \; \sim \hspace{6cm}
$$ 
$$ \hspace{3cm}\sim\;\left\{\;
\begin{array}{ccc}
(\frac{\vom\vec{k}}{k_0})^2  \left[ (\frac{g_n}{k_F^{d-1}})^{-1} |\frac{\vec{k}}{k_F \lambda_n}|^
\alpha + c  |\frac{\vec{k}}{k_0}|^2 \right]^{-1} & , &\mbox{for} |\frac{\vec{k}}{k_0}|
\rightarrow 0 \\
 \left[ (\frac{g_n}{k_F^{d-1}})^{-1} |\frac{\vec{k}}{k_F \lambda_n}|^
\alpha + c' \right]^{-1} & , &\mbox{for} |\frac{k_0}{\vec{k}}|
\rightarrow 0 
\end{array}
\right. \quad .
$$
\\
\vspace{.3cm} 
For $\, 0 \leq \alpha \leq 2\, ,$ the rhs is a bounded function of $ k .$\\ 
Under the assumption that all parameters $ {\cal P}_n $ (except the scale parameter 
$ \lambda_n $) characterizing the effective action $ S_n $ tend to finite values, as 
$ n \rightarrow \infty ,$ it follows that, in the limit $ \lambda_n \rightarrow \infty ,$
the effect of interactions on the bose propagator disappears, as for the system with
short-range interactions.\\
 Hence, in the limit $ \lambda_n \rightarrow \infty ,$ the electron
propagator $ G $ tends to the standard LFL form.

This result is due to the  screening of the bare, long-range interaction potential
$\hat{V}(|\vec{p}|) .$ In \eq{67}, the bare potential  $ \hat{V}(|\frac{\vec{k}}
{\lambda_n}|) $ is replaced by an effective $ RPA-$potential
$$
g_n \hat{V}_n^{\mbox{eff}}(k_0,\vec{k})
\;  :=\;   \frac{g_n \hat{V}(|\frac{\vec{k}}{\lambda_n}|)}{1 \; + \;
 g_n 
\hat{V}(|\frac{\vec{k}}{\lambda_n}|) {\bf \Pi}^{00}_{n}(k)}
 \;=\; 
\frac{1}{g_n^{-1} \hat{V}^{-1}(|\frac{\vec{k}}{\lambda_n}|) \;+ \;
{\bf \Pi}^{00}_{n}(k)  }\;.
$$
\\
In the static limit, $ |\frac{k_0}{\vec{k}}| \rightarrow 0 ,$ we obtain
\\ 
$$
\lim_{|\frac{k_0}{\vec{k}}| \rightarrow 0} g_n \hat{V}_n^{\mbox{eff}}(k_0,\vec{k})
\; = \; \frac{1}{ (\frac{g_n}{k_F^{d-1}})^{-1} (\frac{k_F \lambda_n}{|\vec{k}|})^\alpha \;+ \;c  }\quad,
$$
\\
i.e., in this limit, the effective potential is short-ranged. 
\vspace{.8cm}
\pagebreak

\noindent ii) {\bf Tomographic Luttinger Liquid}
\\

\noindent With \eq{50}, one obtains
\\
$$
\langle \; \hbo(k) \; \hbo(k') \; \rangle_n \; = \;
(2\pi)^{d+1} \delta^{(d+1)}(k+k') \frac{1}{k_0^2 + (v_{Fn}\vom\vec{k})^2} 
\left[ 1 + \frac{g_n}{2\pi}  \right]^{-1} \quad.
$$
\\
As there is no dependence on the perpendicular momenta, one can immediately
read off the effective dynamics for the modes $ \hbo^\| $  propagating along $ \rom $ :
\\
$$
 S^{TL}_n (\bo^\|) \; = \;   \frac{1}{2} \int_{{\bf R}\times[-\frac{k_F}{2},\frac{k_F}{2}]}
 \dbar k_0  
\dbar k_\| \;\cdot \hspace{8cm}
$$
$$\hspace{4cm}\cdot\,  \hbo^\|(-k_0,-k_\|)\,\left[ 1 + \frac{g_n}{2\pi}  \right]\,
\left( k_0^2 + (v_{Fn} k_\|)^2  
\right) \,\hbo^\|(k_0,k_\|)  \quad .
$$

The result for the quasi-electron propagator can be found with the help of formulae
(A18) and (A19) of appendix A :
\be
G_{\pm \om,n} (\xi) \; = \; \delta^{(d-1)}_{\om,k_F}(\vec{\xi}_\perp)
 \cdot \frac{\mp i}{2\pi}\; e^{\pm i \arg(\zeta)}\;
\frac{Z_n}{|\zeta|^{1+\eta}} \quad,
\label{Z67}
\ee
where
$$
\xi\,=\, (\zeta,\vec{\xi}_\perp) \qquad, \mbox{with} \qquad 
\zeta\,:=\,(\xi_0 , \frac{\xi_\|}{v_{Fn}})\;,
$$
and
$$
\delta^{(d-1)}_{\om,k_F}(\vec{\xi}_\perp)\; := \; \int_{\bar{Q}^{d-1}
} \dbar^{\,d-1}k_\perp \; e^{i \vec{k}_\perp  \vec{\xi}_\perp} \quad.
$$
The exponent $ \eta $ is given by
$$
\eta \; = \;  \frac{ 2 (\frac{g_n}{\pi})^2}{1 +  2 (\frac{g_n}{\pi})^2} 
\quad . 
$$
This interacting system describes a {\sl Luttinger Liquid}.\\
The flow of the parameter $Z_n$ in \eq{Z67} can be derived by assuming
that, along each direction $ \vom ,$ the system is scale invariant :

\noindent Recall that
\be
G( \lambda_n \xi )\; \sim \; \sum_\om\;e^{i k_F \lambda_n \vom \vec{\xi}}\;\frac{1}{\lambda_n^d}\; G_{\om,n}( \xi  ) \;\sim\;
\int d\om\;e^{i k_F \lambda_n \vom \vec{\xi}}\;\frac{1}{\lambda_n}\; G^\|_{\om,n}( \zeta )\;.
\label{ZZZ67}
\ee
This equation and the assumption that the theory is scale invariant along each direction $ \vom $ implies
the following matching-property for the Green functions $\, G^\|_{\om,n}( \zeta )\,:$
\be
  G^\|_{\om,n}( M \zeta ) \quad\approx\quad
\frac{1}{M}  G^\|_{\om,n+1}( \zeta )
\label{ZZ67}\quad.
\ee
By inserting \eq{Z67} :
$$
\frac{Z_n}{|M \zeta|^{1+\eta}} \quad \sim \quad
\frac{1}{M}\;\frac{Z_{n+1}}{|\zeta|^{1+\eta}}\quad,
$$
it follows that
$$
\frac{Z_{n+1}}{Z_n}\;\sim\;\frac{1}{M^\eta} \qquad\mbox{or}\qquad Z_n \;\sim
\;\frac{1}{(M^n)^\eta} \quad,
$$
i.e., for this system, the residue $Z_n$  of the one-particle pole vanishes, as $n\rightarrow \infty .$
\vspace{.8cm}

\noindent iii) {\bf Long-range, transverse current-current interactions}
\\

Such interactions have been introduced in \eq{52}, with $ \;g_n V(|\vec{p}|) 
= c_n  |\frac{k_F}{\vec{p}}|^\alpha \; ,\;c_n :=  \frac{g^T_n v_{Fn}^2}{k_F^{d-1}}\;, $  and lead to
the following bosonic propagator (for $ \rom = [\om'] $) :
$$
\langle \; \hbo(k) \; \hbo(k') \; \rangle_n \; = \;
(2\pi)^{d+1} \delta^{(d+1)}(k+k') \frac{1}{k_0^2 + (v_{Fn}\vom\vec{k})^2}\; \cdot \hspace{5cm}
$$
\be
\hspace{1.5cm} \cdot \; 
\left\{ 1 - \frac{2}{\lambda_n^{d-1}}  \frac{1}{2\pi} (\frac{k_F}{2\pi})^{d-1}
\frac{g_n \hat{V}(|\frac{\vec{k}}{\lambda_n}|)}{1 + g_n \hat{V}(|\frac{\vec{k}}{\lambda_n}|)
\, {\bf \Pi}^t_n(k)}\, 
\frac{k_0^2}{k_0^2 + (v_{Fn} \vom\vec{k})^2}
 \left[ \vom^2 - \frac{(\vom\vec{k})^2}{\vec{k}^2}
\right]
\right\} \quad,
\label{68}
\ee
with
$$
{\bf \Pi}^t_n(k) \; = \; \frac{2}{d-1} \frac{1}{2\pi} (\frac{k_F}{2\pi})^{d-1}
\frac{1}{\lambda_n^{d-1}} \sum_{\om \in {\cal S}^+} \frac{k_0^2}{k_0^2 + (v_{Fn}\vom\vec{k})^2}
  \left[ \vom^2 - \frac{(\vom\vec{k})^2}{\vec{k}^2}
\right] \quad.
$$
One finds that $\; \lim_{n \rightarrow \infty}{\bf \Pi}^t_n(k)  =
{\bf \Pi}^t_{\ast}(k)\; $ , with $ {\bf \Pi}^t_{\ast}(k) $ defined in \eq{36}.\\
Using that  $\; \lim_{|\frac{k_0}{\vec{k}}| \rightarrow 0}{\bf \Pi}^t_n(k)  \sim    
|\frac{k_0}{\vec{k}}|\;$  , one can see (cf. appendix C) that the contribution from the interactions 
to the 
propagator yields a singular $\vec{k}_\perp$-dependence, as $ \;|\vec{k}_\perp|
\rightarrow 0 \;$ and $\; \left(|k_\||,|k_0|\right) \ll |\vec{k}_\perp| \; $ . As discussed
above, we then have to average over the perpendicular momenta $ \vec{k}_\perp $,
in order to determine the (effective) action, $ S^T_n$ , for the modes 
$ \hbo^\| $   propagating  along $ \rom $ . 

\noindent The somewhat tedious calculations are 
deferred to appendix C.  Here, we only describe our results.\\ 
We obtain two different regimes, 
depending on the exponent $ \alpha $ which characterizes the  singularity of the interaction
potential in momentum space by $  \hat{V}(|\vec{p}|) \sim \frac{1}{|\vec{p}|^\alpha} : $ 

\begin{itemize}
\item
 For $\; 0 \leq \alpha < d-1 \;,$ the bosonization calculation yields 
\be
G_{\om,n}(\xi)
 \quad \sim \quad \delta^{(d-1)}_{\om,k_F}(\vec{\xi}_\perp)\; (\frac{-i}{2\pi})\; 
e^{i\arg(i\xi_0+\frac{\xi_\|}{v_{Fn}})}\; \frac{Z_n}{\sqrt{(v_{Fn}\xi_0)^2+\xi_\|^2}} \quad.
\label{69}
\ee
 The function
$\; \delta^{(d-1)}_{\om,k_F}(\vec{\xi}_\perp) \; $ has been defined in \eq{Z67}. The influence of
the interaction on the electron propagator is suppressed by a factor $ \lambda_n^{-[(d-1)-\alpha]} ,$
as $ \lambda_n \rightarrow \infty .$\\
From the matching condition (\ref{ZZ67}), it follows that
$$
 Z_{n+1}\; \approx\; Z_n \qquad \mbox{and} \qquad v_{Fn+1} \; \approx \; v_{Fn} \quad.
$$
The system tends to a LFL.
\\

\item
For $\; d-1 < \alpha \leq 2 \;,$ we display the result in the region 
$ \;|\xi_0| \ll \frac{|\xi_\||}{v_{Fn}} \;:$

$$
G_{\om,n}(\xi) 
 \quad \sim \quad \delta^{(d-1)}_{\om,k_F}(\vec{\xi}_\perp)\; (\frac{-i}{2\pi})\; 
e^{i\arg(i\xi_0+\frac{\xi_\|}{v_{F_n}})}\; Z_n \;\cdot \hspace{7cm}
$$
\be
\hspace{1cm} \cdot \;
\left\{
\begin{array}{lcl} 
\exp \left( - u_{\mbox{\tiny I}} k_F c_n^{\frac{d}{1+\alpha}}   |\lambda_n \frac{\xi_\|}{v_{Fn}}|^
{\frac{\alpha-(d-1)}{1+\alpha}}  \right) & , & \mbox{for} \;
 |\xi_\||  \gg c_n v_{Fn} \lambda_n^\alpha \quad,
 \\
\exp \left( - u_{\mbox{\tiny II}} \frac{k_F}{\lambda_n^{d-1}} \frac{|\xi_\||}{v_{Fn}}\right) & , & \mbox{for} \;
c_n v_{Fn} \lambda_n^\alpha \gg |\xi_\||  \gg  \frac{v_{Fn} \lambda^{d-1}_n}
{ k_F}\quad,
\\
\frac{1}{|\xi_\||} & , &\mbox{for} \;\frac{v_{Fn} \lambda_n^{d-1}}{ k_F} \gg   |\xi_\||  \gg
\frac{1}{k_F}   \quad.
\end{array}
\right.
\label{70}
\ee
The constant $ c_n $ stands for $ \frac{ g_n v_{Fn}^2}{k_F^{d-1}} ,$ and
$ u_{\mbox{\tiny I}}, u_{\mbox{\tiny II}} $ are positive constants depending on the
 dimension $ d$ of space  and the exponent $ \alpha $ 
 (cf. appendix C). We have neglected all
 terms which  are of lower order in an expansion in $ \lambda_n^{-1} $ than the leading
terms displayed.
Two interpretations are consistent with this result :

\begin{enumerate}
\item
The parameters ${\cal P}_n $ characterizing the effective action $ S_n $ 
are all of order 1 -- except for the scale parameter
$\lambda_n $ -- , i.e., they do not flow under RG transformations. Then, the system tends to a 
LFL, as $ \lambda_n \rightarrow \infty :$ for finite arguments $|\xi_\|| ,$ the propagator $\,G_{\om,n}\,$
has the standard LFL form. It deviates from this form only for very large arguments 
$\;|\xi_\|| \gg \frac{v_{Fn} \lambda_n^{d-1}}{k_F}\;,$ where the results of a bosonization
calculation are not reliable (remember that,
 because of the linearization of the pieces of the Fermi surface contained in the boxes
$ \overline{Q}_\om ,$ the results of  the present calculation are only reliable for arguments 
$ |\xi_\|| $ smaller
than $  \lambda_n) .$
\\

\item
The parameters of the set 
${\cal P}_n $  
flow under RG transformations,
i.e., they are functions of the scale parameter $\lambda_n .$\\
The electron propagator
deviates from the standard LFL form for finite arguments $|\xi_\|| ,$ if the condition
$$
c_n  v_{Fn} \lambda_n^\alpha \; \sim \; O(\frac{1}{k_F})
$$
is satisfied. 
Further, the requirement that the system is form-invariant under scale 
transformations implies the matching condition (\ref{ZZ67}). This condition can be satisfied if
$$
c_n^{\frac{d}{1+\alpha}} \, v_{Fn}^{\frac{d-1-\alpha}{1+\alpha}}\; \sim\; const.\quad.
$$
Then it follows that
$$
c_n \; \sim \; \lambda_n^{-\frac{\alpha (\alpha-d+1)}{1+\alpha}} \qquad\mbox{and}\qquad
v_{Fn} \; \sim \; \lambda_n^{-\frac{\alpha d}{1+\alpha}}\quad.
$$
The flow of the parameter $Z_n$ cannot be derived in the same way : 
for this, one
would have to know subleading corrections to the argument of the exponential in \eq{70}.\\
Hence the system displays "Non Landau Liquid" behaviour, as
suggested in \cite{2}.
\end{enumerate}

The result of the bosonization calculation permits two consistent scenarios. In order
to be able to decide which one is realized in the physical system one has  to determine
the flow of the parameters  ${\cal P}_n $ under RG transformations. Preliminary
calculations indicate that the second case is realized (cf. also \cite{2,3,4,5}).
\\
\pagebreak
\item
For the critical value $\; \alpha_c = d-1\;$ of the exponent $ \alpha ,$ we find for
the electron propagator in the region $ \;|\xi_0| \ll \frac{|\xi_\||}{v_{Fn}} \;:$

$$
G_{\om,n}(\xi) 
 \quad \sim \quad \delta^{(d-1)}_{\om,k_F}(\vec{\xi}_\perp)\; (\frac{-i}{2\pi})\; 
e^{i\arg(i\xi_0+\frac{\xi_\|}{v_{F_n}})}\; Z_n \;\cdot \hspace{7cm}
$$
\be
\hspace{1cm} \cdot \;
\left\{
\begin{array}{lcl} 
\exp \left( - v\, k_F c_n   \ln^2|\xi_\||  \right) & , & \mbox{for} \;
 |\xi_\||  \gg c_n v_{Fn} \lambda_n\quad,
\\
\frac{1}{|\xi_\||} & , &\mbox{for} \;c_n v_{Fn} \lambda_n \gg   |\xi_\||  \gg
\frac{1}{k_F}   \quad,
\end{array}
\right.
\label{71}
\ee

\noindent where $v$ is a constant of order one. \\
As for $ \alpha > d-1 ,$ we can distinguish two possible scenarios, consistent with
the results of the bosonization calculation :\\
If the parameters $ c_n, v_{Fn} $ are not functions of the scale parameter $\lambda_n ,$
the propagator has the standard LFL form, as $ \lambda_n \rightarrow \infty .$\\
In order to obtain a non-LFL behaviour, the product $ c_n v_{Fn} \lambda_n $ of parameters
must be of order one. Then, the matching condition (\ref{ZZ67}) requires that $ c_{n+1} \approx
c_n .$ By the first condition, it follows that $ v_{Fn} \sim \frac{1}{\lambda_n} .$\\
Again, in order to decide which one of the two scenarios is realized, the flow of the
parameters under RG transformations must be investigated. The calculations in \cite{2,3,4,5} point
in the second direction.

\end{itemize}

\vspace{1cm}

\renewcommand{\theequation}{\Alph{section}.\arabic{equation}}
\renewcommand{\thesection}{\Alph{section}}

\setcounter{section}{1}
\setcounter{equation}{0}
\section*{Appendix A : Bosonization of the Electron Propagator in $1+1$ Dimensions} 

In this appendix, we review the bosonization of a relavistic $1+1$ dimensional electron
system by using functional integrals. Our aim is to summarize the  procedure for
calculating the
 electron propagator. The general ideas of the bosonization technique have been 
presented in \cite{17,18}.

We want to investigate a $1+1$ dimensional electron system in euclidean space-time,
whose  action is given by
\be
S^0(\overline{\Psi},\Psi) \; = \; \int d^2\xi \; \overline{\Psi}(\xi)\;\gamma^\mu\partial_
\mu\; \Psi(\xi) \quad.
\label{A1}
\ee
Quantization is accomplished by using functional integrals. The field $\Psi$ denotes a
two-component Grassmann field, and $\overline{\Psi} := \Psi^\ast\gamma^0 $ ,
where $\Psi^\ast$ is an independent Grassmann field. Choosing the chiral
representation of the $\gamma$-matrices, i.e. , 
$$
\gamma^0 = \sigma_1 \qquad \qquad \gamma^1 = \sigma_2 \qquad \qquad
\gamma^5 = -i\gamma^0\gamma^1 = \sigma_3\qquad \qquad,
 $$
the two components $\;\left(\begin{array}{c}
\psi_1\\ \psi_2\end{array} \right) \; $ of $\Psi$ are the
antiholomorphic and holomorphic modes, resp., which are the euclidean
analogues of left- and right-movers, resp. . \\
We shall perturb the system by a current-current interaction of the form
\be
V(\overline{\Psi},\Psi) \; = \; - \frac{g}{2} \int d^2\xi\;j^\mu(\xi)j^\mu(\xi)\quad,
\label{A2}
\ee
where the one dimensional currents, $j^\mu$, are defined by
$$
j^\mu(\xi) \; = \; -i : \overline{\Psi}(\xi) \gamma^\mu \Psi(\xi) :  \quad \mu = 0,1
\quad.
$$
This is  the  so called (massless) Thirrring model, or,  in the context of solid state
physics, the Tomonaga-Luttinger model.\\
Before studying the interacting system, we  establish the bosonization
formulas for the free system. Identifying the fermionic current density $j^\mu$ with the bosonic 
expressions
\be
j^\mu(\overline{\Psi},\Psi;\xi) \; \leftrightarrow \; j^\mu(\varphi;\xi) \;=\;
\frac{i}{\sqrt{\pi}} \varepsilon^{\mu\nu} \partial_\nu \varphi(\xi)\quad,
\label{A3}
\ee
one can reproduce the Green functions for the currents
\\
$$
\langle j^{\mu_1}(\xi_1) \cdots  j^{\mu_n}(\xi_n) \rangle^0_F \; = \; \int
\frac{{\cal D}(\overline{\Psi},\Psi)}{Z_F} \; e^{- S^0(\overline{\Psi},\Psi)}\;
 j^{\mu_1}(\xi_1) \cdots  j^{\mu_n}(\xi_n) \hspace{3cm}
$$
\be
\hspace{2cm} = \;
\langle j^{\mu_1}(\varphi,\xi_1) \cdots  j^{\mu_n}(\varphi,\xi_n) \rangle^0_B \; = \; \int
\frac{{\cal D}(\varphi)}{Z_B} \; e^{-\tilde{S}^0(d\varphi)}\;
 j^{\mu_1}(\varphi,\xi_1) \cdots  j^{\mu_n}(\varphi,\xi_n)  \quad,
\label{A4}
\ee
\\
where the action $ \tilde{S}(d\varphi)$
\footnote{In this appendix, we write the action $ S(\varphi) $ as a functional, 
$ \tilde{S}(d\varphi) ,$ of  derivatives of the bosonic field.}
  of the free, massless bose field $\varphi$ is given by
\\
\be
\tilde{S}^0(d\varphi) \; = \;  \frac{1}{2} \int d^2\xi \; \partial_\mu\varphi(\xi) 
\partial_\mu\varphi(\xi) \;= \; \frac{1}{2} \int d\varphi(\xi)\wedge\ast\,
d\varphi(\xi)\quad.
\label{A5}
\ee
\\
Here, we have introduced the one form $\; d\varphi = \partial_\mu\varphi\;d\xi^\mu\;$,
 and $\ast $ denotes the Hodge star operation.  We use  standard notations of the
calculus with differential forms. In \cite{18}  one finds a brief summary of the main
definitions.
( Note that,  in \cite{18},  we use slightly modified conventions;
especially $\;\phi \rightarrow \varphi := \frac{\phi}{2\sqrt{\pi}}\;$ and
$\; j'^\mu \rightarrow j^\mu = i j'^\mu \;$. \nolinebreak)

In order to express the Green functions of the electric fields
 $\;\left(\begin{array}{c} \psi_1\\ \psi_2\end{array} \right) \; $ 
in the bosonic theory, one has to introduce disorder fields $ D(\underline{y},\underline{
q}) $ .\\
For  non-zero integers, $\;\underline{q} = \{ q^{(1)}, ... ,q^{(n)} \}\;$, satisfying
$ \sum_{i=1}^{n} q^{(i)} = 0 $, we define the one form
\be
\Pi_{\underline{\eta},\underline{q}}(\xi)\; := \; \sum_{i=1}^{n}
\Pi_{\eta^{(i)},q^{(i)}}(\xi)
\label{A6}
\ee
with
\begin{eqnarray*}
\Pi_{\eta^{(i)},q^{(i)}}(\xi)
& = & \Pi_{\,\nu}^{\eta^{(i)},q^{(i)}}(\xi) \; d\xi^\nu
\; := \; -\sqrt{\pi}\, q^{(i)}\,\left[\,\ast\, d \,\Delta^{-1} \,\delta^{(2)}_{\eta^{(i)}} 
\,\right](\xi)\\
&=&\frac{1}{2\sqrt{\pi}} q^{(i)}\; \frac{\xi^\mu-\eta^{(i) \mu}}{|\xi-\eta^{(i)}|}\; \varepsilon_{\mu\nu}
d\xi^\nu \qquad \qquad \mu,\nu = 0,1 \quad.
\end{eqnarray*}
\\
\noindent Here, we use that
\be
\partial_\mu\partial_\mu \Delta^{-1} (\xi)\; =\; - \delta^{(2)}(\xi) \qquad
\mbox{, i.e. , }\qquad \Delta^{-1}(\xi) \;=\; -\frac{1}{2\pi} \ln|\xi| \quad.
\label{A7}
\ee
\\
In physics terminology, $\Pi_{\eta,q} $ is the vector potential of a magnetic
vortex of charge $ q $ at the point $ \eta = (\eta_0,\eta_1) $ . 
The field strength is given by
$$
d \,\Pi_{\eta,q} (\xi) \; = \; \sqrt{\pi}\, q\,\ast\,\delta(\xi-\eta) \quad,
$$
If $D$ is a two-dimensional domain then
$$
\frac{1}{\sqrt{\pi}} \int_D\; d \,\Pi_{\eta,q} (\xi)  \; = \; \left\{
\begin{array}{ccl}
q & , & \mbox{if}\; \eta \in D\\
0 & , & \mbox{else}
\end{array} \right. \quad.
$$
Locally on $ {\bf R}^2 \setminus \{\eta\} $ , the one form  $\Pi_{\eta,q} $ can
be written as the derivative of a function $ \alpha_{\eta,q}(\xi) $ , i.e. ,
\be
\Pi_{\eta,q} (\xi) \; = \; d\,\alpha_{\eta,q}(\xi) \quad.
\label{A8}
\ee
By identifying the euclidean space-time with the complex plane, $\;
\xi = (\xi_0,\xi_1)  \mapsto \overline{\xi} = i\xi_0+\xi_1 \; $ , the function 
$ \alpha_{\eta,q}(\xi) $ can be represented as
\be
\alpha_{\eta,q}(\xi) \; = \; - \frac{q}{2\sqrt{\pi}} \arg(\overline{\xi}-\overline{\eta}) \quad.
\label{A9}
\ee
Note that this function is not globally defined on {\bf C} ; we identify the branch-cut
with the negative, real axis.

The expectation value of a disorder field
$\;
D(\underline{\eta},\underline{q})  = \prod_{i=1}^{n} \; D(\eta^{(i)},q^{(i)})\;
$
is given by
\be
\langle\; D(\underline{\eta},\underline{q})\; \rangle_B \; = \; \left\{\; \frac{\int
{\cal D}\varphi \; e^{-\tilde{S} (d\varphi + \Pi_{\underline{\eta},\underline{q}} )}}{
\int
{\cal D}\varphi \; e^{-\tilde{S}( d\varphi )}}\; \right\}_{\mbox{ren}} \quad.
\label{A10}
\ee
On the r.h.s. of \eq{A10}, a multiplicative renormalization is necessary, in order to
eliminate an infinite "self-energy".

Correlation functions involving disorder fields $ D(\underline{\eta},\underline{q}) $ 
and functionals, $ {\cal F}(d\varphi,\varphi) ,$  depending on the bosonic field $ \varphi $
and its derivative $ d\varphi $ are defined by
\\
\be
\langle\; D(\underline{\eta},\underline{q})\;{\cal F}(d\varphi,\varphi)\; \rangle_B \; 
= \; \left\{\; \frac{\int
{\cal D}\varphi \;{\cal F}(d\varphi+\Pi_{\underline{\eta},\underline{q}}\,,\,\varphi+
\alpha_{\underline{\eta},\underline{q}})\; e^{-\tilde{S} (d\varphi + \Pi_{\underline{\eta},\underline{q}} )}}{
\int
{\cal D}\varphi \; e^{-\tilde{S}( d\varphi )}}\; \right\}_{\mbox{ren}} \quad.
\label{A11}
\ee

\noindent For the electron fields, the  following identifications hold :
\\
\be
\begin{array}{cclcc}
\psi_1(\xi) & \longleftrightarrow & \frac{1}{(2\pi)^{\frac{1}{4}}}\; D(\xi,1)\;
:e^{i\sqrt{\pi} \varphi(\xi)}: & =: & \psi_1(\varphi;\xi)\;,\\&&&&\\
\psi_2(\xi) & \longleftrightarrow & \frac{1}{(2\pi)^{\frac{1}{4}}}\; D(\xi,1)\;
:e^{-i\sqrt{\pi} \varphi(\xi)}: & =: & \psi_2(\varphi;\xi)\;,\\&&&&\\
\psi_1^\ast(\xi) & \longleftrightarrow & \frac{1}{(2\pi)^{\frac{1}{4}}}\; D(\xi,-1)\;
:e^{-i\sqrt{\pi} \varphi(\xi)}: & =: & \psi_1^\ast(\varphi;\xi)\;,\\&&&&\\
\psi_2^\ast(\xi) & \longleftrightarrow & \frac{1}{(2\pi)^{\frac{1}{4}}}\; D(\xi,-1)\;
:e^{i\sqrt{\pi} \varphi(\xi)}: & =: & \psi_2^\ast(\varphi;\xi)\;,
\end{array}
\label{A12}
\ee
\\
with the convention that, in a product composed of several  $\psi_\alpha '$s
, we write the disorder fields to the left of all functionals depending on
$\varphi$ . 

\noindent The normal order of the exponentials is defined by
\be
:e^{i\int d^2\xi \; \varphi(\xi) f(\xi)}: \; = \;e^{i\int d^2\xi \; \varphi(\xi) f(\xi)}
\; e^{\frac{1}{2} \int d^2\xi\;f(\xi)\,
\left(\Delta + m_0^2\right)^{-1}\,
f(\xi)} \quad,
\label{A13}
\ee
where $\;\Delta = -  \partial_\mu \partial_\mu\;,$  $\; f $ denotes a test function,  and $ m_0 $ is a positive constant.

\noindent It is a staightforward calculation (cf.\cite{17,18}) to verify that
$$
\langle\; \prod_{i=1}^n \psi^\sharp_{\beta_i}(\xi_i)\;\rangle^0_F \quad =
\quad \langle\; \prod_{i=1}^n \psi^\sharp_{\beta_i}(\varphi ; \xi_i)\;\rangle^0_B \quad.
$$ 
{\bf Remark:} We use the prescription that the function $\;\langle\; \prod_{i=1}^n \psi^
\sharp_{\beta_i}(\xi_i)\;\rangle \;$ is the analytic continuation of the function
$ \; \langle\;\hat{T}\;\prod_{i=1}^n \psi^\sharp_{\beta_i}(\xi_i)\;\rangle \; $, where
$ \hat{T} $ denotes the time-ordering operator for Grassmann fields, i.e.,
$$
\hat{T} \left[\;\psi^\sharp(\xi_1) \cdots \psi^\sharp(\xi_n)\;\right] \;
:= \; \mbox{sgn}(\sigma)\; \psi^\sharp(\xi_{\sigma(1)}) \cdots \psi^\sharp(\xi_
{\sigma(n)}) \quad,
$$
where $ \sigma $ is the permutation such that $\; \xi_{\sigma(1)}^0 >
\xi_{\sigma(2)}^0 > \cdots > \xi_{\sigma(n)}^0 \;$ .

\noindent We shall make use of the following basic identities (cf. \cite{17}) :
\\
$$
\int \frac{{\cal D}\varphi}{Z} \; e^{-\frac{1}{2} \int d\xi\;\partial_\mu\varphi  
\partial_\mu\varphi} \prod_{j=1}^n :e^{i \varepsilon_j \varphi(\xi_j)}:\; =
\hspace{6cm}
$$
$$
\hspace{2cm} =\; \left\{
\begin{array}{lcc}
0 & , & \mbox{if} \sum_{j=1}^n \varepsilon_j \neq 0\;, \\
\prod_{i=1}^n c(m_0)^{\frac{\varepsilon_j^2}{4\pi}}\; e^{-\frac{1}{2\pi} \sum_
{1\leq i \leq n} \varepsilon_i \varepsilon_j \ln(\frac{1}{|\xi_i - \xi_j|})} & , &
\mbox{if}  \sum_{j=1}^n \varepsilon_j = 0\;,
\end{array}
\right. 
$$
\\
where the constant $m_0$ can be choosen such that $ c(m_0) = \frac{1}{2\pi} $ ,
$$
-\frac{1}{2} \int d^2\zeta\; \Pi^\mu_{\underline{\xi},\underline{\eta}}(\zeta)
\Pi^\mu_{\underline{\xi},\underline{\eta}}(\zeta) \;= \hspace{9cm}
$$
$$
\hspace{2cm} \; = -\frac{\pi}{2} \left[ \;\sum_{i=1}^n \Delta^{-1}(\xi_i,\xi_i) + \sum_{i=1}^n
\Delta^{-1}(\eta_i,\eta_i)\; \right] \hspace{4cm}
$$
$$
\hspace{2.8cm} -\frac{1}{2} \left[\; \sum_{1\leq i<j\leq n}\ln|\xi_i-\eta_j| -
\sum_{1\leq i<j\leq n}\ln|\xi_i-\xi_{j}| - \sum_{1\leq i<j\leq n}\ln|\eta_i-\eta_{j}|
\;\right]\;,
$$
and
$$
\int d^2\zeta\;\Pi^\mu_{\underline{\xi},\underline{\eta}}(\zeta)\; \partial_\mu\varphi(\zeta)\; = \; 0
\quad , \hspace{7.5cm}
$$
where the vortices located at the points $ \underline{\xi} := (\xi_1, \cdots , \xi_n) $
carry the charge $ q = -1 $ and the ones located at the points $ \underline{\eta}
:= (\eta_1, \cdots , \eta_n) $
 carry the charge $ q = 1 .$ 

As an application, we calculate the Green function $ \;\langle \psi^\ast_
2(\xi) \psi_2(\eta) \rangle^0 \;$ for the free system. Let $ \xi_0 > \eta_0 .$
Then
\begin{eqnarray*}
\langle\;\psi^\ast_2(\xi) \psi_2(\eta)\;\rangle^0_F & = &
\langle\;\psi^\ast_2(\varphi;\xi) \psi_2(\varphi;\eta)\;\rangle^0_B\\
&=& \frac{1}{\sqrt{2\pi}} \langle\;D(\xi,-1) D(\eta,1) :e^{i\sqrt{\pi} \varphi(\xi)}:
:e^{-i\sqrt{\pi} \varphi(\eta)}:\;\rangle^0_B\\
&=& \frac{1}{\sqrt{2\pi}} \Bigg[\; e^{ i \sqrt{\pi} \left( \alpha_{\xi,\eta}(\xi) -
\alpha_{\xi,\eta}(\eta) \right)} \;\cdot \hspace{4cm} \\
&&\hspace{2cm}\cdot\; \int \frac{{\cal D}\varphi}{Z}\; e^{- \tilde{S}^0\left(
d\varphi+\Pi_{\xi,\eta} , d\varphi+\Pi_{\xi,\eta} \right)} \;:e^{i\sqrt{\pi} \varphi(\xi)}:
:e^{-i\sqrt{\pi} \varphi(\eta)}: \; \Bigg]_{\mbox{ren}}\\
&=& \frac{1}{\sqrt{2\pi}} \; e^{-\frac{i}{2} [ \arg(\overline{\eta}-\overline{\xi}) + \arg(\overline
{\xi}-\overline{\eta}) ]}\; e^{-\frac{1}{2} \ln|\xi-\eta|}\; \cdot \hspace{3cm}\\
&&\hspace{2cm}\cdot\; \int \frac{{\cal D}\varphi}{Z}\; e^{- \tilde{S}^0\left(
d\varphi , d\varphi \right)} \;:e^{i\sqrt{\pi} \varphi(\xi)}:
:e^{-i\sqrt{\pi} \varphi(\eta)}:
\end{eqnarray*}
\be
= \; \frac{1}{2\pi}\; \frac{i}{i(\xi_0-\eta_0) + (\xi_1-\eta_1)} \quad.
\hspace{2.5cm}
\label{A14}
\ee 
One observes that 
$$
\langle\;\psi_2(\eta) \psi^\ast_2(\xi)\;\rangle \; =
-\; \langle\;\psi^\ast_2(\xi) \psi_2(\eta)\;\rangle\quad.
$$
Similarly one obtains
\be
\langle\;\psi^\ast_1(\xi) \psi_1(\eta)\;\rangle \; = \;
\frac{1}{2\pi}\; \frac{-i}{-i(\xi_0-\eta_0) + (\xi_1-\eta_1)}
\label{A15}
\ee
and
$$
\langle\;\psi_1(\xi) \psi_2(\eta)\;\rangle \;=\;
\langle\;\psi_1(\xi) \psi_2^\ast(\eta)\;\rangle \; =\;
\langle\;\psi_2(\xi) \psi_1^\ast(\eta)\;\rangle \;=\;
\langle\;\psi^\ast_1(\xi) \psi_2^\ast(\eta)\;\rangle \; = \; 0 \quad.
$$
These are the correct formulas for the free system ( compare, for example 
appendix D of \cite{13}, where the same conventions are used ).

One can show that  the same bosonization identities hold for the interacting
system, with the interaction $ V $ given by \eq{A3} (cf. \cite{18}). The only difference
is that  expectation values of bosonic operators are taken with respect
to the interacting system, i.e.
\be
\langle\; (\; \cdot\; )\; \rangle^V_B\; := \; \int \frac{{\cal D}\varphi}{Z_V} \; e^{- [\;\frac{1}
{2}\int d\xi\;\partial_\mu\varphi(\xi)  \partial_\mu\varphi(\xi)
\;+\; V(d \varphi)\; ]} \; (\; \cdot\; ) \quad,
\label{A16}
\ee
where the functional $ V(\varphi) $ is obtained from the functional $ V(\overline{
\psi};\psi) $ by using the identities (\ref{A3}), i.e. , 
\be
V(d \varphi) \; = \; \frac{g}{2\pi} \int d^2\xi\; \partial_\mu\varphi(\xi) 
\partial_\mu\varphi(\xi) \quad .
\label{A17}
\ee
We repeat the calculation of the Green function  $ \;\langle \psi^\ast_
2(\xi) \psi_2(\eta) \rangle^V \;$ for the interacting system : 
\begin{eqnarray*}
\langle\;\psi^\ast_2(\xi) \psi_2(\eta)\;\rangle^V_F & = &
\langle\;\psi^\ast_2(\varphi;\xi) \psi_2(\varphi;\eta)\;\rangle^V_B\\
&=& \frac{1}{\sqrt{2\pi}} \langle\;D(\xi,-1) D(\eta,1) :e^{i\sqrt{\pi} \varphi(\xi)}:
:e^{-i\sqrt{\pi} \varphi(\eta)}:\;\rangle^V_B\\
&=& \frac{1}{\sqrt{2\pi}} \Bigg[\;\int \frac{{\cal D}\varphi}{Z_V}\; e^{-
(1+\frac{g}{\pi}) \tilde{S}^0\left(
d\varphi+\Pi_{\xi,\eta} , d\varphi+\Pi_{\xi,\eta} \right)}\; \cdot  \hspace{4cm} \\
&&\hspace{2cm}\cdot\;  \;:e^{i\sqrt{\pi} \left(
\varphi(\xi)+\alpha_{\xi,\eta}(\xi) \right)}:
:e^{-i\sqrt{\pi} \left( \varphi(\eta)+\alpha_{\xi,\eta}(\eta) \right)}: \; \Bigg]_
{\mbox{ren}}\\
&=& \frac{i}{\sqrt{2\pi}} \; e^{-i \arg(\overline{\xi}-\overline{\eta}) }
\; e^{-\frac{1}{2} (1+\frac{g}{\pi}) \ln|\xi-\eta|}\; \cdot \hspace{3cm}\\
&&\hspace{2cm}\cdot\; \int \frac{{\cal D}\varphi}{Z_V}\; e^{-
(1+\frac{g}{\pi}) \tilde{S}^0\left(
d\varphi , d\varphi \right)} \;:e^{i\sqrt{\pi} \varphi(\xi)}:
:e^{-i\sqrt{\pi} \varphi(\eta)}:
\end{eqnarray*}
\be
= \; \frac{i}{2\pi}\; e^{-i \arg(\overline{\xi}-\overline{\eta}) }\;
 \frac{1}{|\xi-\eta|^{1+\beta}} \quad,
\hspace{2.45cm}
\label{A18}
\ee
where
$$
\beta \; = \;  \frac{2 \left(\frac{2g}{\pi}\right)^2}{1 + 2 \left(\frac{2g}{\pi}\right)^2}
\quad.
$$
Similarly,
\be 
\langle\;\psi^\ast_1(\xi) \psi_1(\eta)\;\rangle^V \; =
\;   \frac{-i}{2\pi}\; e^{i \arg(\overline{\xi}-\overline{\eta}) }\;
 \frac{1}{|\xi-\eta|^{1+\beta}} \quad.
\label{A19}
\ee

The exponent $\beta$ describes the decay of the elecron propagator for large
arguments. The dependence on the coupling constant is a characteristic feature
of Luttinger Liquids.

\setcounter{section}{2}
\setcounter{equation}{0}
\section*{Appendix B : Consistency of bosonization formulas  }

In this appendix, we show on the example of $\; \langle\;
\overline{\jmath}_{[\om_1]}(\xi)\;\overline{\jmath}_{[\om_2]}(\eta)\;\rangle^0\; $
that the bosonization formulas (\ref{57}) -- (\ref {61}) for the electron fields
$ \psi_\om^\sharp $ imply the bosonization formulas (\ref{37}) -- (\ref{Z39})  for
the electron currents $ j_{[\om]} $ and $ \overline{\jmath}_{[\om]} .$
The generalization  that this statement holds for arbitrary
products of  current densities evaluated in the non-interacting ground state is 
staightforward. 
\\
\\
{\bf Remark :}\\ 
The calculation of  expectation values of products of current
densities in the {\sl interacting} ground state is organized as a perturbation
expansion in powers of the set of coupling constants $ g_n(\vom\cdot\vom') $ and
the inverse scale parameter $\lambda_n^{-1} .$  For  systems with 
 interactions of the form given in \eq{42} (i.e. interactions which can be expressed
in terms of the current densities $ j_{[\om]} $ and $ \overline{\jmath}_{[\om]}
,$ cf. \eq{ZZ43} ) ,  such an
expansion reduces the calculation to the task of evaluating   
products of  current densities  $ j_{[\om]} $ and $ \overline{\jmath}_{[\om]}
$ in the non-interacting ground-state  -- where our statement applies.
 \\
\\
By inserting \eq{Z30}, one obtains for the expression
$\; \langle\;
\overline{\jmath}_{[\om_1]}(\xi)\;\overline{\jmath}_{[\om_2]}(\eta)\;\rangle^0\; :$
$$
 \langle\;
\overline{\jmath}_{[\om_1]}(\xi)\;\overline{\jmath}_{[\om_2]}(\eta)\;\rangle^0\; =
\hspace{11cm} 
$$
$$
\hspace{1.5cm}
\sum_{\om_1',\om_2' \in {\cal S}^+}\; 
\Bigg\{\;\; e^{i k_F \lambda (\vom_1-\vom_1')\vec{\xi}}\, e^{i k_F
\lambda (\vom_2-\vom_2') \vec{\eta}}\;\langle\;
:\!\psi^\ast_{\om_1'}(\xi)\psi_{\om_1}(\xi)\!:\;:\!\psi^\ast_{\om_2'}(\eta)
\psi_{\om_2}(\eta)\!:\;\rangle^0
$$
$$
\hspace{3.5cm}
+\; e^{i k_F \lambda (\vom_1-\vom_1')\vec{\xi}}\, e^{i k_F
\lambda (\vom_2'-\vom_2) \vec{\eta}}\;\langle\;
:\!\psi^\ast_{\om_1'}(\xi)\psi_{\om_1}(\xi)\!:\;:\!\psi^\ast_{\om_2}(\eta)
\psi_{\om_2'}(\eta)\!:\;\rangle^0
$$
$$
\hspace{3.5cm}
+\; e^{i k_F \lambda (\vom_1'-\vom_1)\vec{\xi}}\, e^{i k_F
\lambda (\vom_2-\vom_2') \vec{\eta}}\;\langle\;
:\!\psi^\ast_{\om_1}(\xi)\psi_{\om_1'}(\xi)\!:\;:\!\psi^\ast_{\om_2'}(\eta)
\psi_{\om_2}(\eta)\!:\;\rangle^0
$$
$$
\hspace{3.5cm}
+\; e^{i k_F \lambda (\vom_1'-\vom_1)\vec{\xi}}\, e^{i k_F
\lambda (\vom_2'-\vom_2) \vec{\eta}}\;\langle\;
:\!\psi^\ast_{\om_1}(\xi)\psi_{\om_1'}(\xi)\!:\;:\!\psi^\ast_{\om_2}(\eta)
\psi_{\om_2'}(\eta)\!:\;\rangle^0\;\;\Bigg\}
$$
$$
\hspace{3cm} = \;\hspace{10cm}
$$
$$
\hspace{2cm}
\Bigg\{\;\hspace{2cm}
e^{i k_F \lambda (\vom_1-\vom_2)(\vec{\xi}-\vec{\eta})}\;\langle\;
\psi^\ast_{\om_2}(\xi)\psi_{\om_2}(\eta)\;\rangle^0\;
\langle\;\psi_{\om_1}(\xi)
\psi^\ast_{\om_1}(\eta)\;\rangle^0
$$
$$
\hspace{3cm}
+\; \delta_{\om_1,\om_2}\;\sum_{\om_1' \in {\cal S}^+}\; 
 e^{i k_F \lambda (\vom_1-\vom_1')(\vec{\xi}-\vec{\eta})}\;\langle\;
\psi^\ast_{\om_1'}(\xi)\psi_{\om_1'}(\eta)\;\rangle^0\;
\langle\;\psi_{\om_1}(\xi)
\psi^\ast_{\om_1}(\eta)\;\rangle^0
$$
$$
\hspace{3cm}
+\; 
\delta_{\om_1,\om_2}\;\sum_{\om_1' \in {\cal S}^+}\; 
 e^{i k_F \lambda (\vom_1'-\vom_1)(\vec{\xi}-\vec{\eta})}\;\langle\;
\psi^\ast_{\om_1}(\xi)\psi_{\om_1}(\eta)\;\rangle^0\;
\langle\;\psi_{\om_1'}(\xi)
\psi^\ast_{\om_1'}(\eta)\;\rangle^0
$$
$$
\hspace{3.5cm}
+ \qquad \;\,
e^{i k_F \lambda (\vom_2-\vom_1)(\vec{\xi}-\vec{\eta})}\;\langle\;
\psi^\ast_{\om_1}(\xi)\psi_{\om_1}(\eta)\;\rangle^0\;
\langle\;\psi_{\om_2}(\xi)
\psi^\ast_{\om_2}(\eta)\;\rangle^0\hspace{1cm}\Bigg\}
$$

In order to simplify these expressions we can
apply a "discretized" version of lemma (3.21)  in \cite{13} :

\noindent  {\bf Lemma :}

 For a  decomposition of   the surface
$ {\cal S}^{d-1}_1 $ of the d-dimensional unit sphere (with $d > 1$) into congruent,
quadratic patches with sides of length $ \frac{1}{\lambda} $ we define the set $ {\cal
M}
$ of d-dimensional unit
vectors $ \;\vom_i\;,\;i = 1, ... , N=\frac{Vol ({\cal S}^{d-1}_1)}{\lambda^{d-1}}
\;$ pointing to the centers of these patches. Given a vector $ \vec{\sigma}_j \in {\cal
M} $ and some test function $ f(\vom_i;\vec{\xi}) $, one finds, in the limit $ \lambda
\rightarrow \infty ,$ the following asymptotic formula :
  
$$
\int_{{\bf R}^d} d^d\xi \sum_{\vom_i \in {\cal M}}\;f(\vom_i;\vec{\xi}) \;\left[\;
\delta^{d-1}_{\om_i,k_F}\left(\vec{\xi}_\perp(\vom_i)\right)\; e^{i  k_F \lambda
\vec{\xi} (\vom_i-\vec{\sigma}_j)}\;\right] \hspace{6cm}
$$
\be
\hspace{2cm} = \; \int_{{\bf R}^d} d^d\xi \sum_{\vom_i \in {\cal
M}}\;f(\vom_i;\vec{\xi})
\;\left[\;
\delta_{\om_i,\sigma_j}\;\delta^{d-1}_{\sigma_j,\Lambda}\left(\vec{\xi}_\perp(\vec{
\sigma}_j)\right)\quad+\quad O(\frac{1}{\lambda}) \;\right]\;,
\label{B1}
\ee
where 
$$
\qquad \vec{\xi}_\perp(\vec{\sigma})\; := \; \vec{\xi} - (\vec{\xi}\cdot\vec{\sigma})\,
\vec{\sigma} 
$$
and
$$
\delta^{d-1}_{\sigma_j,\Lambda}\left(\vec{\xi}_\perp(\vec{
\sigma}_j)\right) \;:=
\int_{[\frac{-\Lambda}{2},\frac{\Lambda}{2}]^{d-1}}\dbar^{\,d-1}k_\perp\;e^{i \vec{
k}_\perp \vec{\xi}_\perp(\vec{
\sigma}_j)}\quad.
$$
Here, $ \Lambda \ll \lambda k_F $  denotes an arbitrary momentum which can be
chosen to be equal to $ k_F .$
\\
\\

\noindent The proof of this lemma is analoguous to the proof of the continuous
version of the lemma presented in appendix C of \cite{13} .
\\ 
By applying this lemma  to the expressions appearing above and by inserting the
bosonization formulas (\ref{57}) -- (\ref {61}) for the electron fields $
\psi_\om^\sharp $, we obtain the following formula :
$$
 \langle\;
\overline{\jmath}_{[\om_1]}(\xi)\;\overline{\jmath}_{[\om_2]}(\eta)\;\rangle^0\;
\approx
\; 
- 4\, (\frac{k_F}{2\pi})^{d-1}\;\delta_{\om_1,\om_2}\;\delta^{d-1}_
{\om_1,k_F}\left(\vec{\xi}_\perp(\vom_1)\right)\;\cdot\hspace{5cm}
$$
\be
\hspace{.5cm}
\langle\;
\psi^\ast_{\om_1\parallel}(\xi_0,\xi_\|;\varphi_{[\om_1]}^{\|})\psi_{\om_1\parallel}
(\eta_0,\eta_\|;\varphi_{[\om_1]}^{\|})\;
\rangle_{S^0(\varphi_{[\om_1]}^\|)}
\; 
\langle\;\psi_{\om_1\|}(\xi_0,\xi_\|;\varphi_{[\om_1]}^\|)
\psi^\ast_{\om_1\|}(\eta_0,\eta_\|;\varphi_{[\om_1]}^\|)\;\rangle_
{S^0(\varphi_{[\om_1]}^\|)}\;,
\label{B2}
\ee
where  $\;\psi^\sharp_{\om\parallel}(\xi_0,\xi_\|;\varphi_{[\om]}^{\|}) \;$ is a short-hand
notation for the bosonized expressions of the radial electron fields
$\;\psi^\sharp_{\om\parallel}(\xi_0,\xi_\|)\;$ specified in \eq{Z60}.
 The explicit form of the radial electron propagators
is given in eqs. (\ref{A14}) and (\ref{A15}).\\
It is a straightforward calculation to verify that \eq{B2}  is identical to the following
formula:
$$
 \langle\;
\overline{\jmath}_{[\om_1]}(\xi)\;\overline{\jmath}_{[\om_2]}(\eta)\;\rangle^0\;
\approx
\hspace{11cm}
$$
$$
\hspace{2cm}
- 4\, (\frac{k_F}{2\pi})^{d-1}\;\delta_{\om_1,\om_2}\;\delta^{d-1}_
{\om_1,k_F}\left(\vec{\xi}_\perp(\vom_1)\right)\;\cdot\;
\left\langle\;\overline{\partial}^{\om_1}\varphi_{[\om_1]}^\|(\xi_0,\xi_\|)\;
\overline{\partial}^{\om_2}\varphi_{[\om_2]}^\|(\eta_0,\eta_\|)
\;\right\rangle_
{S^0(\varphi_{[\om_1]}^\|)}
$$
\be
\hspace{1cm}= \; \left( \frac{-2}{\sqrt{\pi}}
(\frac{k_F}{2\pi})^{\frac{d-1}{2}}\right)^2
\; 
\left\langle\;\overline{\partial}^{\om_1}\varphi_{[\om_1]}(\xi)\;
\overline{\partial}^{\om_2}\varphi_{[\om_2]}(\eta)
\;\right\rangle_
{S^0(\{\varphi_{[\om_1]}\})}\quad,
\label{B3}
\ee
where we use \eq{A7} for the propagator of the radial bose field 
$ \bo^\|(\xi_0,\xi_\|) ,$ and \eq{ZZ61} to deduce the second part of the equation.\\
Formula (\ref{B3}) reproduces the bosonization identity for the current
density $ \overline{\jmath}_\om ,$ cf. eqs. (\ref{37}) - ( \ref{Z39}).\\
Hence, we have verified our claim thus showing the consistency
of the bosonization formulas for the electron fields and the current densities.

\setcounter{section}{3}
\setcounter{equation}{0}

\section*{Appendix C :  Calculation of the Electron Propagator for a System with
Long-Range, Transverse Current-Current Interactions}

In order to apply \eq{63} to the calculation of the propagator 
$$
G^{\|\alpha}_{\om,n}(\xi_0-\eta_0 , \xi_\|-\eta_\|)
 \; := \; - \left\langle\;\psi_{\om\alpha}(\xi_0,\xi_\|)\;\psi^\ast_{\om\alpha}(\eta_0,\eta_\|)
\;\right\rangle^\|_n 
$$
of the radial electron field $ \psi^\sharp_{\om\alpha} $, we have to determine the 
effective action $ S^T_n(\bo^\|) $  of the bosonic field $ \bo^\| $ propagating along the direction
$ \rom $.

The propagator $ \langle\;\hbo(k)\;\hbo(k')\;\rangle_n $ of the bosonized interacting
system is given in \eq{68}. It can be written as
\be
\langle\;\hbo(k)\;\hbo(k')\;\rangle_n \; = \; (2\pi)^{d+1} \, \delta^{(d+1)}(k+k')\;
\frac{1}{k_0^2 + (v_{Fn} \vom\vec{k})^2} \; {\cal K}^n_{\rom}(k) \quad,
\label{C.1}
\ee
with
$$
{\cal K}^n_{\rom}(k) \;=\hspace{13cm} 
$$
\be
\qquad
\frac{1 + \frac{2}{\lambda_n^{d-1}} 
\frac{1}{2\pi} (\frac{k_F}{2\pi})^{d-1}
  g_n \hat{V}(|\frac{\vec{k}}{\lambda_n}|) \left[ \frac{1}{d-1}
\sum\limits_{\om'\in{\cal S}^+} \frac{k_0^2}{k_0^2+(v_{Fn} \vom'\vec{k})^2} \left( (\vom')^2 -
\frac{(\vom'\vec{k})^2}{\vec{k}^2} \right) (1 - \delta_{\om,\om'}) \right]}
{1 + \frac{2}{\lambda_n^{d-1}} \frac{1}{2\pi} (\frac{k_F}{2\pi})^{d-1}
g_n  \hat{V}(|\frac{\vec{k}}{\lambda_n}|) \frac{1}{d-1}
\sum\limits_{\om'\in{\cal S}^+} \frac{k_0^2}{k_0^2+(v_{Fn} \vom'\vec{k})^2} \left( (\vom')^2 -
\frac{(\vom'\vec{k})^2}{\vec{k}^2} \right)  }\quad.
\label{C.2}
\ee
One observes that the function $ {\cal K}^n_{\rom}(k) $ is bounded by
$ \; 0 \leq {\cal K}^n_{\rom}(k) \leq 1 \; $, for arbitrary $k$ .\\
In explicit calculations, expression (C.2) for $ {\cal K}^n_{\rom}(k) $ is not convenient.
We replace it by 
\be
\widetilde{\cal K}^n_{\rom}(k)\;=\; \frac{1 +   g_n \hat{V}(|\frac{\vec{k}}{\lambda_n}|)\;
{\bf \Pi}^t_{n}(k)}
{1 +  g_n \hat{V}(|\frac{\vec{k}}{\lambda_n}|) \left[
 {\bf \Pi}^t_{n}(k) + \frac{2}{\lambda_n^{d-1}}  \frac{1}{2\pi} (\frac{k_F}{2\pi})^{d-1}
 \frac{k_0^2}{k_0^2+(v_{Fn} \vom\vec{k})^2} \left( (\vom)^2 -
\frac{(\vom\vec{k})^2}{\vec{k}^2} \right) \right]  }
\label{C.3}
\ee
which has the same asymptotics for small momenta $k$ as $ {\cal K}^n_{\rom}(k) $ .
The function $ {\bf \Pi}^t_{n}(k) $ has been defined in \eq{Z36}, with $\;v_F \rightarrow
v_{Fn} .$\\
The action $ S^T_n(\bo) $ of the bosonic field $ \hbo(k) $ which reproduces the 
propagator (\ref{C.1}) -- but with the function $ {\cal K}(k) $ replaced by $ 
\widetilde{\cal K}(k) $ --
is given by
\be
 S^T_n(\bo) \; = \; \frac{1}{2} \int_{{\bf R}\times\overline{Q}^d}\dbar^{\,d+1}k\;
\hbo(-k)\;\left[\;\left(k_0^2 + (v_{Fn} \vom\vec{k})^2\right)\;\left(\widetilde{\cal K}^n
_{\rom}(k)\right)^{-1}\; \right]\;
\hbo(k) \quad.
\label{C.4}
\ee
We define
\be
 \left(\widetilde{\cal K}^n_{\rom}(k)\right)^{-1} \; := \; 1 + T^n_{\rom}(k) \quad,
\label{C.Z4}
\ee
with
$$
T^n_{\rom}(k) \; := \; \frac{\frac{2}{\lambda_n^{d-1}} \frac{1}{2\pi} 
(\frac{k_F}{2\pi})^{d-1}
  \frac{k_0^2}{k_0^2+(v_{Fn} \vom\vec{k})^2} \left( (\vom)^2 -
\frac{(\vom\vec{k})^2}{\vec{k}^2} \right)}
{g_n^{-1} \hat{V}^{-1}(|\frac{\vec{k}}{\lambda_n}|) + 
{\bf \Pi}^t_{n}(k)} \quad.
$$
\noindent The effective action $ S^T_n(\bo^\|) $ of the modes propagating along the
direction $ \rom $ is deduced from the action $ S^T_n(\bo) $ by averaging over the
momenta perpendicular to $ \rom $ . 
One obtains : 
\be
 S^T_n(\bo^\|) \; = \; \frac{1}{2} \int_{\bf R}\dbar k_0 \int_\frac{-k_F}{2}^
\frac{k_F}{2}\dbar k_\| \;
\hbo(-k)\,\left[\,\left(k_0^2 + (v_{Fn} k_\|)^2\right) \left( 1 + t^n_{\rom}(k_0,k_\|)
\right)\,\right]\,
\hbo(k) \quad,
\label{C.5}
\ee
with
$$
 t^n_{\rom}(k_0,k_\|) \; := \; \frac{1}{k_F^{d-1}} \int_{\overline{Q}^{d-1}}
d^{d-1}k_\perp \; T^n_{\rom}(k) \quad.
$$
For small momenta $ (k_0,k_\|) $ and a large scale factor $ \lambda_n $, the calculation 
yields the following asymptotic formula
 for $  t^n_{\rom}(k_0,k_\|) $ :
$$
 t^n_{\rom}(k_0,k_\|)\quad \sim\quad\tilde{ t}^n_{\rom}(k_0,k_\|)
\hspace{8cm} 
$$
\be
\hspace{1cm}  := \; \left\{
\begin{array}{lcl}
a(d)\,\frac{k_0^2}{k_0^2 +(v_{Fn} k_\|)^2}\,\frac{k_F}{|k_0|}\;\frac{1}{\lambda_n^{d-1}} & , & 
\mbox{for}\, 1 \gg |c_n  k_0|  \gg \frac{1}{\lambda_n^\alpha}  \\
&&\\
a(d)\,\frac{k_0^2}{k_0^2 +(v_{Fn} k_\|)^2}\, k_F c_n\, \lambda_n^{\alpha-(d-1)}\,
\frac{2}{1+\alpha}
  \; \cdot &&\\
\hspace{0.1cm}\;\cdot\;\Gamma(\frac{d}{1+\alpha})\,
\Gamma(\frac{1+\alpha-d}
{1+\alpha}) 
\left[\, (\frac{\lambda_n^\alpha c_n |k_0|}{ \pi^{d-1} 2^{d-2}})^\frac{d-1-\alpha}{1+\alpha} - 1\,
\right] & ,&
\mbox{for}\, \frac{1}{\lambda_n^\alpha} \gg |c_n k_0|  
\qquad,
\end{array} \right.
\label{C.6}
\ee 
where $ d = 2,3 $ is the dimension of space, $ a(d) $ is a positive constant of
order unity,  and the constants $\; c_n := \frac{v_{Fn}^2}{k_F^{d-1}} g_n   > 0 \;$ and
 $ \alpha $, with $ 0 \leq \alpha \leq 2 $ ,
parametrize the interaction potential  : $\;g_n \hat{V}(|\vec{p}|) = 
c_n (\frac{k_F}{|\vec{p}|})^\alpha \; $ . \\
\noindent For $\;1 \gg |c_n k_0|  \gg \frac{1}{\lambda_n^\alpha}\;$, one observes 
that $ \tilde{t}^n $
takes values between $ \frac{1}{\lambda_n^{d-1}} $ and $ \lambda_n^{\alpha-(d-1)} .$\\ 
 If  $ \alpha < d - 1 $, it follows that  the interactions 
can be neglected in the limit $ \lambda_n \rightarrow \infty .$ 
If $ \alpha > d-1 $, the interactions
become important as soon as $ k_0 $ is of order $ \frac{1}{\lambda_n^{d-1}} $ or
smaller.\\
\noindent In the range  $\; \frac{1}{\lambda_n^\alpha} \gg |c_n k_0| \;$, \eq{C.6} can be
 rewritten as
$$
\tilde{ t}^n_{\rom}(k_0,k_\|) \; =\; c(d)\,\frac{k_0^2}{k_0^2 + (v_{Fn} k_\|)^2}\,\cdot
 \hspace{9cm}
$$ 
\be
\hspace{2.5cm} \cdot\; \left\{
\begin{array}{lcl}
b(d,\alpha) \frac{3}{1+\alpha}\,\frac{\pi}{\sin(\pi\frac{d}{1+\alpha})} k_F 
 c_n^{\frac{d}{1+\alpha}} \, (\frac{|k_0|}{\lambda_n})^\frac{d-1-\alpha}
{1+\alpha} & , & \mbox{for}\, \alpha > d-1\\&&\\
- \frac{2}{1+\alpha} \, k_F c_n \, \ln(\frac{\lambda_n^{d-1} g_n |k_0|}{k_F \pi^{d-1} 2^{d-2}}) 
& , & \mbox{for} \, \alpha = d-1\\
&&\\
b(d,\alpha)\,  k_F c_n \, \lambda_n^{\alpha-(d-1)} &, & \mbox{for}\, d-1 > \alpha
\geq 0
\qquad.
\end{array} \right.
\label{C.7}
\ee 
For $ \alpha < d-1 $ , the effect of the interactions is suppressed by a
factor $ \frac{1}{\lambda_n^{d-1-\alpha}} $. For $ \alpha \geq d-1 $, one observes a
singular $k_0$ - dependence of the function $ \tilde{t} .$ 

Using the methods presented in appendix A, the calculation of the radial electron
propagators is straightforward. Following \eq{63}, the large-distance asymptotics
of, e.g.,  the propagator $ G^{\|1}_{\om,n} $ is obtained by evaluating the expression
$$
\left\langle\;\psi^\ast_{\om1}(\zeta)\;\psi_{\om1}(0)\;
\right\rangle^\|_n\; \sim\; 
\frac{1}{\sqrt{2\pi}}\;\bigg[\;e^{- i \sqrt{v_{Fn} \pi}\left( \alpha^\rom_{\zeta,0}
(\zeta) -
\alpha^\rom_{\zeta,0}(0) \right)} \;\cdot\hspace{7cm}
$$
\be\hspace{3cm}\cdot\;  \int{\cal D}\bo^\|\;e^{- \tilde{S}_n^T
( \partial_\mu\bo^\| + \Pi_{\rom\mu}^{\zeta,0} )}
\;:e^{-i \sqrt{v_{Fn} \pi} \bo^\|(\zeta)}:\;:e^{i \sqrt{v_{Fn} \pi} \bo^\|(0)}:\;\bigg]\quad,
\label{C.8}
\ee
with $ \zeta = (\xi_0,\frac{\xi_\|}{v_{Fn}}) .$  For each ray $\rom ,$ the vector potential
$ \Pi_{\rom\mu}^{\zeta,0} $ and the function $ \alpha^{\rom}_
{\zeta,0} $  are given
in eqs.(\ref{A6}) and (\ref{A9})  in appendix A. 
However, in contrast to appendix A,  
 we here display the Fermi velocity $ v_{Fn} $  explicitly.
Then, on the rhs. of  eqs.(\ref{A6}) and (\ref{A9}), there is  an additional
factor $ \frac{1}{\sqrt{v_{Fn}}} .$
Note also the supplementary factor $ \sqrt{v_{Fn}} $
in the exponent of the vertex operators. 

\noindent We again 
 write the action $ S_n^T(\bo^\|) $ as a functional, $ \tilde{S}_n
^T(\partial_\mu\bo^\|) ,$ 
of the derivatives $\partial_\mu\bo^\|  .$

\noindent One has that
$$
\tilde{S}_n^T( \partial_\mu\bo^\| + \Pi^{\rom\mu}_{\zeta,0} ) \; = \;
S_n^T(\bo^\|) \; + \;\tilde{S}_n^T(\Pi^{\rom}_{\zeta,0}) \quad,
$$
where the action $ S_n^T(\bo^\|) $ is defined in \eq{C.6}. The contribution
$ \tilde{S}_n^T(\Pi^{\rom}_{\zeta,0}) $ of the disorder fields is given by

\vspace{-.5cm}
\be
 \tilde{S}_n^T(\Pi^{\rom}_{\zeta,0}) \; = \; 
- \pi \int_{\bf R} \dbar k_0 \int_{\frac{-k_F}{2}}^{\frac{k_F}{2}}
\dbar k_\| \; [ e^{i k\zeta} - 1 ]\;\frac{v_{Fn}}{k_0^2 + (v_{Fn} k_\|)^2}\;
 \left( 1 + t^n_{\rom}(k) 
\right) \quad,
\label{C.9}
\ee

\noindent and the contribution of the vertex-operators in \eq{C.8} amounts to 
 $$  \int\frac{{\cal D}\bo^\|}{Z_n^T}\;e^{- S^T_n
(\partial_\mu\bo^\| )}
\;:e^{-i \sqrt{v_{Fn} \pi} \bo^\|(\zeta)}:\;:e^{i \sqrt{v_{Fn} \pi} \bo^\|(0)}:\;=
 \hspace{5cm}
$$
\be
\hspace{2cm}=\; \frac{1}{\sqrt{2\pi}}\; \exp \left[ \pi \int_{{\bf R}\times k_F} 
\dbar^{\,2}k \; [ e^{i k\zeta} - 1 ] \;\left( \frac{v_{Fn}}{k_0^2 + 
(v_{Fn} k_\|)^2}\;
\frac{1}{1 +  t^n_{\rom}(k)} \right)\;\right] \quad.
\label{C.10}
\ee
\noindent
We display the behaviour of the fermionic Green function  $ G^{\|1}_{\om,n}(\zeta) $ in the
following two regions :
\\
$$
\mbox{I}\;=\; \left\{ \;\zeta = (\xi_0,\frac{\xi_\|}{v_{Fn}}) \; :\; c_n \lambda_n^\alpha \ll  |\zeta| \;
\right\}\;,
$$
$$
\mbox{II}\;=\; \left\{\; \zeta = (\xi_0,\frac{\xi_\|}{v_{Fn}}) \; :\; |\zeta| \ll  c_n \lambda_n^\alpha\;
\mbox{and}\;  
|\xi_0| \ll |\frac{\xi_\|}{v_{Fn}}|  \; \right\}\; .
$$

\vspace{.2cm}
\noindent In  region I , the propagator $ G^{\|1}_{\om,n}(\zeta) $  is obtained by
evaluating the integrals in eqs. (\ref{C.9}) and (\ref{C.10}) by using the asymptotic form
(\ref{C.8}) :
\\
$$
\left\langle\;\psi^\ast_{\om1}(\zeta)\;\psi_{\om1}(0)\;\right\rangle^\|_n
 \quad \sim \quad  (\frac{-i}{2\pi}) \; 
e^{i\arg(i\xi_0+\frac{\xi_\|}{v_{Fn}})}\;  \cdot \hspace{7cm}
$$
\be
\;\cdot \;
\left\{
\begin{array}{lcc}
\exp \Bigg(-\overline{u}_{\mbox{\tiny I}}  k_F \, c_n^{\frac{d}{1+\alpha}} \lambda_n^{\frac{\alpha-(d-1)}
{1+\alpha}} \; \cdot &&\\
\hspace{1cm}\cdot\;\bigg\{
 |\zeta|^{\frac{\alpha-(d-1)}{1+\alpha}} \frac{1+\alpha}{\alpha-(d-1)}
 \cos\left[ \frac{\alpha-(d-1)}
{1+\alpha} \arctan(|\frac{v_{Fn} \xi_0}{\xi_\|}|) \right]\; -  & & \\
\hspace{2,5cm} - \;|\frac{\xi_\|}{v_{Fn}}| |\zeta|^{\frac{-d}{1+\alpha}} \cos 
\left[ \frac{-d}{1+\alpha}
\arctan(|\frac{v_{Fn} \xi_0}{\xi_\|}|) \right]\quad\bigg\}
 \bigg) & , & 2 \geq \alpha > d-1\\
& & \\
&&\\
\exp \left( -v\, k_F c_n \ln^2|\zeta| \right) & , & \alpha = d-1\\
& & \\
&&\\ \frac{1}{\sqrt{|v_{Fn} \zeta|}} \;
\left[ (1+w\frac{k_F c_n }{\lambda_n^{d-1-\alpha}})
 \xi_\|^2 + (v_{Fn}\xi_0)^2\right]^{-
\frac{1}{4} (\sqrt{1 + w\frac{k_F c_n}{\lambda_n^{d-1-\alpha}}})^{-1}}\; \cdot
 &&\\
&&\\
\hspace{3.5cm}\cdot\;\left[\xi_\|^2 + (v_{Fn}\xi_0)^2\right]^{-
\frac{w}{8}\frac{k_F c_n}{\lambda_n^{d-1-\alpha}}}
  & , & d-1 > \alpha > 0 \quad ,
\end{array}
\right.
\label{C.11}
\ee
\\
where
\\
$$
\overline{u}_{\mbox{\tiny I}}\;=\; c(d,\alpha)\, \frac{2}{1+\alpha}\, \frac{\pi}
{\sin \pi(\frac{d}{1+\alpha})}\,\Gamma(\frac{d}{1+\alpha})\,  \qquad \mbox{and} \qquad
v\;= \; c(d,\alpha=d-1)\, \frac{1}{d}\quad.
$$

{\bf Remark} : As it stands, the limit of the propagator for $\, \alpha \downarrow
d-1 \,$ is singular and does not reproduce the formula for $\, \alpha = d-1 \,$ .
The reason is that, for $\, \alpha > d-1 \,,$ there is a subleading term, not
displayed in \eq{C.11}, which becomes  leading for $\, \alpha = d-1 \,$ .
Idem for   $\, \alpha \uparrow d-1 \,.$
\\
\\
In region II, one finds :
\begin{itemize}
\item for $\; d-1 < \alpha \leq 2 \;$
$$
\left\langle\;\psi^\ast_{\om1}(\zeta)\;\psi_{\om1}(0)\;\right\rangle^\|_n
 \quad \sim \quad  (\frac{-i}{2\pi})\; 
e^{i\arg(i\xi_0+\frac{\xi_\|}{v_{Fn}})}\; \cdot \hspace{7cm}
$$
\be
\hspace{3cm} \cdot \;
\left\{
\begin{array}{lcl} 
e^{- u_{\mbox{\tiny II}} \frac{k_F}{\lambda_n^{d-1}} \frac{|\xi_\||}{v_{Fn}}}& , & \mbox{for} \;
c_n v_{Fn} \lambda_n^\alpha \gg |\xi_\|| \gg \frac{v_{Fn} \lambda_n{d-1}}{ k_F}\quad,
\\
\frac{1}{|\xi_\||} & , &\mbox{for} \;\frac{v_{Fn} \lambda_n^{d-1}}{ k_F} \gg
   |\xi_\|| \gg
\frac{1}{k_F}   \quad.
\end{array}
\right.
\label{C.12}
\ee

\item for $\; 0 < \alpha <  d-1\;$
\be
\left\langle\;\psi^\ast_{\om1}(\zeta)\;\psi_{\om1}(0)\;\right\rangle^\|_n
 \quad \sim \quad  (\frac{-i}{2\pi})\; 
e^{i\arg(i\xi_0+\frac{\xi_\|}{v_{Fn}})}\; \frac{1}{|\xi_\||} \quad,\quad
\mbox{for} \; |\xi_\|| \gg
\frac{1}{k_F} \quad.
\label{C.13}
\ee
\end{itemize}

\noindent {\bf Remark} : In eqs.(\ref{C.12}) and (\ref{C.13}), we have omitted subleading
terms in an expansion in $\frac{1}{\lambda_n} .$

\end{document}